\documentclass[reprint,twocolumn,superscriptaddress,showpacs,nofootinbib,notitlepage,preprintnumbers,aps,prd,floatfix]{revtex4-2}
\usepackage[utf8]{inputenc}
\usepackage{graphicx}
\usepackage{latexsym,amsmath,amssymb,lmodern,float,url}
\usepackage{qcircuit}
\usepackage{natbib}
\usepackage{color}
\usepackage{microtype}
\usepackage{bbold}
\usepackage{slashed}
\usepackage{multirow}
\usepackage{tikz}
\usepackage{xcolor}
\usepackage{mathrsfs}  
\usetikzlibrary{shapes}
\usepackage{adjustbox}
\usepackage{todonotes}
\usepackage{braket}

\usepackage[colorlinks=true,backref=false, linktocpage=true,
citecolor=blue,urlcolor=blue,linkcolor=blue,pdfpagemode=UseOutlines]{hyperref}
\usepackage{cleveref}

\hypersetup{%
 bookmarksnumbered=true,
 pdftitle = {},
 pdfsubject = {},
 pdfauthor = {},
 pdfkeywords = {}
}

\let\Re\undefined

\newcommand{\bi}{\mathbb{BI}}
\newcommand{\bo}{\mathbb{BO}}
\newcommand{\btt}{\mathbb{BT}}
\newcommand{\ds}{\Sigma(36\times3)}
\newcommand{\dsa}{\Sigma(72\times3)}
\newcommand{\dsb}{\Sigma(216\times3)}
\newcommand{\dsc}{\Sigma(360\times3)}

\DeclareMathOperator{\Tr}{Tr}
\DeclareMathOperator{\tr}{Tr}
\DeclareMathOperator{\Re}{Re}

\newcommand{\one}{\mathbf{1}}

\newcommand{\three}{\mathbf{3}}

\definecolor{vermillion}{RGB}{213,94,0}
\definecolor{cblue}{RGB}{0,114,178}
\definecolor{corange}{RGB}{230,159,0}
\definecolor{blgreen}{RGB}{0,158,115}
\definecolor{repurple}{RGB}{204,121,167}

\definecolor{mycolor}{RGB}{0,255,0}

\begin{document}
 \preprint{FERMILAB-PUB-24-0132-SQMS-T, TUM-HEP-1506/24}
 
\title{Primitive Quantum Gates for an \texorpdfstring{$SU(3)$}{SU(3)} Discrete Subgroup: \texorpdfstring{$\Sigma(36\times3)$}{S(36x3)}}
\author{Erik J. Gustafson}
\email{egustafson@usra.edu}
\affiliation{Superconducting and Quantum Materials System Center (SQMS), Batavia, Illinois, 60510, USA.}
\affiliation{Fermi National Accelerator Laboratory, Batavia,  Illinois, 60510, USA}
\affiliation{Quantum Artificial Intelligence Laboratory (QuAIL),
NASA Ames Research Center, Moffett Field, CA, 94035, USA}
\affiliation{USRA Research Institute for Advanced Computer Science (RIACS), Mountain View, CA, 94043, USA}
\author{Yao Ji}
\email{yao.ji@tum.de}
\affiliation{Physik Department T31, James-Franck-Stra\ss e 1, Technische Universit\"at M\"unchen, D-85748 Garching, Germany}
\author{Henry Lamm}
\email{hlamm@fnal.gov}
\affiliation{Superconducting and Quantum Materials System Center (SQMS), Batavia, Illinois, 60510, USA.}
\affiliation{Fermi National Accelerator Laboratory, Batavia, Illinois, 60510, USA}
\author{Edison M. Murairi}
\email{emm712@gwu.edu}
\affiliation{Department of Physics, The George Washington University, Washington, DC  20052}
\author{Sebastian Osorio Perez}
\affiliation{Fermi National Accelerator Laboratory, Batavia, Illinois, 60510, USA}
\author{Shuchen Zhu}
\email{sz424@georgetown.edu} 
\affiliation{Department of Computer Science, Georgetown University, Washington, DC 20057, USA}
\date{\today}
\begin{abstract}
We construct the primitive gate set for the digital quantum simulation of the 108-element $\Sigma(36\times3)$ group. This is the first time a nonabelian crystal-like subgroup of $SU(3)$ has been constructed for quantum simulation. The gauge link registers and necessary primitives -- the inversion gate, the group multiplication gate, the trace gate, and the $\Sigma(36\times3)$ Fourier transform -- are presented for both an eight-qubit encoding and a heterogeneous three-qutrit plus two-qubit register. For the latter, a specialized compiler was developed for decomposing arbitrary unitaries onto this architecture. 
\end{abstract}
\maketitle
\section{Introduction}

Classical computers face significant challenges in simulating lattice gauge theories due to inherent exponentially large Hilbert spaces with the lattice volume. Monte Carlo simulations in Euclidean time are generally used to circumvent this problem. However, this approach also fails when we are interested in the real time dynamics of the system or in the properties of matter at finite density due to the sign problem~\cite{Aarts_2016,Philipsen:2007aa,forcr2010simulating,10.1143/PTP.110.615,Gattringer:2016kco,Alexandru:2018ddf,Alexandru:2020wrj,Troyer:2004ge}. 

Quantum computers provide a natural way of simulating lattice gauge theories. Yet, they are currently limited to a small number of qubits and circuit depths. Gauge theories contain bosonic degrees of freedom and have a continuous symmetry, e.g. Quantum chromodynamics (QCD) with $SU(3)$ local symmetry. Storing a faithful matrix representation of $SU(3)$ to double precision would require $\mathcal{O}(10^3)$ qubits per link -- far beyond accessibility to near-term quantum computers. Moreover, these qubits being noisy significantly limits the circuit depths that can be reliably performed on these devices. Therefore, studying lattice gauge theories with current and near-future quantum computers requires efficient digitization methods of gauge fields as well as optimized computational subroutines. Finally, it is important to note that a choice of digitization method affects the computation cost. 

To this end, several digitization methods have been proposed in the past decade to render the bosonic Hilbert space finite. Traditionally, most approaches have considered only qubit devices. However, due to recent demonstrations of qudit gates, there has been an increasing interest in qudit-based digitization methods~\cite{Gustafson:2021jtq,Gustafson:2021qbt,Popov:2023xft,kurkcuoglu2022quantum,Calajo:2024qrc,Gonzalez-Cuadra:2022hxt, illa2024qu8its, Zache:2023cfj}. 
One digitization method utilizes the representation (electric field) basis, and impose a cut off on the maximal representation~\cite{Unmuth-Yockey:2018ugm,Unmuth-Yockey:2018xak, Klco:2019evd, Farrell:2023fgd, Farrell:2024fit, Illa:2024kmf,Ciavarella:2021nmj, Bazavov:2015kka, Zhang:2018ufj,PhysRevD.99.114507,Bauer:2021gek,Grabowska:2022uos,Buser:2020uzs,Bhattacharya:2020gpm,Kavaki:2024ijd,Calajo:2024qrc,Murairi:2022zdg,Zohar:2015hwa,Zohar:2012xf,Zohar:2012ay,Zohar:2013zla}. Another recent proposal~\cite{Zache:2023dko,Zache:2023cfj}, the $q$-deformed formulation  renders a finite dimension by replacing the continuous gauge group with a so-called quantum group~\cite{Biedenharn1995}. 
 Moreover, the loop-string-hadron formulation explicitly enforces gauge invariance on the Hilbert space \cite{davoudi2024scattering,Raychowdhury:2018osk,Kadam:2023gli,Davoudi:2020yln,Mathew:2022nep} before truncation. 
 Each of the above methods is extendable to the full gauge group, i.e. it has an infinite-dimensional limit. 
Other methods begin with different formulations or perform different approximations exist such as light-front quantization~\cite{Kreshchuk:2020dla,Kreshchuk:2020aiq,Kreshchuk:2020kcz}, conformal truncation~\cite{Liu:2020eoa}, strong-coupling and large-$N_c$ expansions~\cite{Fromm:2023bit,Ciavarella:2024fzw}.

Another approach is to try and formulate a finite-dimensional Hilbert space theory with continuous local gauge symmetry which is in the same universality class as the original theory.  For example, the author of Ref.~\cite{HORN1981149} constructed a $SU(2)$ gauge theory where each link Hilbert space is five-dimensional. A generalization to $SU(N)$, however, was not obtained due to a spurious $U(1)$ symmetry. Later, different finite-dimensional formulations were found for $SU(2)$ gauge theories~\cite{ORLAND1990647}, the smallest of which being four-dimensional. Recently, a method inspired from non-commutative geometry was used to construct a $SU(2)$ gauge theory in 16-dimensional Hilbert space on each link as well as a generalization to $U(N)$ gauge theory~\cite{Alexandru:2023qzd}. Another finite-dimensional digitization known as quantum link models uses an ancillary dimension to store a quantum state~\cite{Brower:1997ha}. This method can be extended to an arbitrary $SU(N)$, and has been further investigated in Refs.~\cite{Singh:2019jog,Singh:2019uwd,Wiese:2014rla,Brower:1997ha,Brower:2020huh,Mathis:2020fuo,Halimeh:2020xfd, budde2024quantum, osborne2024quantum, Osborne:2023rzx,Luo:2019vmi}. Although this approach may greatly simplify the cost of digitization, establishing the universality class is non-trivial~\cite{Alexandru:2023qzd,ORLAND1990647,PhysRevD.105.114509}. 

\begin{figure}[!th]
\centering
    \includegraphics[width=0.9\linewidth]{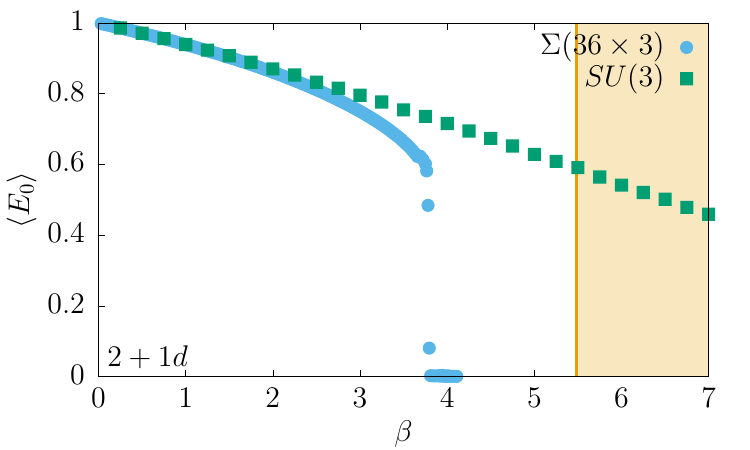}
    \includegraphics[width=0.9\linewidth]{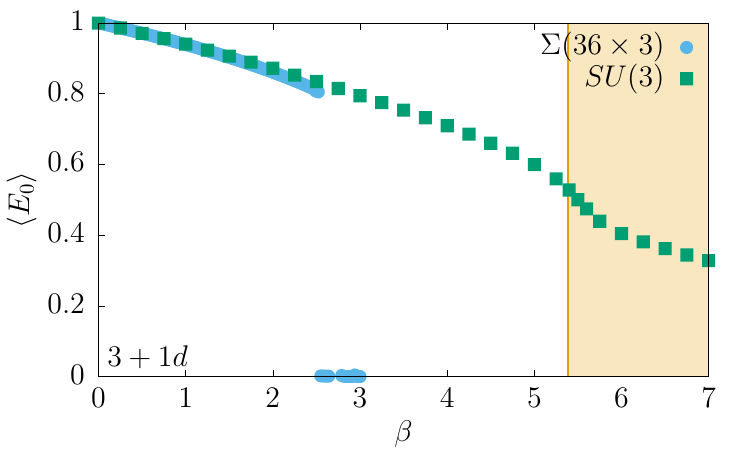}
    \caption{\label{fig:freezing}Euclidean calculations of lattice energy density $\langle E_0\rangle$ of $\ds$ as measured by the expectation value of the plaquette as a function of Wilson coupling $\beta$  on $8^d$ lattices for (top) $(2+1)$-dimensions (bottom) $(3+1)$-dimensions. The shaded region indicates $\beta\geq \beta_f$.} 
    \label{fig:lattice-energy}
\end{figure}

Another promising approach to digitization is the discrete subgroup approximation~\cite{Bender:2018rdp,Hackett:2018cel,Alexandru:2019nsa,Yamamoto:2020eqi,Ji:2020kjk,Haase:2020kaj,Carena:2021ltu,Armon:2021uqr,Gonzalez-Cuadra:2022hxt,Charles:2023zbl,Irmejs:2022gwv,Gustafson:2020yfe, Bender:2018rdp,Hackett:2018cel,Alexandru:2019nsa,Yamamoto:2020eqi,Ji:2020kjk,Haase:2020kaj,Carena:2021ltu,Armon:2021uqr,Gonzalez-Cuadra:2022hxt,Charles:2023zbl,Irmejs:2022gwv, Hartung:2022hoz,Carena:2024dzu,Zohar:2014qma,Zohar:2016iic, davoudi2024scattering}. This method was explored early on in Euclidean lattice field theory to reduce computational resources in Monte Carlo simulations of gauge theories. Replacing $U(1)$ by $\mathbb{Z}_N$ was considered in~\cite{Creutz:1979zg,Creutz:1982dn}. Extensions to the crystal-like subgroups of $SU(N)$ were made in Refs.~\cite{Bhanot:1981xp,Petcher:1980cq,Bhanot:1981pj,Hackett:2018cel,Alexandru:2019nsa,Ji:2020kjk,Ji:2022qvr,Alexandru:2021jpm,Carena:2022hpz}, including with fermions~\cite{Weingarten:1980hx,Weingarten:1981jy}. Theoretical studies revealed that the discrete subgroup approximation corresponds to continuous groups broken by a Higgs mechanism~\cite{Kogut:1980qb,romers2007discrete,Fradkin:1978dv,Harlow:2018tng,Horn:1979fy}. We additionally provide properties of certain $SU(3)$ discrete groups in Tab. \ref{tab:groupcomp}. On the lattice, this causes the discrete subgroup to poorly approximate the continuous group below a \emph{freezeout} lattice spacing $a_f$ (or beyond a coupling $\beta_f$), see Fig.~\ref{fig:lattice-energy}.

The discrete group approximation has several significant advantages over many other methods discussed above. It is a finite mapping of group elements to integers that preserves a group structure; therefore it avoids any need for expensive fixed- or floating-point quantum arithmetic. The inherent discrete gauge structure further allows for coupling the gauge redundancy to quantum error correction \cite{rajput2021quantum,Carena:2024dzu}. Additionally, while other method in principle need to increase both circuit depth and qubit count to improve the accuracy of the Hilbert space truncation, the discrete group approximation only needs to include additional terms into the Hamiltonian \cite{Carena:2022kpg,Alexandru:2021jpm}

\begin{table}
\caption{\label{tab:groupcomp} Parameters of a crystal-like subgroups of $SU(3)$. $\Delta S$ is the gap between $\mathbb{1}$ and the nearest neighbors to it. $\mathcal{N}$ is number of group elements that neighbor the $\mathbb{1}$.}
\begin{center}
\begin{tabular}
{c| c c c c c}
\hline\hline
$G$ & $\Delta S$ & $\mathcal{N}$ & $\beta^{2+1d}_f$ & $\beta^{3+1d}_f$ \\
\hline
$\ds$&$\frac{2}{3}$&18&3.78(2)&2.52(3)\\
$\dsa$&$\frac{2}{3}$&54&---&3.2(1)\footnote{\label{bhanot}from \cite{Bhanot:1981xp}}\\
$\dsb$&$1-\frac{1}{3}\left(\cos \frac{\pi}{9}+\cos \frac{2\pi}{9}\right)$&24&---&3.43(2)\footref{bhanot}\\
$\dsc$&$\frac{5-\sqrt{5}}{6}$&72&---&3.935(5)\footnote{from \cite{Alexandru:2019nsa}}\\
\hline\hline
\end{tabular}
\end{center}
\end{table}

In this work, we consider the smallest crystal-like subgroup of a $SU(3)$ with a $\mathbb{Z}_3$ center -- $\ds$ which has 108 elements.  These elements can be naturally encoded into a register consider of 8 qubits or 3 qutrits \& 2 qubits. A number of smaller nonabelian subgroups of $SU(2)$ have been considered previously: the $2N$-element dihedral groups $D_N$~\cite{Bender:2018rdp,Lamm:2019bik,Alam:2021uuq,Fromm:2022vaj}, the 8-element $\mathbb{Q}_8$~\cite{Gonzalez-Cuadra:2022hxt}, the crystal-like 24-element $\btt$~\cite{Gustafson:2022xdt}, and the crystal-like 48 element $\bo$~\cite{Gustafson:2023kvd}.  From Fig.~\ref{fig:lattice-energy}, we observe that freezeout occurs far before the scaling regime.  This implies that the Kogut-Susskind Hamiltonian (which can be derived from the Wilson action) is insufficient for $\ds$ to approximate $SU(3)$, but classical calculations suggest with modified or improved Hamiltonians $H_{I}$ may prove sufficient for some groups~\cite{Alexandru:2019nsa,Ji:2020kjk,Ji:2022qvr,Alexandru:2021jpm}.
 
This paper is organized as follows. In Sec.~\ref{sec:group}, the group theory needed for $\Sigma(36\times3)$ are summarized and the digitization scheme is presented. Sec.~\ref{sec:gates} demonstrates the quantum circuits for the four primitive gates required for implementing the group operations: the inversion gate, the multiplication gate, the trace gate, and the Fourier transform gate. Using these gates, Sec.~\ref{sec:resources} presents a resource estimates for simulating $3+1d$ $SU(3)$.  We conclude and discuss future work in Sec.~\ref{sec:conclusion}.

\section{Properties of \texorpdfstring{$\Sigma(36\times3)$}{S(36x3)}}
\label{sec:group}
$\Sigma(36\times 3)$ is a discrete subgroup of SU(3) with 108 elements. The group elements, $g$, of $\Sigma(36\times3)$ can be written in the following ordered product or otherwise known as strong generating set.
That is, all the group elements, $g$, can be enumerated as a product of left or right transversals such that
\begin{equation}
    \label{eq:orderedproduct}
    g = \omega^p C^q E^r V^{2s + t},
\end{equation}
where $0\leq p,~q,~r\leq 2$ and $0\leq s,~t\leq 1$. This indicates that either 8 qubits (2 each for $p$ , $q$, and $r$, and one each for $s$ and $t$) or 3 qutrits ($p$, $q$, and $r$) and 2 qubits ($s$ and $t$) will be required to store the group register. 
Because the indices $p,~q,$ and $r$ take on values between 0 and 2, there exists an ambiguity in mapping the three-level states to a pair of two-level systems. 
We use the mapping $|0\rangle_3=|00\rangle_2$, $|1\rangle_3=|01\rangle_2$, and $|2\rangle _3= |10\rangle_2$ with the $|11\rangle_2$ state being forbidden. 
Throughout this work, we will use $|\rangle_3$ to denote a three-level state and $|\rangle_2$ to denote a two-level state when there is the possibility of ambiguity. In this way, the index $p$ is decomposed in binary as $p = p_0 + 2p_1$ and encoded as the state $|p\rangle_3 = |p_1p_0\rangle_2$. This process is done similarly for $q$ and $r$.

The strong generating set shown in Eq.~\ref{eq:orderedproduct} explicitly builds the presentation of the group from subgroups. In this way primitive gates for smaller discrete groups can be used as building blocks to construct efficient primitive gates of larger groups~\cite{Pueschel:1998zzo, Gustafson:2022xdt, Gustafson:2023kvd}. 
In the case of $\ds$, the subgroups of interest are as follows:
$\omega^p$ generates the subgroup $\mathbb{Z}_3$; $\omega^p C^{q}$ generates the subgroup $\mathbb{Z}_3\times\mathbb{Z}_3$; $\omega^p C^q E^r$ generates the subgroup $\Delta(27)$; $\omega^p C^q E^r V^{2s}$ generates $\Delta(54)$.
Detailed information regarding these subgroups can be found in Ref. \cite{grimus2010principal}.

As we proceed with constructing primitive gates (see Sec.~\ref{sec:primitive-gates}),the following ``reordering" relations are useful:
\begin{align}
\label{eq:basicreorderingrules}
    EC = & \omega CE,~VC=EV,~\text{and}~VE=C^2V.
\end{align}
One can extend the relations above to derive the generalized reordering relations:
\begin{align}
    V^{2s + t}C^q = & C^{(1+s)q(1 - t)}E^{t(1+s)q}V^{2s + t},\notag\\
    E^r C^q = & \omega^{rq}C^qE^r,\notag\\
    V^{2s + t}E^r =& C^{t(s+1)(3-r)}E^{(1-t)r(1+s)}V^{2s + t}.
    \label{eq:group-structure}
\end{align}

It is useful to have the irreducible representations, irreps, of $\ds$ for deriving a quantum Fourier transformation (see Sec.~\ref{sec:primitive-gates}).
This group has 14 irreducible representations (irreps). There are four one-dimensional (1d) irreps, eight three-dimensional irreps (3d), and two four-dimensional (4d) irreps. 
The 1d irreps are:
\begin{equation}
    \label{eq:1dirrep}
    \rho^{(1)}_a(g) = i^{a(2s + t)},~0\leq a\leq3.
\end{equation}
The eight 3d irreps can be written as
\begin{equation}
 \rho^{(3)}_{a, b}(g) = (-1)^{abt}\omega^{(1+b)p}C^{(1+b)q}E^{r}(i^aV)^{2s + (-1)^bt},
 \end{equation}
where $0\leq a\leq 3$ and $0\leq b \leq 1$ and the matrices $\omega$, $C$, $E$, and $V$ are given by 
\begin{align}
    \omega = e^{2\pi i / 3}&,~C=\text{Diag}(1,\omega,\omega^2),\notag\\
    ~E =  \begin{pmatrix}
        0&1&0\\
        0&0&1\\
        1&0&0
        \end{pmatrix}&\text{, and}~V=\frac{1}{\sqrt{3}i}\begin{pmatrix}
        1&1&1\\
        1&\omega&\omega^2\\
        1&\omega^2&\omega
        \end{pmatrix}.
\end{align}
The irrep $\rho_{0, 0}^{(3)}$ corresponds to the faithful irrep that resembles the fundamental irrep of $SU(3)$.
The 4d irreps are given by Eq.~(\ref{eq:orderedproduct}) with 
\begin{equation}
\begin{split}
    \rho^{(4)}_{b}(\omega) = & \mathbb{1},~~\rho^{(4)}_b(C) = \text{Diag}(\omega^b, \omega,\omega^{2b}, \omega^2),\\
    \rho^{(4)}_{b}(E) = & \text{Diag}(\omega, \omega^{2b}, \omega^2, \omega^b),\\
    \rho^{(4)}_b(V) = & \begin{pmatrix}
        0 & 1 & 0 & 0\\
        0 & 0 & 1 & 0\\
        0 & 0 & 0 & 1\\
        1 & 0 & 0 & 0\\
    \end{pmatrix}.
\end{split}
\end{equation}
In addition for conciseness we provide the character table from Ref \cite{grimus2010principal} in Tab. \ref{tab:charactertable}, which will be useful in constructing the trace gate.

The Hamiltonians we particularly are targeting are the pure gauge theory Kogut Susskind Hamiltonian,
\begin{align}
    H_{KS} = & \sum_{\square}Tr(g_1 g_2 g_3^{\dagger} g_4^{\dagger})|g_1 g_2 g_3 g_4\rangle \langle g_1 g_2 g_3 g_4|\notag\\
    & + \sum_{l} \sum_{g_l, h_l} e^{\beta Tr(g_l h_l^{\dagger})}|g_l\rangle \langle h_l|, 
\end{align}
where $\sum_{\square}$ indicates a sum over all of the plaquettes with $g_1,~...,g_4$ elements of the plaquettes. Additionally, the second term is the kinetic term where the sum over $l$ is a sum over all links. There is generally a freedom in the construction of the electric term. In Appendix~\ref{app:electric-term}, we provide a straightforward construction based on the procedure outlined in Ref.~\cite{PhysRevD.107.114513}. 
The second Hamiltonian we consider is the improved Hamiltonian, $H_I$, which was highlighted in Ref. \cite{Carena:2022kpg}. This includes terms with the six link rectangles and an extended electric field operator. The desire to consider improved Hamiltonians comes from the fact that there are both reduced lattice spacing errors, i.e., the discretization artifacts are moved to $\mathcal{O}(a^4)$, and the $\beta_f$ is a larger value. 

\begin{table}
\caption{Character table of $\Sigma(36\times3)$ with $\omega=e^{2\pi i/3}$~\cite{grimus2010principal}. Size indicates the number of elements in the group while Ord. (order) indicates the number of times the operator can be multiplied before yielding the identity.
    \label{tab:charactertable}}
\begin{tabular}{c||ccc|c|c|ccc|ccc|ccc}
Size & 1 & 1 & 1 & 12 & 12 & 9 & 9 & 9 &
9 & 9 & 9 & 9 & 9 & 9 \\
Ord. & 1 & 3 & 3 & 3 & 3 & 2 & 6 & 6 & 4 & 12 & 12 & 4
& 12 & 12 \\ 
\hline \hline
$\one^{(0)}$ & $1$ & $1$ & $1$ & $1$ & $1$ & $1$ & $1$ & $1$ & $1$ & $1$ & $1$ & $1$ & $1$ & $1$ \\
$\one^{(1)}$ & $1$ & $1$ & $1$ & $1$ & $1$ 
& $\text{-}1$ & $\text{-}1$ & $\text{-}1$
& $i$ & $i$ & $i$ 
& $\text{-}i$ & $\text{-}i$ & $\text{-}i$ \\
$\one^{(2)}$ & $1$ & $1$ & $1$ & $1$ & $1$ 
& $1$ & $1$ & $1$
& $\text{-}1$ & $\text{-}1$ & $\text{-}1$ 
& $\text{-}1$ & $\text{-}1$ & $\text{-}1$ \\
$\one^{(3)}$ & $1$ & $1$ & $1$ & $1$ & $1$ 
& $\text{-}1$ & $\text{-}1$ & $\text{-}1$
& $\text{-}i$ & $\text{-}i$ & $\text{-}i$ & 
$i$ & $i$ & $i$ \\
\hline
$\three^{(0)}$ & $3$ & $3 \omega$ & $3 \omega^2$ &
$0$ & $0$ &
$\text{-}1$ & $\text{-}\omega$ & $\text{-}\omega^2$ &
$1$ & $\omega$ & $\omega^2$ &
$1$ & $\omega$ & $\omega^2$ \\
$\three^{(1)}$ & $3$ & $3 \omega$ & $3 \omega^2$ &
$0$ & $0$ &
$1$ & $\omega$ & $\omega^2$ &
$i$ & $i \omega$ & $i \omega^2$ &
$\text{-}i$ & $\text{-}i \omega$ & $\text{-}i \omega^2$ \\
$\three^{(2)}$ & $3$ & $3 \omega$ & $3 \omega^2$ &
$0$ & $0$ &
$\text{-}1$ & $\text{-}\omega$ & $\text{-}\omega^2$ &
$\text{-}1$ & $\text{-}\omega$ & $\text{-}\omega^2$ &
$\text{-}1$ & $\text{-}\omega$ & $\text{-}\omega^2$ \\
$\three^{(3)}$ & $3$ & $3 \omega$ & $3 \omega^2$ &
$0$ & $0$ &
$1$ & $\omega$ & $\omega^2$ &
$\text{-}i$ & $\text{-}i \omega$ & $\text{-}i \omega^2$ &
$i$ & $i \omega$ & $i \omega^2$ \\
$\three^{(0)*}$  & $3$ & $3 \omega^2$ & $3 \omega$ &
$0$ & $0$ &
$\text{-}1$ & $\text{-}\omega^2$ & $\text{-}\omega$ &
$1$ & $\omega^2$ & $\omega$ &
$1$ & $\omega^2$ & $\omega$ \\
$\three^{(1)^*}$ & $3$ & $3 \omega^2$ & $3 \omega$ &
$0$ & $0$ &
$1$ & $\omega^2$ & $\omega$ &
$\text{-}i$ & $\text{-}i \omega^2$ & $\text{-}i \omega$ &
$i$ & $i \omega^2$ & $i \omega$ \\
$\three^{(2)*}$  & $3$ & $3 \omega^2$ & $3 \omega$ &
$0$ & $0$ &
$\text{-}1$ & $\text{-}\omega^2$ & $\text{-}\omega$ &
$\text{-}1$ & $\text{-}\omega^2$ & $\text{-}\omega$ &
$\text{-}1$ & $\text{-}\omega^2$ & $\text{-}\omega$ \\
$\three^{(3)*}$ & $3$ & $3 \omega^2$ & $3 \omega$ &
$0$ & $0$ &
$1$ & $\omega^2$ & $\omega$ &
$i$ & $i \omega^2$ & $i \omega$ &
$\text{-}i$ & $\text{-}i \omega^2$ & $\text{-}i \omega$ \\
\hline
$\mathbf{4}$ & $4$ & $4$ & $4$ &
$1$ & $\text{-}2$ &
$0$ & $0$ & $0$ &
$0$ & $0$ & $0$ &
$0$ & $0$ & $0$ \\
$\mathbf{4}^\prime$ & $4$ & $4$ & $4$ &
$\text{-}2$ & $1$ &
$0$ & $0$ & $0$ &
$0$ & $0$ & $0$ &
$0$ & $0$ & $0$ \\
\end{tabular}
\end{table}

\section{Basic Gates \label{sec:basic-gates}}

In this work, we consider gate sets for both qubit and hybrid qubit-qutrit systems. Our qubit decompositions use the well-known fault tolerant Clifford + T gate set \cite{Chuang:1996hw}. 
This choice is informed by the expectation that quantum simulations for lattice gauge theories will ultimately require fault tolerance to achieve quantum advantage~\cite{Jordan:2011ci,Jordan:2011ne,Gustafson:2022xdt,Gustafson:2023kvd,Ciavarella:2021nmj}. Throughout, we adopt the notation $\oplus_m$ to mean addition mod $m$. 

For conciseness, we use a larger than necessary gate set which we will later decomposes in terms of T-gate to obtain resource costs. The single qubit gates used are the Pauli rotations, $R_\alpha(\theta) = e^{i\theta \alpha / 2}$, where $\alpha=X,Y,Z$. 
We also consider four entangling operations: SWAP, CNOT, and  multicontrolled C$^n$NOT, and the controlled SWAP (CSWAP). The two-qubit operations can be written as
\begin{align}
&\text{SWAP}~|a\rangle |b\rangle =  |b\rangle |a\rangle,\\
&\text{CNOT}~|a\rangle |b\rangle =  |a\rangle |b\oplus_2 a\rangle.
\end{align}
while the multicontrolled generalizations are
\begin{equation}
\text{C}^n\text{NOT}~ \Big(\prod_{n}|a_n\rangle\Big)|b\rangle = \Big(\prod_{n}|a_n\rangle\Big) |b \oplus_2 \prod_n a_n\rangle,
\end{equation}
\begin{equation}
\text{CSWAP}|a\rangle |b\rangle  |c\rangle =|a\rangle  |b(1\oplus_2 a) \oplus_2 ac\rangle  |c(1\oplus_2 a) \oplus_2 ba\rangle.
\end{equation}

The hybrid encoding uses a more novel set of single, double, and triple qudit gates. The single-qudit two level rotations we consider are denoted by $R^{\alpha}_{b, c}(\theta)$, where $\alpha=\lbrace X,~Y,~Z\rbrace$ and indicate a Pauli-style rotation between levels $b$ and $c$. The subscripts will be omitted to indicate that the operation is performed on a qubit rather than a qutrit state. Additionally, we account for the primitive two-qudit gate,
\begin{equation}
\label{eq:qutritcnot}
    C^a_b X^{c}_{d,e}~,
\end{equation}
which corresponds to the CNOT operation controlled on state $b$ of qubit $a$, and targets qubit $c$ with an $X$ operation between the levels $d$ and $e$. We also for conciseness consider the $\textsc{CSum}$ gate:
\begin{equation}
    \textsc{CSum}^{a, b}|i\rangle_a|j\rangle_b = |i\rangle_a |i \oplus_3 j\rangle_b.
\end{equation}
which is a controlled operation on qubit or qutrit $a$ and targets qutrit $b$. It can be verified that that the $\textsc{CSum}$ (see e.g. Ref.~\cite{Gustafson:2021qbt, Di:2011cvl}) gate is related to the $C^{a}_{b}X^{c}_{d, e}$ gates by
 \begin{align}
     \textsc{CSum}^{a, b} &= \left(C^a_1 X^b_{0,1} C^a_1 X^b_{1,2}\right) \left(C^a_2 X^b_{1,2} C^a_2 X^b_{0,1}\right).
 \end{align}
 
Finally, we consider multi-controlled versions of both of these gates. The gate 
$C^a_bC^c_dX^e_{f,g}$ corresponds to multicontrolled generalization of Eq. (\ref{eq:qutritcnot}). The second multiqudit gate is the \textsc{CCSum} which acts as follows
\begin{equation}
    \textsc{CCSum}^{a,b,c}|i\rangle_a|j\rangle_b|k\rangle_c = |i\rangle_a|j\rangle_b|k\oplus_3 ij\rangle_c.
\end{equation}

 A unique artifact of this choice of quantum gates is that one can decompose multi-controlled gates using the traditional Toffoli staircase  decomposition \cite{2019arXiv190401671B,Chuang:1996hw,PhysRevA.52.3457}.  
 
\section{Primitive Gates \label{sec:primitive-gates}}
\label{sec:gates}
We present the primitive gates for a pure gauge theory in the following subsections using the methods developed in previous papers on the binary tetrahedral, $\mathbb{BT}$, and binary octahedral, $\mathbb{BO}$ groups~\cite{Lamm:2019bik,Alam:2021uuq,Gustafson:2022xdt}. Using this formulation confers at least two benefits: first, it is possible to design algorithms in a theory- and hardware-agnostic way; second, the circuit optimization is split into smaller, more manageable pieces. This construction begins with defining for a finite group $G$ a $G$-register by identifying each group element with a computational basis state $\ket{g}$. Then, Ref.~\cite{Lamm:2019bik} showed that Hamiltonian time evolution can be performed using a set of primitive gates. These primitive gates are: inversion $\mathfrak U_{-1}$, multiplication $\mathfrak U_{\times}$, trace $\mathfrak U_{\rm Tr}$, and Fourier transform $\mathfrak U_{F}$~\cite{Lamm:2019bik}.

The inversion gate, $\mathfrak U_{-1}$, is a single register gate that takes a group element to its inverse:
\begin{equation}
\mathfrak U_{-1} \ket{g} = \ket{g^{-1}}\text.
\end{equation}

The group multiplication gate acts on two $G-$registers. It takes the target $G-$register and changes the state to the left-product with the control $G-$register:
\begin{equation}
    \mathfrak U_{\times} \ket{g}\ket{h} = \ket{g} \ket{gh}.
\end{equation}
Left multiplication is sufficient for a minimal set as right multiplication can be implemented using two applications of $\mathfrak U_{-1}$ and $\mathfrak U_{\times}$, albeit optimal algorithms may take advantage of an explicit construction~\cite{Carena:2022kpg}.

The trace of products of group elements appears in lattice Hamiltonians.  We can implement these terms by combining $\mathfrak U _{\times}$ with a single-register trace gate:
\begin{equation}
\mathfrak U_{\Tr}(\theta) \left|g\right> = e^{i \theta \Re\Tr g} \left|g\right>.
\label{eqn:trace-gate}
\end{equation}

\begin{figure}
    \centering
    \includegraphics[width=\linewidth]{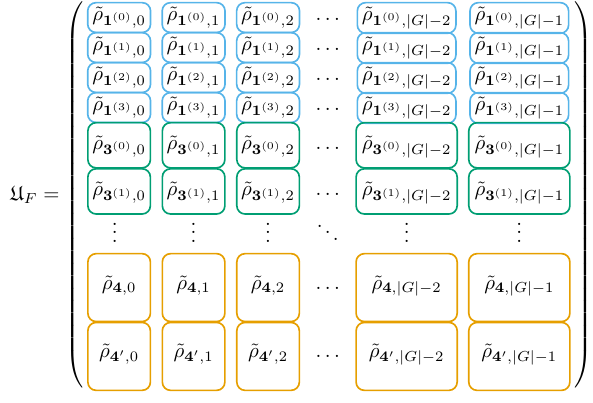}
    \caption{Construction of $\mathfrak U_{F}$ from Eq.~(\ref{eqn:Fourier-group}) using column vectors  $\tilde\rho_{i,j}=\sqrt{d_\rho/|G|}\rho_{i,j}$ where $\rho_{i,j}=\rho_i(g_j)$.}
    \label{fig:qft_cartoon}
\end{figure}

The next gate required is the group Fourier transform $\mathfrak U_F$. The Fourier transform of a finite $G$ is defined as
\begin{eqnarray}
\hat{f}(\rho) = \sqrt{\frac{d_{\rho}}{|G|}} \sum_{g \in G} f(g) \rho(g),
\label{eqn:Fourier-group}
\end{eqnarray}
where $\vert G \vert$ is the size of the group, $d_{\rho}$ is the dimensionality of the representation $\rho$, and $f$ is a function over $G$. The gate that performs this acts on a single $G$-register with some amplitudes $f(g)$ which rotate it into the Fourier basis:
\begin{equation}
\label{eq:uft}
\mathfrak U_F \sum_{g \in G} f(g)\left|g \right>
=
\sum_{\rho \in \hat G} \hat f(\rho)_{ij} \left|\rho,i,j\right>.
\end{equation}
The second sum is taken over $\rho$, the irreducible representations of $G$; $\hat f$ denotes the Fourier transform of $f$. After performing the Fourier transform, the register is denoted as a $\hat G$-register to indicate the change of basis. A schematic example of this gate is show in Fig.~\ref{fig:qft_cartoon}.
A related final gate is $\mathfrak{U}_{Ph}$ which induces the phases corrresponding to the eigenvalue sof the kinetic term of the Hamiltonian. 

While it is possible to pass the matrix constructed in Fig.~\ref{fig:qft_cartoon} into a transpiler, more efficient methods for constructing these operators exist \cite{ffttocome,Pueschel:1998zzo,Hoyer:1997qc,10.1145/258533.258548,KAWANO2016219,moore2006generic}. While method varies in their actualization, the underlying spirit is the same as for the discrete quantum Fourier transformation. The principle method involves building the quantum Fourier transformation up through a series of subgroups. In \cite{moore2006generic}, it was shown that instead of the exponential $\mathcal{O}(4^n)$ scaling for traditional transpilation, the quantum Fourier transform scales like $\mathcal{O}(\text{polylog}(|2^n|))$, where polylog indicates a polynomial of logarithms. 

In the rest of this paper, we will construct each of these primitive gates, and evaluate the overall cost. For each gate, we will start with a pure qubit system. Then, we will consider a register with three qutrits and two qubits as suggested by the group presentation in Eq.~\ref{eq:orderedproduct}. 

\subsection{Inversion Gate \label{sec:inverse-gate}}

\begin{figure*}
\includegraphics[width=0.85\linewidth]{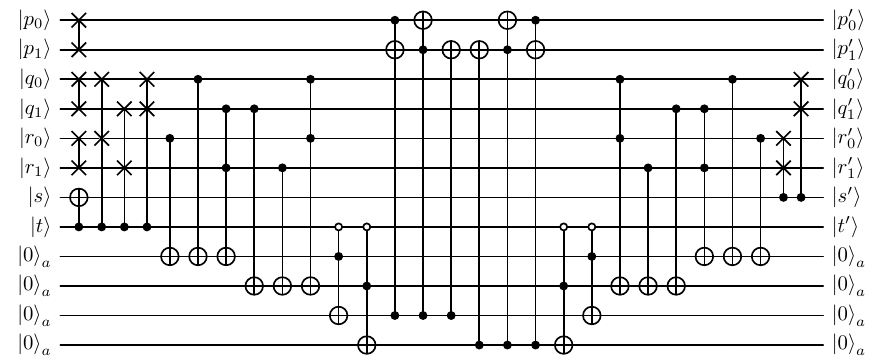}
    \caption{T-gate optimized version of $\mathfrak{U}_{-1}$ for $\Sigma(36\times3)$. The letter indicates the generator and subscript indicates the qubit in the generator register. This implementation requires 119 T-gates and 4 ancilla.} 
    \label{fig:s108inversion3}
\end{figure*}

\begin{figure}
\includegraphics[width=0.7\linewidth]{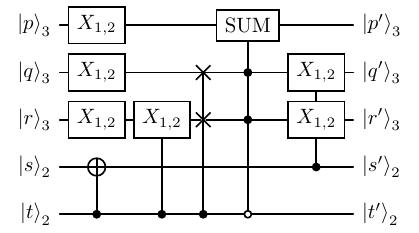}    \caption{$\mathfrak{U}_{-1}$, for $\Sigma(36\times3)$ using qutrit-qubit encoding.}
    \label{fig:quditinversion}
\end{figure}

For the construction of $\mathfrak{U}_{-1}$, we first write the inverse of the group element $g$ as
\begin{equation}
    \label{eq:inverse1}
    g^{-1} = \omega^{2p} V^{t + 2t} V^{2s} E^{2r} C^{2q} = \omega^{p'}C^{q'}E^{r'}V^{2s' + t'}.
\end{equation}
where the permutations rules are found to be
\begin{align}
\label{eq:inversionrules}
    p' = & 2p \oplus_3 qr \oplus_3 2qrt,\notag\\
    q' = & 2(q \oplus_3 qs \oplus_3 2qt \oplus_3 rt \oplus_3 2qst \oplus_3 rst),\notag\\
    r' = & 2(r \oplus_3 rs \oplus_3 2qt \oplus_3 2rt \oplus_3 2qst \oplus_3 2rst),\notag\\
    s' = & s \oplus_2 t,\notag\\
    t' = & t.
\end{align}

A detailed derivation of the permutation rules and the associated $\mathfrak{U}_{-1}$ is found in App.~\ref{app:Alternate_Inversion} along with two other forms of $\mathfrak{U}_{-1}$ which use fewer ancilla.  The idealized qubit circuit is shown in Fig.~\ref{fig:s108inversion3} and requires 119 T-gates and 4 clean ancilla.\footnote{A \emph{clean} ancilla is in state $\ket{0}$. \emph{Dirty} ancillae have unknown states.} The qubit-qutrit hybrid encoding $\mathfrak{U}_{-1}$ is found in Fig.~\ref{fig:quditinversion}.

\subsection{Multiplication Gate \label{sec:multiplication-gate}}
\begin{figure*}
    \includegraphics[width=1\linewidth]{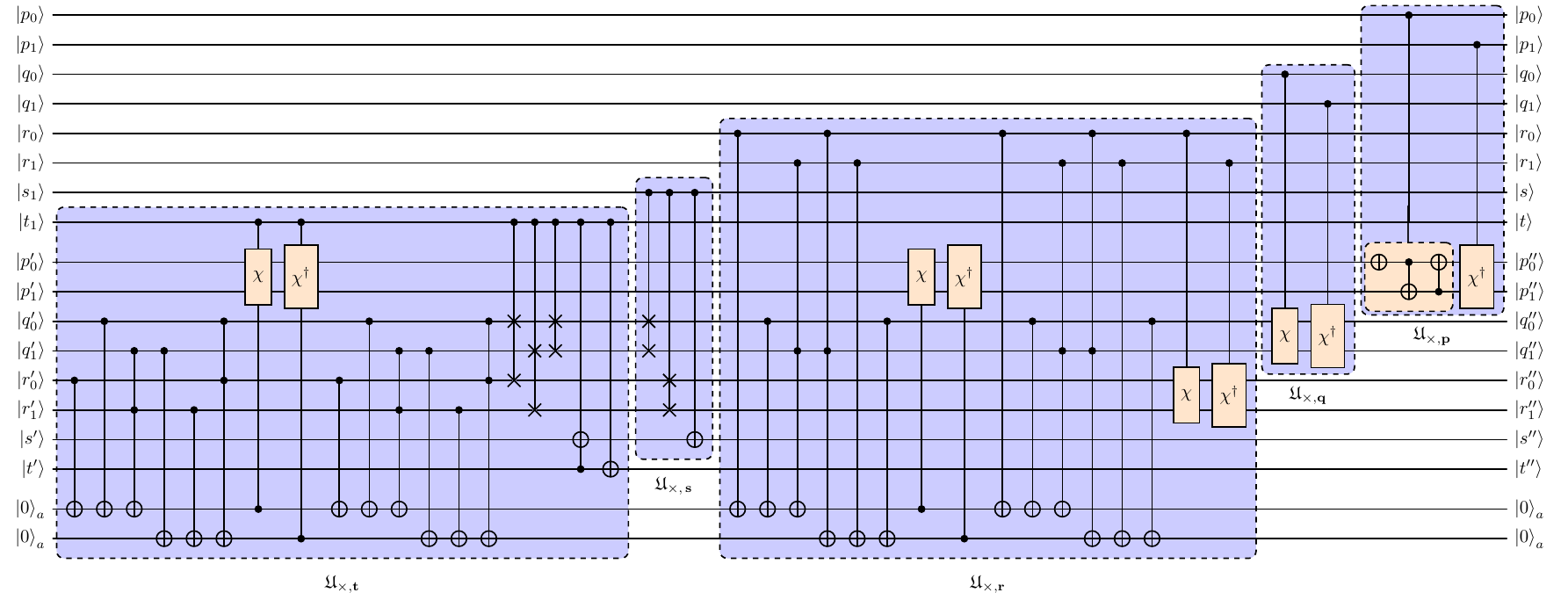}
    \caption{Qubit implementation of $\mathfrak U_{\times}$ using the permutation gate $\chi$ and its inverse (both shaded orange) using 2 ancillae and has a cost of 308 T-gates.}
    \label{fig:s108multiply}
\end{figure*}

\begin{figure}
    \centering
    \includegraphics[width=1\linewidth]{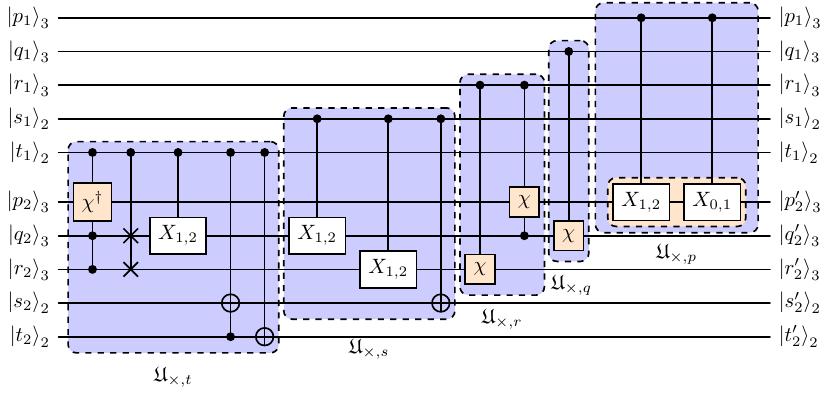}
    \caption{Qutrit+Qubit implementation of $\mathfrak U_{\times}$ using the permutation gate $\chi=X^{(1,2)}X^{(0, 1)}$ and its inverse (both shaded orange).}
    \label{fig:qutrit_mult}
\end{figure}

The multiplication gate $\mathfrak{U}_\times$ takes two $G$-registers storing two group elements $g = \omega^pC^qE^rV^{2s+t}$ and $h=\omega^{p'}C^{q'}E^{r'}V^{2s'+t'}$ and stores into the second register the group element $gh= \omega^{p''}C^{q''}E^{r''}V^{2s'' + t''}$.  Using the reordering relations of Eq.~(\ref{eq:group-structure}) one can derive that
\begin{align}
\label{eq:productrules}
    p'' = &p\oplus_3 p'\oplus_3 q'r\oplus_3 q'rs\oplus_3 2q'rt\oplus_3 2q'r't\notag\\
    &\oplus_3 2rr't\oplus_3 2q'rst\oplus_3 2rr'st,\notag\\
    q'' = &q+q'+sq'\oplus_3 2tq'\oplus_3 2stq'\oplus_3 2tr'\oplus_3 2str',\notag\\
    r'' = &r\oplus_3 tq'\oplus_3 stq'\oplus_3 r'\oplus_3 sr'\oplus_3 2tr'\oplus_3 2str',\notag\\
    s'' = &(s \oplus_2 s' \oplus_2 t t'),\notag\\
    t'' = &(t\oplus_2 t').
\end{align}
These rules are rather clunky and in order to write a systematic multiplication gate we decompose $\mathfrak{U}_{\times}$ into the following product,
\begin{equation}
    \mathfrak{U}_{\times} = \mathfrak{U}_{\times,\omega}\mathfrak{U}_{\times, C} \mathfrak{U}_{\times, E}\mathfrak{U}_{\times, V^2}\mathfrak{U}_{\times, V},
\end{equation}
where $\mathfrak{U}_{\times, O}$ indicate multiplying the state of the $O$ generator register from the $g_1$ register onto the $g_2$ register. We provide a detailed discussion of the breakdown of the rules in App. \ref{app:multiplication_discussion}. The breakdown of using this method and the product rules from Eq. (\ref{eq:productrules}) yields the circuits composed in Fig.~\ref{fig:s108multiply} and~\ref{fig:qutrit_mult} for the two encodings. 

\subsection{Trace Gate}
There are two principle methods one could derive $\mathfrak{U}_{Tr}$.
One method is to define a Hamiltonian of the form:
\begin{equation}
    \label{eq:traceham}
    \hat{H}_{Tr} = \sum_{g} Tr(g) |g \rangle \langle g|. 
\end{equation}
Then, the trace operator can be written as $\mathfrak{U}_{Tr}\left(\theta\right) = \exp{\left(-\mathrm{i}\,\theta\, H_{{\rm Tr}}\right)}$.
This operator corresponds to the phasing of the magnetic plaquette operator when $g$ corresponds to a closed Wilson loop.
To obtain the matrix form of $\hat{H}_{Tr}$, one may fix a basis $\left\{|g_1\rangle, {...}, |g_{|G|}\rangle\right\}$ where $|G| = 108$ is the order of the group. In this basis, $\Hat{H}_{Tr}$ is diagonal, and each diagonal entry is given by $H_{i, i} = Tr(g_i)$. 
 
To obtain a quantum circuit realizing $\mathfrak{U}_{{\rm Tr}}$, we use the tree-traversal algorithm developed in~\cite{Murairi:2022}\footnote{The codes accompanying the cited publication can be found on \href{https://github.com/emm71201/QC-Hamiltonian-Compilation}{this Github repository}.} which was shown to yield an exact circuit with asymptotically optimal CNOT gates count. 
The circuit obtained has 130  CNOT gates and 111 $R_z$ gates. Additional methods are also found in Ref. \cite{Hadfield_2021}. 
A second method for deriving this gate involves mapping group elements to their respective trace classes.
$\Sigma(36\times3)$ has 14 conjugacy classes that map to 10 different trace classes.
If we only require the real part of the trace then this grouping reduces the 10 trace classes to 7 trace classes.
The seven valid traces are ${\rm ReTr}(g) = \lbrace 3, -\frac{3}{2}, 0, \pm 1, \pm \frac{1}{2}\rbrace$, which we can be labelled using three bits $\left(v_0, v_1, v_2\right)$ as shown in Tab.~\ref{tab:real-trace-map}.

\begin{table}
\begin{tabular}{c|ccccccc}
\hline\hline
$\Re\tr(g)$ & $3$ & $-\frac{3}{2}$ & $-1$ & $-\frac{1}{2}$ & 0 & 1 & $\frac{1}{2}$ \\[0.3ex]\hline
$v_0$& 0 & 1 & 0 & 1& 0 & 0  &1 \\
$v_1$& 0 & 0 & 1 & 0& 1 & 0  &1 \\
$v_2$& 0 & 0 & 1 & 1& 0 & 1  &1
\\\hline\hline
\end{tabular}
\caption{Map $(p,q,r,s,t)\mapsto (v_0,v_1,v_2)$ via ${\rm{Re} Tr}(g)$.}
\label{tab:real-trace-map}
\end{table}

This map can be represented as three boolean functions, one for each of the variables $v_0$, $v_1$ and $v_2$. For quantum computation, it is convenient to write boolean functions in the so-called exclusive-or sum of products (ESOP) form~\cite{4145683,10.1145/988952.988971}. Then, the function can be mapped to a quantum circuit in a straightforward manner since each term in the ESOPs corresponds to a Toffoli gate. For each function, we start with their minterm forms~\cite{mishchenko2001fast}. Then, we use the exorcism algorithm to find a simpler ESOP expression for each of the three functions. After factorizations, we show the final expressions in the following equation:

\begin{align}
    v_0 = & \Bar{p}_0\left[1 \oplus_2 \Bar{t}\left(q_0r_1 \oplus_2 q_1r_0\right) \oplus_2 \Bar{s}\left(\Bar{q}_0\Bar{r}_0 \oplus_2 \Bar{q}_1\Bar{r}_1\right)\right]\notag\\
    &\oplus_2 \Bar{p}_1\left[t \oplus_2 s\left(\Bar{q}_1\Bar{r}_0 \oplus_2 \Bar{q}_0\Bar{r}_1\right)\right]\notag\\
    &\oplus_2 \Bar{p}_1\Bar{t}(q_0 \oplus_2 \Bar{q}_1)(r_0 \oplus_2 \Bar{r}_1),\notag\\
    v_1 = &\Bar{t} \oplus_2 (q_0 \oplus_2 \Bar{q}_1)(r_0 \oplus_2 \Bar{r}_1)\Bar{s}\Bar{t},\notag\\
    \label{eq:vmap}
    v_2 = &s \oplus \Bar{s}t.
\end{align}

$\mathfrak{U}_{Tr}$ can be decomposed as $\mathfrak{U}_{Tr}(\theta) = V\, \mathfrak{U}^{\prime}_{Tr} (\theta) V^\dagger $ where $V$ is a unitary operator realizing the map $\left(p,q,r,s,t\right) \mapsto \left(v_0, v_1, v_2\right)$. This yields $\mathfrak{U}^{\prime}_{Tr} (\theta)\equiv e^{\mathrm{i} \theta H^\prime }$, where 
\begin{align}
    H^\prime =& \frac{3}{16} III + \frac{3}{16} IIZ + \frac{3}{16} IZI + \frac{5}{16} IZZ\notag\\
    &+ \frac{1}{16} ZII + \frac{9}{16} ZIZ + \frac{9}{16} ZZI + \frac{15}{16} ZZZ.
\end{align}

\begin{figure}
    \centering
    \includegraphics[width=\linewidth]{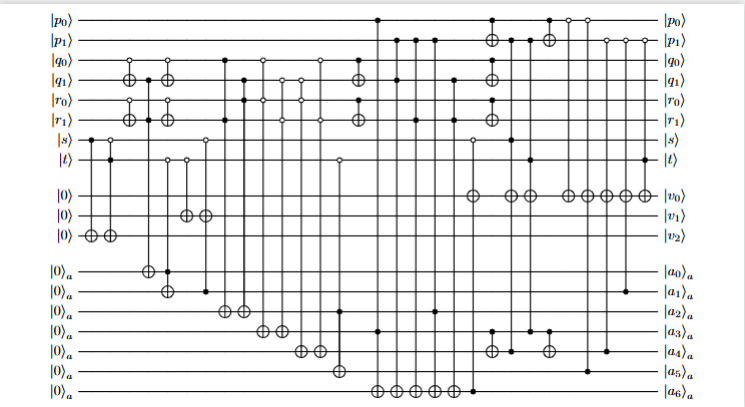}
    \caption{Quantum Circuit of the map $\left(p,q,r,s,t\right) \mapsto \left(v_0, v_1, v_2\right)$ from the group to the seven real trace classes ${\rm ReTr(g)} = \lbrace 3, -1.5, 0, \pm 1, \pm 0.5\rbrace$. This requires 15 Toffoli gates and thus 105 T gates}
    \label{fig:tr-gate-v}
\end{figure}

\begin{figure}
    \centering
    \includegraphics[width=\linewidth]{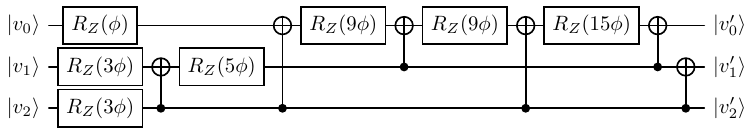}
    \caption{Quantum circuit of the operator $\mathfrak{U}^\prime_{Tr}(\theta)$ where we have set $\phi\equiv -\frac{\theta}{8}$.}
    \label{fig:trace-uprime}
\end{figure}

Figure~\ref{fig:tr-gate-v} shows the quantum circuit of the operator $V$ realizing the map $\left(p,q,r,s,t\right) \mapsto \left(v_0, v_1, v_2\right)$. Finally, the circuit of $\mathfrak{U_{Tr}}^\prime(\theta)$ is shown in Fig.~\ref{fig:trace-uprime}.


\begin{figure}
    \centering
    \includegraphics[width=1\linewidth]{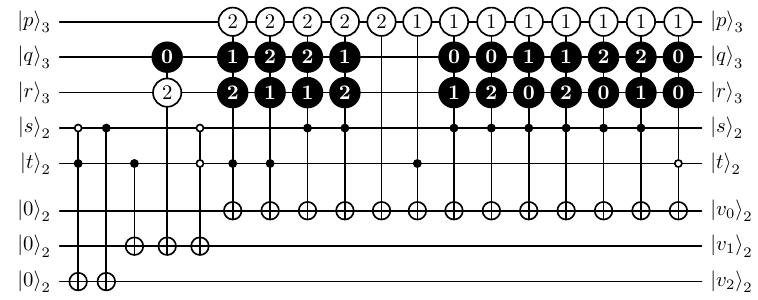}
    \caption{Quantum Circuit of the map $\left(p,q,r,s,t\right) \mapsto \left(v_0, v_1, v_2\right)$ from the group to the seven real trace classes ${\rm ReTr(g)} = \lbrace 3, -1.5, 0, \pm 1, \pm 0.5\rbrace$ for qutrit+qubit.  In analogy to the closed and open circle notation for control qubits, the black and white qutrit controls represent controlled-on or controlled-on-others e.g. a white 1 is a control qutrit which applies a gate if in the state $\ket{0}$ or $\ket{2}$.  }
    \label{fig:tr-gate-v-qutrit}
\end{figure}

\subsection{Fourier Transform Gate \label{sec:qft}}

The standard $n$-qubit quantum Fourier transform (QFT)~\cite{nielsen2000quantum} corresponds to the quantum version of the fast Fourier transform of $\mathbb{Z}_{2^n}$. 
Quantum Fourier transforms, $\mathfrak U_{QFT}$, over some nonabelian groups are known~\cite{Hoyer:1997qc,10.1145/258533.258548,Pueschel:1998zzo,moore2006generic,Alam:2021uuq}. However, for all the crystal-like subgroups of interest to high energy physics $\mathfrak U_{QFT}$ is currently unknown~\cite{childs2010quantum} and there is not a clear algorithmic way to construct $\mathfrak U_{QFT}$ in general.  Therefore, we instead construct a suboptimal $\mathfrak U_{F}$ from Eq.~(\ref{eqn:Fourier-group}) using the irreps of Sec.~\ref{sec:group}. The structure of $\mathfrak U_{F}$ is ordered as follows. The columns index $|g\rangle$ from $\ket{0}$ to $\ket{256}$ according to Eq.~(\ref{eq:orderedproduct}).  We then index the irreducible representation $\rho_i$ ordered sequentially from $i=1$ to $i=14$.

Since $\ds$ has 108 elements, on a qubit device $\mathfrak U_{F}$ must be embedded into a larger $2^8\times2^8=256\times 256$ matrix\footnote{An explicit construction of the matrix $\mathfrak U_{F}$ is found in the supplementary information}. While the matrix was then passed to the \textsc{Qiskit} v0.43.1 transpiler, and an optimized version of $\mathfrak U_{F}$ needed 30956 CNOTs, 2666 $R_X$, 32806 $R_Y$, and 55234 $R_Z$ gates; the Fourier gate is the most expensive qubit primitive.  As will be discussed in Sec.~\ref{sec:resources}, $\mathfrak U_F$ dominates the total simulation costs and future work should be devoted to finding a $\ds$ $\mathfrak U_{QFT}$ .

For the hybrid qubit-qutrit implementation, $\mathfrak U_{F}$ is of dimensions $108 \times 108$. To obtain a quantum circuit, we built a qubit-qutrit compiler, see Ref.~\cite{hcompiler}. The outline of the compilation is discussed in Appendix~\ref{app:qubit-qutrit-compiler}. The compiler uses the gates discussed in Sec.~\ref{sec:basic-gates}, and Table~\ref{tab:qft-hybrid-cost} shows its resulting gates count. The final component is a phasing corresponding to the gauge kinetic term. This involves decomposing the diagonal phasing operation into a sum of Pauli matrices or equivalents for qudit based systems. For the pure qubit based system this decomposition involves 256 $R_z$ rotations and 254 CNOT gates. The gate cost on a mixed qubit-qutrit device is shown in Table~\ref{tab:electric-gate-cost}.

\begin{table}
\begin{tabular}{l | c c c c }
\hline\hline
Basic Gate &$\mathfrak{U}_{-1}$ &$\mathcal{U}_{\times}$&$\mathcal{U}_{Tr}$&$\mathcal{U}_{F}$   \\
\hline

$R_\alpha$         &0 &0 &7& 92,568\\
$R^{\alpha}_{b,c}$ &3 &0 &0& 35,310\\
{\small CNOT}      &1 &2 &9& 16,656 \\
C$^2$NOT           &0 &1 &2&0\\
C$^3$NOT           &0 &0 &0&0\\
C$^4$NOT           &0 &0 &0&0\\
$C_2X_{b,c}$       &2 &3&0& 67,260 \\
$C_3X_{b,c}$       &12&20&1& 8,988 \\
$C_3C_2X_{b,c}$    &8 &12&13&0\\
$C_3C_3X_{b,c}$    &4 &36 &45&0\\
\hline\hline
\end{tabular}
\caption{Gate cost of primitive gates for $\ds$ for a qutrit-qubit architecture.  The costs for $\mathcal{U}_{F}$ were obtained with the hybrid compiler described in Appendix~\ref{app:qubit-qutrit-compiler}. $C_dX_{b,c}$ refers to an $X$ rotation between states $b,c$ of a qutrit controlled by a qudit with $d$ levels}
\label{tab:qft-hybrid-cost}
\end{table}
\section{Resource Costs}
\label{sec:resources}

The relatively deep circuits presented above strongly suggest that simulating $\ds$ will require error correction and longer coherence times on quantum devices. The preclusion of universal transversal sets of gates stated in the Eastin-Knill theorem~\cite{Eastin_2009} requires compromises be made.  In most error correcting codes, the Clifford gates are designed to be transversal \cite{Chuang:1996hw,1996PhRvA..54.1098C,PhysRevLett.77.793,1996RSPSA.452.2551S,PhysRevA.54.4741}.  This leaves the nontransversal T gate as the dominant cost of fault-tolerant algorithms~\cite{Chuang:1996hw,1997RuMaS..52.1191K}. Beyond these standard codes, novel universal sets exist with transversal $\btt,\bo,\bi$ and He(3)\footnote{The Heisenberg group of dimension 3, which is a noncrystal-like subgroup of SU(3)} gates~\cite{Kubischta:2023nlb,Denys:2023syu,Jain:2023deu,Denys:2022iyj,Herbert:2023qgu} which warrant exploration for use in lattice gauge theory.

\begin{table}[]
    \centering
    \caption{Number of physical T gates and clean ancilla required to implement logical gates for (top) basic gates taken from~\cite{Chuang:1996hw} (bottom) primitive gates for $\ds$.}
    \label{tab:tgatecost}
    \begin{tabular}{ccc}
    \hline\hline
         Gate & T gates & Clean ancilla\\
         \hline
         C$^2$NOT & 7 & 0\\
         C$^3$NOT & 21 & 1\\
         C$^4$NOT & 35 & 2\\
         CSWAP & 7 & 0\\
         $R_Z$ & 1.15 $\log_2(1/\epsilon)$ & 0\\
         \hline
         $\mathfrak{U}_{-1}$ & 119 & 4\\
         $\mathfrak{U}_{\times}$ & 308 & 2\\
         $\mathfrak{U}_{Tr}$ & $378+ 8.05 \log_2(1 / \epsilon)$& 7\\
         $\mathfrak{U}_{F}$ & $185898\log_2(1/\epsilon)$ & 0\\
         $\mathfrak{U}_{Ph}$ & $294.4 \log_2(1 / \epsilon)$ & 0 \\\hline\hline
    \end{tabular}
\end{table}

For this work, we will consider the following decompositions of gates into T gates for our resource estimates. First, while the CNOT is transversal, the Toffoli gate decomposes into six CNOTs and seven T gates~\cite{Chuang:1996hw}.  With this, one can construct any C$^n$NOT gates using $2\lceil\log_2{n}\rceil-1$ Toffoli gates and $n-2$ dirty ancilla qubits which can be reused later~\cite{2019arXiv190401671B,Chuang:1996hw,PhysRevA.52.3457}. For the $R_Z$ gates, we use the repeat-until-success method of~\cite{PhysRevLett.114.080502} which finds these gates can be approximated to precision $\epsilon$ with on average $1.15 \log_2(1/\epsilon))$ T gates (and at worst $-9 + 4 \log_2(1/\epsilon)$~\cite{10.5555/2685188.2685198}). For $R_Y$ and $R_X$, one can construct them with at most three $R_Z$. Putting all together, we can construct gate estimates for $\ds$ (See Tab.~\ref{tab:tgatecost}). 

\begin{table}[ht]
   \caption{Number of primitive gates per link per $\delta t$ neglecting boundary effects as a function of $d$ for $H_{KS}$ and $H_{I}$.}
    \label{tab:primcost}
    \begin{tabular}{c|cccc}
    \hline\hline
    Gate & $\mathfrak U_F$&$\mathfrak U_{\rm Tr}$&$\mathfrak U_{-1}$&$\mathfrak U_{\times}$\\
    \hline
    $e^{-i\delta H_{KS}}$&2&$\frac{1}{2}(d-1)$ & $3(d-1)$ &$6(d-1)$\\
    $e^{-i\delta H_I}$&4&$\frac{3}{2}(d-1)$ & $2+11(d-1)$&$4+26(d-1)$\\
    \hline\hline
    \end{tabular}
\end{table}

While the results in Tab. \ref{tab:tgatecost} are nearly optimal for the $\mathfrak{U}_{\Tr}$, $\mathfrak{U}_{\times}$, and $\mathfrak{U}_{-1}$, the result for $\mathfrak{U}_{F}$ is not. 
In Ref. \cite{ffttocome} the authors show explicit demonstrations of an efficient decomposition of the nonabelian quantum Fourier transform $\mathfrak{U}_{QFT}$ using the methods of \cite{Pueschel:1998zzo} for certain $SU(2)$ and $SU(3)$ subgroups.
Since it is expected that the gate cost for Fourier transforms should scale as a polynomial of logarithms of the group size \cite{moore2006generic}, one can perform a fit from the results in Ref. \cite{ffttocome} to obtain an order of magnitude estimate for $\mathfrak{U}_{QFT}$ of $147 + 75 \rm{log}_2(1 / \epsilon)$ -- a factor of $\sim 2000$ smaller than our $\mathfrak U_{F}$.

Clearly, the cost of simulating $\ds$ depends on $\epsilon$. To optimize the cost, the synthesis error from finite $\epsilon$ should be balanced with other sources of error in the quantum simulation like Trotter error, discretization error, and finite volume error.  These other sources of error are highly problem-dependent, but here we will follow prior works~\cite{Kan:2021xfc,Davoudi:2020yln,Shaw:2020udc} and take a fiducial $\epsilon=10^{-8}$.

Primitive gate costs for implementing $H_{KS}$~\cite{PhysRevD.11.395} and $H_I$~\cite{Carena:2022kpg}, per link per Trotter step $\delta t$ are shown in Tab.~\ref{tab:primcost}.  Using this result, we can determine the total T gate count $N^{H}_T=C^H_T\times d L^d N_t$ for a $d$ spatial lattice simulated for a time $t=N_t\delta t$. We find that for $H_{KS}$
\begin{equation}
    C^{KS}_{T}=2394(d-1)+(371791+4.025d)\log_2\frac{1}{\epsilon}.
\end{equation}
With this, the total synthesis error $\epsilon_T$ can be estimated as the sum of $\epsilon$ from each $R_Z$.  In the case of $H_{KS}$ this is 
\begin{equation}
    \epsilon_T=\frac{1}{2}(646593+7d)d L^d N_t\times \epsilon.
\end{equation}
If one looks to reduce lattice spacing errors for a fixed number of qubits, one can use $H_I$ which would require
\begin{equation}
    C^{I}_{T}=9884d-8414+(744167+12.075d)\log_2\frac{1}{\epsilon},
\end{equation}
where the total synthesis error is
\begin{equation}
    \epsilon_T =\frac{1}{2}(1293179+21d)d L^d N_t \times\epsilon.
\end{equation}
Following \cite{Cohen:2021imf,Kan:2021xfc,Gustafson:2022xdt}, we will make resource estimates based on our primitive gates for the calculation of the shear viscosity $\eta$ on a $L^3=10^3$ lattice evolved for $N_t=50$, and total synthesis error of $\epsilon_T=10^{-8}$. Considering only the time evolution and neglecting state preparation (which can be substantial~\cite{Xie:2022jgj,Davoudi:2022uzo,Avkhadiev:2019niu,Gustafson:2022hjf,peruzzo2014variational,Gustafson:2019mpk,Gustafson:2019vsd,Harmalkar:2020mpd,Gustafson:2020yfe,Jordan:2017lea,PhysRevLett.108.080402,motta2020determining,Clemente:2020lpr,motta2020determining,deJong:2021wsd,Gustafson:2023ayr,Kane:2023jdo}), Kan and Nam estimated $6.5\times10^{48}$ T gates would be required for an pure-gauge $SU(3)$ simulation of $H_{KS}$. This estimate used a truncated electric-field digitization and considerable fixed-point arithmetic -- greatly inflating the T gate cost. Here, using $\ds$ to approximate $SU(3)$ requires $7.0\times10^{12}$ T gates for $H_I$ and $3.5\times10^{12}$ T gates for $H_{KS}$.  The T gate density is roughly 1 per $\ds-$register per clock cycle. Thus $\ds$ reduces the gate costs of~\cite{Kan:2021xfc} by $10^{36}$. Similar to the previous results for discrete groups of SU(2), $\mathfrak{U}_{F}$ dominates the simulations -- being over 99\% of the computation regardless of Hamiltonian.
However \cite{ffttocome} shows that the Fourier transformation for $\mathbb{BT}$ and $\mathbb{BO}$ can be brought down.  Using the estimate for $\ds$, the Fourier gate contribution is reduced to only 51\% of the simulation with a reduced total T gate count of $5.7\times10^{9}$ for $H_{I}$ with $L=10$ . 

\section{Outlook}
\label{sec:conclusion}

This article provided a construction of primitive gates necessary to simulate a pure SU(3) gauge theory via a discrete subgroup $\Sigma(36\times3)$. In addition, we have also estimated the T-gate cost incurred to compute the shear viscosity using the $\Sigma(36\times3)$ group. Notably, we found that our construction improves the T-gate cost upon that of Ref.~\cite{Kan:2021xfc} by \emph{36 orders of magnitude}. This cost reduction comes at the expense of model accuracy. 

For both qubit and hybrid qubit-qutrit implementations, $\mathfrak{U}_{F}$ dominates the cost suggesting that further reductions by identifying a $\mathfrak{U}_{QFT}$ for $\Sigma(36\times3)$. In fact, as demonstrated in Ref.~\cite{ffttocome}, the cost of a $\mathfrak{U}_{QFT}$ versus $\mathfrak{U}_{F}$ can be as large as a factor of $\sim 2000$. 

In addition, the much-improved overall cost due to the use of the $\Sigma(36\times3)$ group supports the need to also study other discrete subgroups of SU(2) and SU(3). To this end, recent studies (e.g. Ref~\cite{Gustafson:2022xdt,Gustafson:2023kvd}) have already constructed primitive gates for some SU(2) discrete subgroups, the binary tetrahedral and binary octahedral. It remains to develop such gates for other subgroups, for example the larger subgroups of SU(3) such as $\Sigma (72 \times 3)$, $\Sigma(216 \times 3)$ and $\Sigma(360 \times 3)$ as well as the $\bi$ group. The larger groups will reduce discretization errors but at the cost of a longer circuit depth.

Finally, beyond pure gauge, approximating QCD requires incorporating fermion fields \cite{Florio:2023kel, Zohar:2016iic, Zohar:2018nvl}. Many methods exist to incorporate staggered and Wilson fermions. It is worth comparing the resource costs for explicit spacetime simulations using staggered versus Wilson fermions not only in terms of T-gates but also spacetime costs using methods such as \cite{Gui:2023coj}.

\begin{acknowledgements}
The authors would like to thank Sohaib Alam and Stuart Hadfield for helpful feedback.
This material is based on work supported by the U.S. Department of Energy, Office of Science, National Quantum Information Science Research Centers, Superconducting Quantum Materials and Systems Center (SQMS) under contract number DE-AC02-07CH11359.
EG was supported by the NASA Academic Mission Services, Contract No. NNA16BD14C and  NASA-DOE interagency agreement SAA2-403602. YJ is grateful for the support of Deutsche Forschungsgemeinschaft
(DFG, German Research Foundation) grant SFB TR 110/2.
EM is supported by the U.S. Department of Energy, Office of Nuclear Physics under Award Numbers DE-SC021143 and DE-
FG02-95ER40907. 
SZ is supported by the National Science Foundation CAREER award (grant CCF-1845125). 
Part of this research was performed while SZ was visiting the Institute for Pure and Applied Mathematics (IPAM), which is supported by the National Science Foundation (Grant No. DMS-1925919). Fermilab is operated by Fermi Research Alliance, LLC under contract number DE-AC02-07CH11359 with the United States Department of Energy. 
\end{acknowledgements}

\appendix 

\section{Derivation of Inversion Gates}
\label{app:Alternate_Inversion}
\begin{figure}
    \centering
    \includegraphics[width=\linewidth]{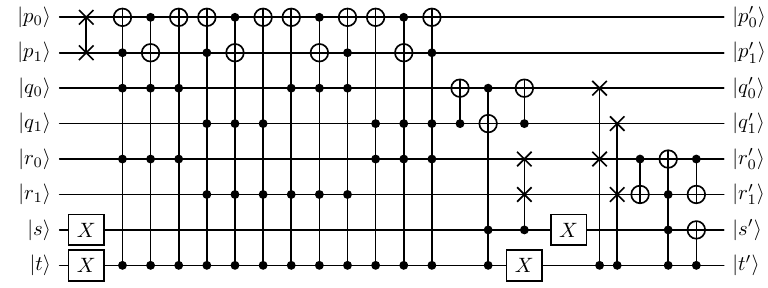}
    \caption{$\mathfrak{U}_{-1}$ for $\Sigma(108)$ which requires 420-T gates.}
    \label{fig:s108inversion1}
\end{figure}

\begin{figure*}
    \includegraphics[width=0.83\linewidth]{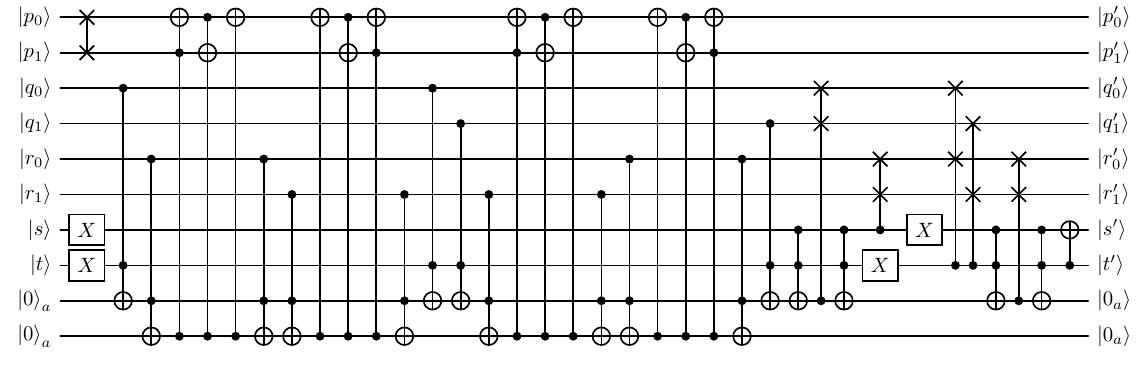}
    \caption{A more $T$-gate optimized version of $\mathfrak{U}_{-1}$ for $\Sigma(108)$ which requires requires 203-T gates and 2 ancilla.}
    \label{fig:s108inversion2}
\end{figure*}

The inversion rules from Eq. (\ref{eq:inversionrules}) are written in a three level notation. 
If one wants to simulate systems using qubits one needs to map these from qudit basis rules to qubit basis rules.
We first begin with the $p'$ rule

\begin{align}
    p' = & 2p \oplus_3 qr(1 - t)\notag\\
    = & 2(p_0 \oplus_3 2p_1) \oplus_3 (q_0 \oplus_3 2q_1)(r_0 \oplus_3 2r_1)(1 - t)\notag\\
    = & 2(p_0 \oplus_3 2p_1) \oplus_3 (1 - t)\big(q_0 r_0 \oplus_3 2 q_1 r_0\notag\\ 
    & \oplus_3 2 r_1 q_0 \oplus_3 4 q_1 r_1\big).\notag
\end{align}
In order to turn this trinary arithmetic into binary arithmetic we need the following transformation axiom:
\begin{align}
\label{eq:shiftbyone}
    p' = &p \oplus_3 1,\notag\\
    p_0' = &p_0 \oplus_2 p_1 \oplus_2 1,\notag\\
    p_1' = &p_0,
\end{align}
and
\begin{align}
\label{eq:shiftbytwo}
    p' = & p \oplus_3 2,\notag\\
    p_0' = & p_1,\notag\\
    p_1' = & p_1 \oplus_2 p_0 \oplus_2 1.
\end{align}
Using this set of transformation rules we find
\begin{align}
    p_0' = &p_1 \oplus_2 (1 \oplus_2 t)\big((1 \oplus_2 p_0)(q_0 r_0 \oplus_2 q_1 r_1)\notag\\
    & \oplus_2 p_0(q_1 r_0 \oplus_2 r_1 q_0)\big),\notag\\
    p_1' = &p_0 \oplus_2 (1 \oplus_2 t)\big((1 \oplus_2 p_1)(q_1 r_0 \oplus_2 q_0 r_1)\notag\\
    & \oplus_2 p_1(q_0 r_0 \oplus_2 q_1 r_1)\big),\notag\\
    q_0' = & q_1(1 \oplus_2 s)(1 \oplus_2 t) \oplus_2 q_0 s (1 \oplus_2 t) \notag\\
    & \oplus_2 r_1(1 \oplus_2 s)t \oplus_2 r_0 s t,\notag\\
    q_1' = & q_0(1 \oplus_2 s)(1 \oplus_2 t) \oplus_2 q_1 s (1 \oplus_2 t) \notag\\
    & \oplus_2 r_0(1 \oplus_2 s)t \oplus r_1 s t,\notag\\
    r_0' = & r_1 (1 \oplus_2 s)(1 \oplus_2 t)\oplus_2 r_0 s (1 \oplus_2t)\notag\\
    & \oplus q_0 t (1 \oplus_2 s) \oplus q_1 t s,\notag\\
    r_1' = & r_0 (1 \oplus_2 s)(1 \oplus_2 t)\oplus_2 r_1 s (1 \oplus_2 t)\notag\\
    & \oplus_2 q_1 t (1 \oplus_2 s) \oplus_2 q_0 t s.
\end{align}
Naively translating these rules as written yields the circuit provided in Fig. \ref{fig:s108inversion1}.
However the resource cost of 420 T-gates can be optimized significantly. By clever use of ancillae one could reduce the T-gate costs down to  203 T-gates using the circuit provided in Fig. \ref{fig:s108inversion2}.

Instead of writing a circuit for the circuit for the whole inversion ruleset of Eq. (\ref{eq:inversionrules}), one instead could use commutation rules to reduce the T-gate costs even further.  
This construction allows the inversion operation to be decomposed into a product of smaller operations:
\begin{equation}
    \label{eq:inversedecomp}
    \mathfrak{U}_{-1} = \mathfrak{U}_{-1}^{C}\mathfrak{U}_{-1}^{E}\mathfrak{U}_{-1}^{V^2}\mathfrak{U}_{-1}^{V}\mathfrak{U}_{-1}^{l}.
\end{equation}
$\mathfrak{U}_{-1}^{l}$ takes each local generator to its inverse:
\begin{align}
    &t \rightarrow t\notag\\
    &s \rightarrow s \oplus_2 t\notag\\
    &r \rightarrow 2 r\notag\\
    &q \rightarrow 2 q\notag\\
    &p \rightarrow 2 p\notag.
\end{align}
The operation $\mathfrak{U}_{-1}^{V}$ involves propagating through the operator $V^{t}$ until it is the right most element. This yields the transformations:
\begin{align}
    &s \rightarrow s\notag\\
    &r \rightarrow r(1 - t) \oplus_3 qt\notag\\
    &q \rightarrow 2rt \oplus_3 q(t - 1)\notag\\
    &p \rightarrow p \oplus_3 rq(1 - t)\notag.
\end{align}
The generators $C$ and $E$ are normal ordered at this point. 
The operation $\mathfrak{U}_{-1}^{V^2}$ has the following transformation rule
\begin{align}
    &r \rightarrow 2rs \oplus_3 r(1 - s)\notag\\
    &q \rightarrow 2qs \oplus_3 q(1 - s)\notag\\
    &p \rightarrow p.\notag
\end{align}
At this point the transformation rules for $C,~E,$ and $\omega$ are trivial.
After all these suboperations are constructed we end up with the inversion operation from the main text provided in Fig. \ref{fig:s108inversion3} and Fig. \ref{fig:quditinversion}.

\section{Derivation of the Multiplication Gate}
\label{app:multiplication_discussion}
The construction of the multiplication gate rules is going to follow in a similar spirit to the derivation of the inversion rules. We first start with two registers corresponding to group elements
\begin{align*}
    g = \omega^{p_1} C^{q_1} E^{r_1} V^{2 s_1 + t_1},
\end{align*}
and
\begin{align*}
    h = \omega^{p_2} C^{q_2} E^{r_2} V^{2 s_2 + t_2},
\end{align*}
with $gh$ given by the product rules of Eq.~(\ref{eq:productrules}).
When we multiply the group elements $g$ and $h$ together, we iteratively move the elements of $g$ over onto $h$. 
This commutation begins by first by moving the $V^{t_1}$ component over to $h$:
\begin{align}
    gh & = \omega^{p_1} C^{q_1} E^{r_1} V^{2 s_1 + t_1} \omega^{p_2} C^{q_2} E^{r_2} V^{2 s_2 + t_2}\notag\\
    & = (\omega^{p_1} C^{q_1} E^{r_1} V^{2 s_1})\omega^{p_2'} C^{q_2'} E^{r_2'} V^{2 s_2' + t_2'}.
\end{align}
Propagating through $V^{t_1}$ gives the following transformations to the elements $p_2$, $q_2$, $r_2$, $s_2$, and $t_2$:
\begin{align}
    p_2 \rightarrow p_2' =  p_2 \oplus_3 2 r_2 q_2 t_1\notag\\
    q_2 \rightarrow q_2' =  q_2(1 - t_1) \oplus_3 2 r_2 t_1\notag\\
    r_2 \rightarrow r_2' =  q_2 t_1 \oplus_3 r_2(1-t_1)\notag\\
    s_2 \rightarrow s_2' =  s_2 \oplus_2 t_1 t_2\notag\\
    t_2 \rightarrow t_2' =  t_2 \oplus_2 t_1.
\end{align}
All together this gives the circuit operation, $\mathfrak{U}_{\times, t}$ in Fig. \ref{fig:qutrit_mult}. 

The next step involves moving the $V^{2s_1}$ operation across such that
\begin{align}
    gh = &(\omega^{p_1} C^{q_1} E^{r_1}) V^{2 s_1} \omega^{p_2'} C^{q_2'} E^{r_2'} V^{2 s_2' + t_2'}\notag\\
    = & (\omega^{p_1} C^{q_1} E^{r_1})\omega^{p_2''} C^{q_2''} E^{r_2''} V^{2 s_2'' + t_2''}.
\end{align}
In this case, the operators now transform under the rules 
\begin{align}
    p_2' \rightarrow p_2'' = &p_2'\notag\\
    q_2' \rightarrow q_2'' = &q_2'(1 - s_1) \oplus_3 2 q_2' s_1\notag\\
    r_2' \rightarrow r_2'' = & r_2'(1 - s_1) \oplus_3 2 r_2' s_1\notag\\
    s_2' \rightarrow s_2'' = & s_2' \oplus s_1\notag\\
    t_2' \rightarrow t_2'' = &t_2''.
\end{align}
It follows immediately then that this is a controlled permutation on the $|1\rangle_3$-$|2\rangle_3$ subspace on the $q$ and $r$ qutrits and a simple CNOT on the $s_2$ register. 

Propagation through of the $E^{r_1}$ then transforms the remaining states on the $h$ register to:
\begin{align}
    p_2'' \rightarrow p_2''' = & p_2'' \oplus_3 q_2'' r_1\notag\\
    q_2'' \rightarrow q_2''' = & q_2''\notag\\
    r_2'' \rightarrow r_2''' = & r_2'' \oplus_3 r_1\notag\\
    s_2'' \rightarrow s_2''' = & s_2''\notag\\
    t_2'' \rightarrow t_2''' = & t_2'',
\end{align}
which gives the expression for $\mathfrak{U}_{\times, E}$ in Fig. \ref{fig:qutrit_mult}.

\section{Qubits Decomposition of $X_{0,1}$, $X_{1,2}$ and $\chi$ Gates}
Since these gates act on qutrits, we will implement them using two qubits. We encode the qubit states as $\ket{q_1q_0}$ where $q_0$ is the least significant bit. That is, the states are ordered as $\ket{00}$, $\ket{01}$, $\ket{10}$ and $\ket{11}$.

\begin{figure}
    \centering
    \includegraphics{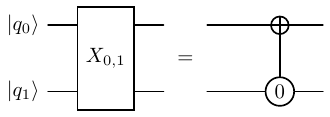}
    \caption{Two-qubit implementation of the $X_{0,1}$ gate.}
    \label{fig:x01-decomposition}
    \includegraphics{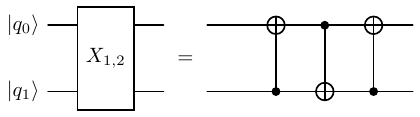}
    \caption{Two-qubit implementation of the $X_{1,2}$ gate.}
    \label{fig:x12-decomposition}
    \includegraphics{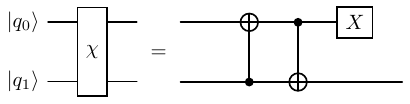}
    \caption{Two-qubit implementation of the $\chi$ gate.}
    \label{fig:chi-decomposition}
\end{figure}

The gate $X_{0,1}$ interchanges the states $\ket{00}$ and $\ket{01}$. It can be implemented as shown in Fig.~\ref{fig:x01-decomposition}. The $X_{1,2}$, on the other hand, can be implemented as a qubit swap gate, see Fig.~\ref{fig:x12-decomposition}. In addition, the $\chi$ gate can be implemented using the circuit in Fig.~\ref{fig:chi-decomposition}.

\section{Electric term of the Hamiltonian}
\label{app:electric-term}
In this section, we will explain the construction of the electric term in the momentum (representation) basis. Such construction for Lie groups can be performed straightforwardly from the Casimir operators, see e.g.~\cite{PhysRevD.11.395}. A generalization to discrete groups can be obtained by using the Laplacian operator on a Cayley graph associated to the group as is done in Ref.~\cite{PhysRevD.107.114513}.

For a brief review of a procedure, we choose $\Gamma$, a subset of the group such that $\Gamma$ is closed under inversion and conjugation. That is $\Gamma^{-1} = \Gamma$ and $g \,\Gamma\, g^{-1} = \Gamma$ for all $g \in G$. In addition, we will choose $\mathbb{1} \notin \Gamma$ as including this element will result in a constant shift of the spectrum. Clearly, there may be several choices of $\Gamma$. However, it is shown in Ref.~\cite{Harlow:2018tng} that the choice $\Gamma = \{g \in G \; | \; {\rm Re Tr}(g) \; {\rm is \; maximal}\,\}$ follows from the Wilson action, and therefore results in a manifestly Lorentz-invariant term in the Hamiltonian. Moreover, such a choice of $\Gamma$ also clearly fulfills the first two conditions. Then, to compute the electric term, we will discard the identity element and choose only those elements with ${\rm max} \left[{\rm Re Tr}(g) \right] = 1$ for the case of $\Sigma(36 \times 3)$. We find that $\Gamma$ consists of 18 elements that generate the whole $\Sigma(36 \times 3)$ group.

Having defined $\Gamma$, the electric term can be computed as
\begin{align}
    H_{E} &= \frac{-g^2}{2}\sum\limits_{\rho,m,n}f(\rho) \ket{\rho,m,n} \,\bra{\rho,m,n}
\end{align}
where the eigenvalues 
\begin{align}
    f(\rho) = |\Gamma| - \frac{1}{{\rm dim}(\rho)} \sum\limits_{g \in \Gamma} {\rm Tr}_\rho(g).
\end{align}
Direct computation of $f(\rho)$ yields the values shown in Table~\ref{tab:electric-eval}.
\begin{table}
\begin{tabular}{l | c c c c c c c c c c c c c c c }
\hline\hline
$\rho$ &1 & 2 & 3 & 4 & 5 & 6 & 7 & 8 & 9 & 10 & 11 & 12 & 13 & 14  \   \\
\hline
$f(\rho)$& 0 & 18 & 36 & 18 & 12 & 18 & 24 & 18 & 18 & $\lambda_1$ & 18 & $\lambda_2$ & 18 & 18 \\
\hline\hline
\end{tabular}
\caption{Eigenvalues of the Electric term. We have defined $\lambda_1 = 2 (9 + \sqrt{3})$ and $\lambda_2 = -2 (-9 + \sqrt{3})$.}
\label{tab:electric-eval}
\end{table}

Having constructed the electric term above $H_{E}$, we can construct a quantum circuit of its time-evolution using the hybrid qubit-qutrit compiler that we will describe in Appendix~\ref{app:qubit-qutrit-compiler}. We obtain the gate cost shown in Table~\ref{tab:electric-gate-cost}.
\begin{table}
\begin{tabular}{l | c c c c c c }
\hline\hline
Basic Gate & $R^{Z}$ & $R^Z_{b,c}$ & CNOT & $C_1X_{b,c}$ & $C_2X_{b,c}$ & Total  \\
\hline
Count& 3 & 98 & 2 & 164 & 20 & 287  \\
\hline\hline
\end{tabular}
\caption{Gate cost for the time evolution due to the electric term shown in Table~\ref{tab:electric-eval}.}
\label{tab:electric-gate-cost}
\end{table}



\section{Qubit-Qutrit Compiler\label{app:qubit-qutrit-compiler}}
This section details the compilation method, and the full codes can be found in Ref.~\cite{hcompiler}. The overarching approach of this compiler is to generalize the qubit Quantum Shannon Decomposition (QSD)~\cite{Shende2006,Drury2008} to apply to a register with qubits and qutrits. We will consider a unitary operator $U$ acting on $n_1$ qubits and $n_2$ qutrits; that is $U$ is of dimension $N \times N$ where $N = 2^{n_1}\times 3^{n_2}$. 

First, let's organize the qudits as $q_1, q_2, {...}, q_{n_1 + n_2}$. We will say that the left-most qudit is the top qudit. The compiler iteratively performs the qubit QSD if the top qudit is a qubit, and otherwise performs our realization of the qutrit QSD. The process eventually terminates when we reach the bottom qudit, in which case, we use either a single-qubit gate decomposition or a single-qutrit gate decomposition depending on whether the bottom qudit is qubit of qutrit. For the single qubit gate, we use the Euler angle parametrization $ZYZ$ and for the single qutrit gate, we use the decomposition given in Ref.~\cite{Di:2011cvl}. 

It is convenient to start with the qubit QSD case. In this case, a Cosine-Sine decomposition (CSD) (see e.g. Refs.~\cite{Paige1994,Chen2013,PhysRevA.92.062317}) is first performed, resulting in 
\begin{equation}
    U = (V_1\oplus V_2)
    \begin{pmatrix}
        C & -S \\
        S & C\\
    \end{pmatrix}
    (W_1\oplus W_2),
\end{equation}
where $V_{1,2}$, $W_{1,2}$ are unitaries with dimension $N/2$. $C$ and $S$ are diagonal matrices e.g $C = \rm{diag}\left(\cos\theta_1, {...}, \cos\theta_{N/2}\right)$ and similarly for $S$. 

Following~\cite{Shende2006}, the next step is to decompose the two block-diagonal unitary matrices:
\begin{align}
    \left(V_1\oplus V_2\right) 
    &= (\mathbb{1} \otimes M) (D \oplus D^\dagger) (\mathbb{1} \otimes N).
\end{align}
where $M$ and $N$ are unitaries acting only on $n_1 - 1$ qubits and $n_2$ qutrits, and $D$ is a diagonal unitary of dimension $N/2$.  Thus, the compilation problem is reduced to decomposing the unitaries $D\oplus D^\dagger$ and the $CS$ into simple gates. The $D\oplus D^\dagger$ can be implemented as a uniformly controlled rotation on the top qudit (see e.g. Ref.~\cite{Shende2006}). The $CS$ matrix, on the other hand, is related to $D \oplus D^\dagger$ by the rotation gate $R_x(\pi/2)$ on the top qubit. This concludes the case the top qudit is a qubit.

In the case that the top qudit is qutrit, we need to find a qutrit realization of the procedure above. The starting point is to perform two CSD as in e.g. Ref.~\cite{Khan2006}. This decomposition reads as 
\begin{widetext}
\begin{align}
    U = \left(V_1 \oplus V_2 \oplus V_3\right)
    &\left(\mathbb{1} \oplus D \oplus D^\dagger\right)
    \left(W_1 \oplus W_2 \oplus W_3\right)
    \begin{pmatrix}
        C & & -S \\
          & \mathbb{1} & \\
        S & &  C
    \end{pmatrix}
    \left(V^\prime_1 \oplus V^\prime_2 \oplus V^\prime_3\right)
    \left(\mathbb{1} \oplus D^\prime \oplus D^{\prime\dagger}\right)
    \left(W^\prime_1 \oplus W^\prime_2 \oplus W^\prime_3\right),
\end{align}
where each block is a unitary of dimension $N/3$. The blocks $C$, $S$ $D$ and $D^\prime$ are defined analogously to the qubit case. 

The rest is to decompose the remaining block diagonal unitaries. By performing the decomposition in Ref.~\cite{Shende2006} twice, we obtain the relation
\begin{align}
    V_1\oplus V_2 \oplus V_3 = \left(\mathbb{1}_{3\times3} \otimes M\right) \left( D \oplus D \oplus D^\dagger \right) \left(\mathbb{1}_{3\times3} \otimes N\right)
    &\left(D^\prime \oplus D^{\prime\dagger} \oplus \mathbb{1}\right)
    \left(\mathbb{1}_{3\times3} \otimes M^\prime\right) \left( D^{\prime\prime} \oplus D^{\prime\prime} \oplus D^{\prime\prime\dagger} \right) \left(\mathbb{1}_{3\times3} \otimes N^\prime\right).
\end{align}
\end{widetext}
The unitaries $\mathbb{1}\oplus D \oplus D^\dagger$, $D \oplus D^\dagger \oplus \mathbb{1}$, $D \oplus D \oplus D^\dagger$ and $CS$ are uniformly controlled rotations. We can focus on the diagonal blocks because as in the qubit case, the $CS$ matrix can be diagonalized with an appropriate $R^{i,j}_x(\pi/2)$ on the top qutrit. Ref.~\cite{PhysRevA.87.012325} outlines the decomposition of qutrits uniformly controlled rotations in terms of single- and two-qutrit gates. A generalization can be obtained simply by limiting a $C^a_b X^c_{i,j}$ gate to $C^a_1 X^c_{i,j}$ when the control qudit is a qubit. 

For $n = 2$ qubits, there exists optimal compilation algorithms (see e.g. Ref.~\cite{PhysRevA.69.032315}). Therefore, when the bottom two qudits are both qubits, we stop the decomposition and use Qiskit transpiler to obtain a quantum circuit.

\bibliography{wise}

\begin{thebibliography}{166}%
\makeatletter
\providecommand \@ifxundefined [1]{%
 \@ifx{#1\undefined}
}%
\providecommand \@ifnum [1]{%
 \ifnum #1\expandafter \@firstoftwo
 \else \expandafter \@secondoftwo
 \fi
}%
\providecommand \@ifx [1]{%
 \ifx #1\expandafter \@firstoftwo
 \else \expandafter \@secondoftwo
 \fi
}%
\providecommand \natexlab [1]{#1}%
\providecommand \enquote  [1]{``#1''}%
\providecommand \bibnamefont  [1]{#1}%
\providecommand \bibfnamefont [1]{#1}%
\providecommand \citenamefont [1]{#1}%
\providecommand \href@noop [0]{\@secondoftwo}%
\providecommand \href [0]{\begingroup \@sanitize@url \@href}%
\providecommand \@href[1]{\@@startlink{#1}\@@href}%
\providecommand \@@href[1]{\endgroup#1\@@endlink}%
\providecommand \@sanitize@url [0]{\catcode `\\12\catcode `\$12\catcode
  `\&12\catcode `\#12\catcode `\^12\catcode `\_12\catcode `\%12\relax}%
\providecommand \@@startlink[1]{}%
\providecommand \@@endlink[0]{}%
\providecommand \url  [0]{\begingroup\@sanitize@url \@url }%
\providecommand \@url [1]{\endgroup\@href {#1}{\urlprefix }}%
\providecommand \urlprefix  [0]{URL }%
\providecommand \Eprint [0]{\href }%
\providecommand \doibase [0]{https://doi.org/}%
\providecommand \selectlanguage [0]{\@gobble}%
\providecommand \bibinfo  [0]{\@secondoftwo}%
\providecommand \bibfield  [0]{\@secondoftwo}%
\providecommand \translation [1]{[#1]}%
\providecommand \BibitemOpen [0]{}%
\providecommand \bibitemStop [0]{}%
\providecommand \bibitemNoStop [0]{.\EOS\space}%
\providecommand \EOS [0]{\spacefactor3000\relax}%
\providecommand \BibitemShut  [1]{\csname bibitem#1\endcsname}%
\let\auto@bib@innerbib\@empty
\bibitem [{\citenamefont {Aarts}(2016)}]{Aarts_2016}%
  \BibitemOpen
  \bibfield  {author} {\bibinfo {author} {\bibfnamefont {G.}~\bibnamefont
  {Aarts}},\ }\bibfield  {title} {\bibinfo {title} {Introductory lectures on
  lattice qcd at nonzero baryon number},\ }\href
  {https://doi.org/10.1088/1742-6596/706/2/022004} {\bibfield  {journal}
  {\bibinfo  {journal} {Journal of Physics: Conference Series}\ }\textbf
  {\bibinfo {volume} {706}},\ \bibinfo {pages} {022004} (\bibinfo {year}
  {2016})}\BibitemShut {NoStop}%
\bibitem [{\citenamefont {Philipsen}(2007)}]{Philipsen:2007aa}%
  \BibitemOpen
  \bibfield  {author} {\bibinfo {author} {\bibfnamefont {O.}~\bibnamefont
  {Philipsen}},\ }\bibfield  {title} {\bibinfo {title} {Lattice qcd at finite
  temperature and density},\ }\href
  {https://doi.org/10.1140/epjst/e2007-00376-3} {\bibfield  {journal} {\bibinfo
   {journal} {The European Physical Journal Special Topics}\ }\textbf {\bibinfo
  {volume} {152}},\ \bibinfo {pages} {29} (\bibinfo {year} {2007})}\BibitemShut
  {NoStop}%
\bibitem [{\citenamefont {de~Forcrand}(2010)}]{forcr2010simulating}%
  \BibitemOpen
  \bibfield  {author} {\bibinfo {author} {\bibfnamefont {P.}~\bibnamefont
  {de~Forcrand}},\ }\href@noop {} {\bibinfo {title} {Simulating qcd at finite
  density}} (\bibinfo {year} {2010}),\ \Eprint
  {https://arxiv.org/abs/1005.0539} {arXiv:1005.0539 [hep-lat]} \BibitemShut
  {NoStop}%
\bibitem [{\citenamefont {Muroya}\ \emph {et~al.}(2003)\citenamefont {Muroya},
  \citenamefont {Nakamura}, \citenamefont {Nonaka},\ and\ \citenamefont
  {Takaishi}}]{10.1143/PTP.110.615}%
  \BibitemOpen
  \bibfield  {author} {\bibinfo {author} {\bibfnamefont {S.}~\bibnamefont
  {Muroya}}, \bibinfo {author} {\bibfnamefont {A.}~\bibnamefont {Nakamura}},
  \bibinfo {author} {\bibfnamefont {C.}~\bibnamefont {Nonaka}},\ and\ \bibinfo
  {author} {\bibfnamefont {T.}~\bibnamefont {Takaishi}},\ }\bibfield  {title}
  {\bibinfo {title} {{Lattice QCD at Finite Density: An Introductory Review}},\
  }\href {https://doi.org/10.1143/PTP.110.615} {\bibfield  {journal} {\bibinfo
  {journal} {Progress of Theoretical Physics}\ }\textbf {\bibinfo {volume}
  {110}},\ \bibinfo {pages} {615} (\bibinfo {year} {2003})}\BibitemShut
  {NoStop}%
\bibitem [{\citenamefont {Gattringer}\ and\ \citenamefont
  {Langfeld}(2016)}]{Gattringer:2016kco}%
  \BibitemOpen
  \bibfield  {author} {\bibinfo {author} {\bibfnamefont {C.}~\bibnamefont
  {Gattringer}}\ and\ \bibinfo {author} {\bibfnamefont {K.}~\bibnamefont
  {Langfeld}},\ }\bibfield  {title} {\bibinfo {title} {{Approaches to the sign
  problem in lattice field theory}},\ }\href
  {https://doi.org/10.1142/S0217751X16430077} {\bibfield  {journal} {\bibinfo
  {journal} {Int. J. Mod. Phys. A}\ }\textbf {\bibinfo {volume} {31}},\
  \bibinfo {pages} {1643007} (\bibinfo {year} {2016})},\ \Eprint
  {https://arxiv.org/abs/1603.09517} {arXiv:1603.09517 [hep-lat]} \BibitemShut
  {NoStop}%
\bibitem [{\citenamefont {Alexandru}\ \emph {et~al.}(2018)\citenamefont
  {Alexandru}, \citenamefont {Bedaque}, \citenamefont {Lamm}, \citenamefont
  {Lawrence},\ and\ \citenamefont {Warrington}}]{Alexandru:2018ddf}%
  \BibitemOpen
  \bibfield  {author} {\bibinfo {author} {\bibfnamefont {A.}~\bibnamefont
  {Alexandru}}, \bibinfo {author} {\bibfnamefont {P.~F.}\ \bibnamefont
  {Bedaque}}, \bibinfo {author} {\bibfnamefont {H.}~\bibnamefont {Lamm}},
  \bibinfo {author} {\bibfnamefont {S.}~\bibnamefont {Lawrence}},\ and\
  \bibinfo {author} {\bibfnamefont {N.~C.}\ \bibnamefont {Warrington}},\
  }\bibfield  {title} {\bibinfo {title} {{Fermions at Finite Density in 2+1
  Dimensions with Sign-Optimized Manifolds}},\ }\href
  {https://doi.org/10.1103/PhysRevLett.121.191602} {\bibfield  {journal}
  {\bibinfo  {journal} {Phys. Rev. Lett.}\ }\textbf {\bibinfo {volume} {121}},\
  \bibinfo {pages} {191602} (\bibinfo {year} {2018})},\ \Eprint
  {https://arxiv.org/abs/1808.09799} {arXiv:1808.09799 [hep-lat]} \BibitemShut
  {NoStop}%
\bibitem [{\citenamefont {Alexandru}\ \emph {et~al.}(2022)\citenamefont
  {Alexandru}, \citenamefont {Ba\ifmmode~\mbox{\c{s}}\else \c{s}\fi{}ar},
  \citenamefont {Bedaque},\ and\ \citenamefont
  {Warrington}}]{Alexandru:2020wrj}%
  \BibitemOpen
  \bibfield  {author} {\bibinfo {author} {\bibfnamefont {A.}~\bibnamefont
  {Alexandru}}, \bibinfo {author} {\bibfnamefont {G.~m.~c.}\ \bibnamefont
  {Ba\ifmmode~\mbox{\c{s}}\else \c{s}\fi{}ar}}, \bibinfo {author}
  {\bibfnamefont {P.~F.}\ \bibnamefont {Bedaque}},\ and\ \bibinfo {author}
  {\bibfnamefont {N.~C.}\ \bibnamefont {Warrington}},\ }\bibfield  {title}
  {\bibinfo {title} {Complex paths around the sign problem},\ }\href
  {https://doi.org/10.1103/RevModPhys.94.015006} {\bibfield  {journal}
  {\bibinfo  {journal} {Rev. Mod. Phys.}\ }\textbf {\bibinfo {volume} {94}},\
  \bibinfo {pages} {015006} (\bibinfo {year} {2022})}\BibitemShut {NoStop}%
\bibitem [{\citenamefont {Troyer}\ and\ \citenamefont
  {Wiese}(2005)}]{Troyer:2004ge}%
  \BibitemOpen
  \bibfield  {author} {\bibinfo {author} {\bibfnamefont {M.}~\bibnamefont
  {Troyer}}\ and\ \bibinfo {author} {\bibfnamefont {U.-J.}\ \bibnamefont
  {Wiese}},\ }\bibfield  {title} {\bibinfo {title} {{Computational complexity
  and fundamental limitations to fermionic quantum Monte Carlo simulations}},\
  }\href {https://doi.org/10.1103/PhysRevLett.94.170201} {\bibfield  {journal}
  {\bibinfo  {journal} {Phys. Rev. Lett.}\ }\textbf {\bibinfo {volume} {94}},\
  \bibinfo {pages} {170201} (\bibinfo {year} {2005})},\ \Eprint
  {https://arxiv.org/abs/cond-mat/0408370} {arXiv:cond-mat/0408370 [cond-mat]}
  \BibitemShut {NoStop}%
\bibitem [{\citenamefont {Gustafson}\ \emph {et~al.}(2021)\citenamefont
  {Gustafson} \emph {et~al.}}]{Gustafson:2021jtq}%
  \BibitemOpen
  \bibfield  {author} {\bibinfo {author} {\bibfnamefont {E.}~\bibnamefont
  {Gustafson}} \emph {et~al.},\ }\bibfield  {title} {\bibinfo {title} {{Large
  scale multi-node simulations of $\mathbb{Z}_2$ gauge theory quantum circuits
  using Google Cloud Platform}},\ }in\ \href
  {https://doi.org/10.1109/QCS54837.2021.00012} {\emph {\bibinfo {booktitle}
  {{IEEE/ACM Second International Workshop on Quantum Computing Software}}}}\
  (\bibinfo {year} {2021})\ \Eprint {https://arxiv.org/abs/2110.07482}
  {arXiv:2110.07482 [quant-ph]} \BibitemShut {NoStop}%
\bibitem [{\citenamefont {Gustafson}(2021)}]{Gustafson:2021qbt}%
  \BibitemOpen
  \bibfield  {author} {\bibinfo {author} {\bibfnamefont {E.}~\bibnamefont
  {Gustafson}},\ }\bibfield  {title} {\bibinfo {title} {{Prospects for
  Simulating a Qudit Based Model of (1+1)d Scalar QED}},\ }\href
  {https://doi.org/10.1103/PhysRevD.103.114505} {\bibfield  {journal} {\bibinfo
   {journal} {Phys. Rev. D}\ }\textbf {\bibinfo {volume} {103}},\ \bibinfo
  {pages} {114505} (\bibinfo {year} {2021})},\ \Eprint
  {https://arxiv.org/abs/2104.10136} {arXiv:2104.10136 [quant-ph]} \BibitemShut
  {NoStop}%
\bibitem [{\citenamefont {Popov}\ \emph {et~al.}(2024)\citenamefont {Popov},
  \citenamefont {Meth}, \citenamefont {Lewenstein}, \citenamefont {Hauke},
  \citenamefont {Ringbauer}, \citenamefont {Zohar},\ and\ \citenamefont
  {Kasper}}]{Popov:2023xft}%
  \BibitemOpen
  \bibfield  {author} {\bibinfo {author} {\bibfnamefont {P.~P.}\ \bibnamefont
  {Popov}}, \bibinfo {author} {\bibfnamefont {M.}~\bibnamefont {Meth}},
  \bibinfo {author} {\bibfnamefont {M.}~\bibnamefont {Lewenstein}}, \bibinfo
  {author} {\bibfnamefont {P.}~\bibnamefont {Hauke}}, \bibinfo {author}
  {\bibfnamefont {M.}~\bibnamefont {Ringbauer}}, \bibinfo {author}
  {\bibfnamefont {E.}~\bibnamefont {Zohar}},\ and\ \bibinfo {author}
  {\bibfnamefont {V.}~\bibnamefont {Kasper}},\ }\bibfield  {title} {\bibinfo
  {title} {{Variational quantum simulation of U(1) lattice gauge theories with
  qudit systems}},\ }\href {https://doi.org/10.1103/PhysRevResearch.6.013202}
  {\bibfield  {journal} {\bibinfo  {journal} {Phys. Rev. Res.}\ }\textbf
  {\bibinfo {volume} {6}},\ \bibinfo {pages} {013202} (\bibinfo {year}
  {2024})}\BibitemShut {NoStop}%
\bibitem [{\citenamefont {Kurkcuoglu}\ \emph {et~al.}(2022)\citenamefont
  {Kurkcuoglu}, \citenamefont {Alam}, \citenamefont {Job}, \citenamefont {Li},
  \citenamefont {Macridin}, \citenamefont {Perdue},\ and\ \citenamefont
  {Providence}}]{kurkcuoglu2022quantum}%
  \BibitemOpen
  \bibfield  {author} {\bibinfo {author} {\bibfnamefont {D.~M.}\ \bibnamefont
  {Kurkcuoglu}}, \bibinfo {author} {\bibfnamefont {M.~S.}\ \bibnamefont
  {Alam}}, \bibinfo {author} {\bibfnamefont {J.~A.}\ \bibnamefont {Job}},
  \bibinfo {author} {\bibfnamefont {A.~C.~Y.}\ \bibnamefont {Li}}, \bibinfo
  {author} {\bibfnamefont {A.}~\bibnamefont {Macridin}}, \bibinfo {author}
  {\bibfnamefont {G.~N.}\ \bibnamefont {Perdue}},\ and\ \bibinfo {author}
  {\bibfnamefont {S.}~\bibnamefont {Providence}},\ }\href@noop {} {\bibinfo
  {title} {Quantum simulation of $\phi^4$ theories in qudit systems}} (\bibinfo
  {year} {2022}),\ \Eprint {https://arxiv.org/abs/2108.13357} {arXiv:2108.13357
  [quant-ph]} \BibitemShut {NoStop}%
\bibitem [{\citenamefont {Calaj\`o}\ \emph {et~al.}(2024)\citenamefont
  {Calaj\`o}, \citenamefont {Magnifico}, \citenamefont {Edmunds}, \citenamefont
  {Ringbauer}, \citenamefont {Montangero},\ and\ \citenamefont
  {Silvi}}]{Calajo:2024qrc}%
  \BibitemOpen
  \bibfield  {author} {\bibinfo {author} {\bibfnamefont {G.}~\bibnamefont
  {Calaj\`o}}, \bibinfo {author} {\bibfnamefont {G.}~\bibnamefont {Magnifico}},
  \bibinfo {author} {\bibfnamefont {C.}~\bibnamefont {Edmunds}}, \bibinfo
  {author} {\bibfnamefont {M.}~\bibnamefont {Ringbauer}}, \bibinfo {author}
  {\bibfnamefont {S.}~\bibnamefont {Montangero}},\ and\ \bibinfo {author}
  {\bibfnamefont {P.}~\bibnamefont {Silvi}},\ }\href@noop {} {\bibinfo {title}
  {{Digital quantum simulation of a (1+1)D SU(2) lattice gauge theory with ion
  qudits}}} (\bibinfo {year} {2024}),\ \Eprint
  {https://arxiv.org/abs/2402.07987} {arXiv:2402.07987 [quant-ph]} \BibitemShut
  {NoStop}%
\bibitem [{\citenamefont {Gonz\'alez-Cuadra}\ \emph {et~al.}(2022)\citenamefont
  {Gonz\'alez-Cuadra}, \citenamefont {Zache}, \citenamefont {Carrasco},
  \citenamefont {Kraus},\ and\ \citenamefont
  {Zoller}}]{Gonzalez-Cuadra:2022hxt}%
  \BibitemOpen
  \bibfield  {author} {\bibinfo {author} {\bibfnamefont {D.}~\bibnamefont
  {Gonz\'alez-Cuadra}}, \bibinfo {author} {\bibfnamefont {T.~V.}\ \bibnamefont
  {Zache}}, \bibinfo {author} {\bibfnamefont {J.}~\bibnamefont {Carrasco}},
  \bibinfo {author} {\bibfnamefont {B.}~\bibnamefont {Kraus}},\ and\ \bibinfo
  {author} {\bibfnamefont {P.}~\bibnamefont {Zoller}},\ }\href@noop {}
  {\bibinfo {title} {{Hardware efficient quantum simulation of non-abelian
  gauge theories with qudits on Rydberg platforms}}} (\bibinfo {year} {2022}),\
  \Eprint {https://arxiv.org/abs/2203.15541} {arXiv:2203.15541 [quant-ph]}
  \BibitemShut {NoStop}%
\bibitem [{\citenamefont {Illa}\ \emph
  {et~al.}(2024{\natexlab{a}})\citenamefont {Illa}, \citenamefont {Robin},\
  and\ \citenamefont {Savage}}]{illa2024qu8its}%
  \BibitemOpen
  \bibfield  {author} {\bibinfo {author} {\bibfnamefont {M.}~\bibnamefont
  {Illa}}, \bibinfo {author} {\bibfnamefont {C.~E.~P.}\ \bibnamefont {Robin}},\
  and\ \bibinfo {author} {\bibfnamefont {M.~J.}\ \bibnamefont {Savage}},\
  }\href@noop {} {\bibinfo {title} {Qu8its for quantum simulations of lattice
  quantum chromodynamics}} (\bibinfo {year} {2024}{\natexlab{a}}),\ \Eprint
  {https://arxiv.org/abs/2403.14537} {arXiv:2403.14537 [quant-ph]} \BibitemShut
  {NoStop}%
\bibitem [{\citenamefont {Zache}\ \emph
  {et~al.}(2023{\natexlab{a}})\citenamefont {Zache}, \citenamefont
  {Gonz\'alez-Cuadra},\ and\ \citenamefont {Zoller}}]{Zache:2023cfj}%
  \BibitemOpen
  \bibfield  {author} {\bibinfo {author} {\bibfnamefont {T.~V.}\ \bibnamefont
  {Zache}}, \bibinfo {author} {\bibfnamefont {D.}~\bibnamefont
  {Gonz\'alez-Cuadra}},\ and\ \bibinfo {author} {\bibfnamefont
  {P.}~\bibnamefont {Zoller}},\ }\bibfield  {title} {\bibinfo {title}
  {{Fermion-qudit quantum processors for simulating lattice gauge theories with
  matter}},\ }\href {https://doi.org/10.22331/q-2023-10-16-1140} {\bibfield
  {journal} {\bibinfo  {journal} {Quantum}\ }\textbf {\bibinfo {volume} {7}},\
  \bibinfo {pages} {1140} (\bibinfo {year} {2023}{\natexlab{a}})}\BibitemShut
  {NoStop}%
\bibitem [{\citenamefont {Unmuth-Yockey}\ \emph {et~al.}(2018)\citenamefont
  {Unmuth-Yockey}, \citenamefont {Zhang}, \citenamefont {Bazavov},
  \citenamefont {Meurice},\ and\ \citenamefont {Tsai}}]{Unmuth-Yockey:2018ugm}%
  \BibitemOpen
  \bibfield  {author} {\bibinfo {author} {\bibfnamefont {J.}~\bibnamefont
  {Unmuth-Yockey}}, \bibinfo {author} {\bibfnamefont {J.}~\bibnamefont
  {Zhang}}, \bibinfo {author} {\bibfnamefont {A.}~\bibnamefont {Bazavov}},
  \bibinfo {author} {\bibfnamefont {Y.}~\bibnamefont {Meurice}},\ and\ \bibinfo
  {author} {\bibfnamefont {S.-W.}\ \bibnamefont {Tsai}},\ }\bibfield  {title}
  {\bibinfo {title} {{Universal features of the Abelian Polyakov loop in 1+1
  dimensions}},\ }\href {https://doi.org/10.1103/PhysRevD.98.094511} {\bibfield
   {journal} {\bibinfo  {journal} {Phys. Rev.}\ }\textbf {\bibinfo {volume}
  {D98}},\ \bibinfo {pages} {094511} (\bibinfo {year} {2018})},\ \Eprint
  {https://arxiv.org/abs/1807.09186} {arXiv:1807.09186 [hep-lat]} \BibitemShut
  {NoStop}%
\bibitem [{\citenamefont {Unmuth-Yockey}(2019)}]{Unmuth-Yockey:2018xak}%
  \BibitemOpen
  \bibfield  {author} {\bibinfo {author} {\bibfnamefont {J.~F.}\ \bibnamefont
  {Unmuth-Yockey}},\ }\bibfield  {title} {\bibinfo {title} {{Gauge-invariant
  rotor Hamiltonian from dual variables of 3D $U(1)$ gauge theory}},\ }\href
  {https://doi.org/10.1103/PhysRevD.99.074502} {\bibfield  {journal} {\bibinfo
  {journal} {Phys.\ Rev.\ D}\ }\textbf {\bibinfo {volume} {99}},\ \bibinfo
  {pages} {074502} (\bibinfo {year} {2019})},\ \Eprint
  {https://arxiv.org/abs/1811.05884} {arXiv:1811.05884 [hep-lat]} \BibitemShut
  {NoStop}%
\bibitem [{\citenamefont {Klco}\ \emph {et~al.}(2020)\citenamefont {Klco},
  \citenamefont {Stryker},\ and\ \citenamefont {Savage}}]{Klco:2019evd}%
  \BibitemOpen
  \bibfield  {author} {\bibinfo {author} {\bibfnamefont {N.}~\bibnamefont
  {Klco}}, \bibinfo {author} {\bibfnamefont {J.~R.}\ \bibnamefont {Stryker}},\
  and\ \bibinfo {author} {\bibfnamefont {M.~J.}\ \bibnamefont {Savage}},\
  }\bibfield  {title} {\bibinfo {title} {{SU(2) non-Abelian gauge field theory
  in one dimension on digital quantum computers}},\ }\href
  {https://doi.org/10.1103/PhysRevD.101.074512} {\bibfield  {journal} {\bibinfo
   {journal} {Phys. Rev. D}\ }\textbf {\bibinfo {volume} {101}},\ \bibinfo
  {pages} {074512} (\bibinfo {year} {2020})}\BibitemShut {NoStop}%
\bibitem [{\citenamefont {Farrell}\ \emph {et~al.}(2023)\citenamefont
  {Farrell}, \citenamefont {Illa}, \citenamefont {Ciavarella},\ and\
  \citenamefont {Savage}}]{Farrell:2023fgd}%
  \BibitemOpen
  \bibfield  {author} {\bibinfo {author} {\bibfnamefont {R.~C.}\ \bibnamefont
  {Farrell}}, \bibinfo {author} {\bibfnamefont {M.}~\bibnamefont {Illa}},
  \bibinfo {author} {\bibfnamefont {A.~N.}\ \bibnamefont {Ciavarella}},\ and\
  \bibinfo {author} {\bibfnamefont {M.~J.}\ \bibnamefont {Savage}},\
  }\href@noop {} {\bibinfo {title} {{Scalable Circuits for Preparing Ground
  States on Digital Quantum Computers: The Schwinger Model Vacuum on 100
  Qubits}}} (\bibinfo {year} {2023}),\ \Eprint
  {https://arxiv.org/abs/2308.04481} {arXiv:2308.04481 [quant-ph]} \BibitemShut
  {NoStop}%
\bibitem [{\citenamefont {Farrell}\ \emph {et~al.}(2024)\citenamefont
  {Farrell}, \citenamefont {Illa}, \citenamefont {Ciavarella},\ and\
  \citenamefont {Savage}}]{Farrell:2024fit}%
  \BibitemOpen
  \bibfield  {author} {\bibinfo {author} {\bibfnamefont {R.~C.}\ \bibnamefont
  {Farrell}}, \bibinfo {author} {\bibfnamefont {M.}~\bibnamefont {Illa}},
  \bibinfo {author} {\bibfnamefont {A.~N.}\ \bibnamefont {Ciavarella}},\ and\
  \bibinfo {author} {\bibfnamefont {M.~J.}\ \bibnamefont {Savage}},\
  }\href@noop {} {\bibinfo {title} {{Quantum Simulations of Hadron Dynamics in
  the Schwinger Model using 112 Qubits}}} (\bibinfo {year} {2024}),\ \Eprint
  {https://arxiv.org/abs/2401.08044} {arXiv:2401.08044 [quant-ph]} \BibitemShut
  {NoStop}%
\bibitem [{\citenamefont {Illa}\ \emph
  {et~al.}(2024{\natexlab{b}})\citenamefont {Illa}, \citenamefont {Robin},\
  and\ \citenamefont {Savage}}]{Illa:2024kmf}%
  \BibitemOpen
  \bibfield  {author} {\bibinfo {author} {\bibfnamefont {M.}~\bibnamefont
  {Illa}}, \bibinfo {author} {\bibfnamefont {C.~E.~P.}\ \bibnamefont {Robin}},\
  and\ \bibinfo {author} {\bibfnamefont {M.~J.}\ \bibnamefont {Savage}},\
  }\href@noop {} {\bibinfo {title} {{Qu8its for Quantum Simulations of Lattice
  Quantum Chromodynamics}}} (\bibinfo {year} {2024}{\natexlab{b}}),\ \Eprint
  {https://arxiv.org/abs/2403.14537} {arXiv:2403.14537 [quant-ph]} \BibitemShut
  {NoStop}%
\bibitem [{\citenamefont {Ciavarella}\ \emph {et~al.}(2021)\citenamefont
  {Ciavarella}, \citenamefont {Klco},\ and\ \citenamefont
  {Savage}}]{Ciavarella:2021nmj}%
  \BibitemOpen
  \bibfield  {author} {\bibinfo {author} {\bibfnamefont {A.}~\bibnamefont
  {Ciavarella}}, \bibinfo {author} {\bibfnamefont {N.}~\bibnamefont {Klco}},\
  and\ \bibinfo {author} {\bibfnamefont {M.~J.}\ \bibnamefont {Savage}},\
  }\href@noop {} {\bibinfo {title} {{A Trailhead for Quantum Simulation of
  SU(3) Yang-Mills Lattice Gauge Theory in the Local Multiplet Basis}}}
  (\bibinfo {year} {2021}),\ \Eprint {https://arxiv.org/abs/2101.10227}
  {arXiv:2101.10227 [quant-ph]} \BibitemShut {NoStop}%
\bibitem [{\citenamefont {Bazavov}\ \emph {et~al.}(2015)\citenamefont
  {Bazavov}, \citenamefont {Meurice}, \citenamefont {Tsai}, \citenamefont
  {Unmuth-Yockey},\ and\ \citenamefont {Zhang}}]{Bazavov:2015kka}%
  \BibitemOpen
  \bibfield  {author} {\bibinfo {author} {\bibfnamefont {A.}~\bibnamefont
  {Bazavov}}, \bibinfo {author} {\bibfnamefont {Y.}~\bibnamefont {Meurice}},
  \bibinfo {author} {\bibfnamefont {S.-W.}\ \bibnamefont {Tsai}}, \bibinfo
  {author} {\bibfnamefont {J.}~\bibnamefont {Unmuth-Yockey}},\ and\ \bibinfo
  {author} {\bibfnamefont {J.}~\bibnamefont {Zhang}},\ }\bibfield  {title}
  {\bibinfo {title} {{Gauge-invariant implementation of the Abelian Higgs model
  on optical lattices}},\ }\href {https://doi.org/10.1103/PhysRevD.92.076003}
  {\bibfield  {journal} {\bibinfo  {journal} {Phys. Rev.}\ }\textbf {\bibinfo
  {volume} {D92}},\ \bibinfo {pages} {076003} (\bibinfo {year} {2015})},\
  \Eprint {https://arxiv.org/abs/1503.08354} {arXiv:1503.08354 [hep-lat]}
  \BibitemShut {NoStop}%
\bibitem [{\citenamefont {Zhang}\ \emph {et~al.}(2018)\citenamefont {Zhang},
  \citenamefont {Unmuth-Yockey}, \citenamefont {Zeiher}, \citenamefont
  {Bazavov}, \citenamefont {Tsai},\ and\ \citenamefont
  {Meurice}}]{Zhang:2018ufj}%
  \BibitemOpen
  \bibfield  {author} {\bibinfo {author} {\bibfnamefont {J.}~\bibnamefont
  {Zhang}}, \bibinfo {author} {\bibfnamefont {J.}~\bibnamefont
  {Unmuth-Yockey}}, \bibinfo {author} {\bibfnamefont {J.}~\bibnamefont
  {Zeiher}}, \bibinfo {author} {\bibfnamefont {A.}~\bibnamefont {Bazavov}},
  \bibinfo {author} {\bibfnamefont {S.~W.}\ \bibnamefont {Tsai}},\ and\
  \bibinfo {author} {\bibfnamefont {Y.}~\bibnamefont {Meurice}},\ }\bibfield
  {title} {\bibinfo {title} {{Quantum simulation of the universal features of
  the Polyakov loop}},\ }\href {https://doi.org/10.1103/PhysRevLett.121.223201}
  {\bibfield  {journal} {\bibinfo  {journal} {Phys. Rev. Lett.}\ }\textbf
  {\bibinfo {volume} {121}},\ \bibinfo {pages} {223201} (\bibinfo {year}
  {2018})},\ \Eprint {https://arxiv.org/abs/1803.11166} {arXiv:1803.11166
  [hep-lat]} \BibitemShut {NoStop}%
\bibitem [{\citenamefont {Bazavov}\ \emph {et~al.}(2019)\citenamefont
  {Bazavov}, \citenamefont {Catterall}, \citenamefont {Jha},\ and\
  \citenamefont {Unmuth-Yockey}}]{PhysRevD.99.114507}%
  \BibitemOpen
  \bibfield  {author} {\bibinfo {author} {\bibfnamefont {A.}~\bibnamefont
  {Bazavov}}, \bibinfo {author} {\bibfnamefont {S.}~\bibnamefont {Catterall}},
  \bibinfo {author} {\bibfnamefont {R.~G.}\ \bibnamefont {Jha}},\ and\ \bibinfo
  {author} {\bibfnamefont {J.}~\bibnamefont {Unmuth-Yockey}},\ }\bibfield
  {title} {\bibinfo {title} {Tensor renormalization group study of the
  non-abelian higgs model in two dimensions},\ }\href
  {https://doi.org/10.1103/PhysRevD.99.114507} {\bibfield  {journal} {\bibinfo
  {journal} {Phys. Rev. D}\ }\textbf {\bibinfo {volume} {99}},\ \bibinfo
  {pages} {114507} (\bibinfo {year} {2019})}\BibitemShut {NoStop}%
\bibitem [{\citenamefont {Bauer}\ and\ \citenamefont
  {Grabowska}(2021)}]{Bauer:2021gek}%
  \BibitemOpen
  \bibfield  {author} {\bibinfo {author} {\bibfnamefont {C.~W.}\ \bibnamefont
  {Bauer}}\ and\ \bibinfo {author} {\bibfnamefont {D.~M.}\ \bibnamefont
  {Grabowska}},\ }\href@noop {} {\bibinfo {title} {{Efficient Representation
  for Simulating U(1) Gauge Theories on Digital Quantum Computers at All Values
  of the Coupling}}} (\bibinfo {year} {2021}),\ \Eprint
  {https://arxiv.org/abs/2111.08015} {arXiv:2111.08015 [hep-ph]} \BibitemShut
  {NoStop}%
\bibitem [{\citenamefont {Grabowska}\ \emph {et~al.}(2022)\citenamefont
  {Grabowska}, \citenamefont {Kane}, \citenamefont {Nachman},\ and\
  \citenamefont {Bauer}}]{Grabowska:2022uos}%
  \BibitemOpen
  \bibfield  {author} {\bibinfo {author} {\bibfnamefont {D.~M.}\ \bibnamefont
  {Grabowska}}, \bibinfo {author} {\bibfnamefont {C.}~\bibnamefont {Kane}},
  \bibinfo {author} {\bibfnamefont {B.}~\bibnamefont {Nachman}},\ and\ \bibinfo
  {author} {\bibfnamefont {C.~W.}\ \bibnamefont {Bauer}},\ }\href@noop {}
  {\bibinfo {title} {{Overcoming exponential scaling with system size in
  Trotter-Suzuki implementations of constrained Hamiltonians: 2+1 U(1) lattice
  gauge theories}}} (\bibinfo {year} {2022}),\ \Eprint
  {https://arxiv.org/abs/2208.03333} {arXiv:2208.03333 [quant-ph]} \BibitemShut
  {NoStop}%
\bibitem [{\citenamefont {Buser}\ \emph {et~al.}(2020)\citenamefont {Buser},
  \citenamefont {Bhattacharya}, \citenamefont {Cincio},\ and\ \citenamefont
  {Gupta}}]{Buser:2020uzs}%
  \BibitemOpen
  \bibfield  {author} {\bibinfo {author} {\bibfnamefont {A.~J.}\ \bibnamefont
  {Buser}}, \bibinfo {author} {\bibfnamefont {T.}~\bibnamefont {Bhattacharya}},
  \bibinfo {author} {\bibfnamefont {L.}~\bibnamefont {Cincio}},\ and\ \bibinfo
  {author} {\bibfnamefont {R.}~\bibnamefont {Gupta}},\ }\href@noop {} {\bibinfo
  {title} {{Quantum simulation of the qubit-regularized O(3)-sigma model}}}
  (\bibinfo {year} {2020}),\ \Eprint {https://arxiv.org/abs/2006.15746}
  {arXiv:2006.15746 [quant-ph]} \BibitemShut {NoStop}%
\bibitem [{\citenamefont {Bhattacharya}\ \emph {et~al.}(2020)\citenamefont
  {Bhattacharya}, \citenamefont {Buser}, \citenamefont {Chandrasekharan},
  \citenamefont {Gupta},\ and\ \citenamefont {Singh}}]{Bhattacharya:2020gpm}%
  \BibitemOpen
  \bibfield  {author} {\bibinfo {author} {\bibfnamefont {T.}~\bibnamefont
  {Bhattacharya}}, \bibinfo {author} {\bibfnamefont {A.~J.}\ \bibnamefont
  {Buser}}, \bibinfo {author} {\bibfnamefont {S.}~\bibnamefont
  {Chandrasekharan}}, \bibinfo {author} {\bibfnamefont {R.}~\bibnamefont
  {Gupta}},\ and\ \bibinfo {author} {\bibfnamefont {H.}~\bibnamefont {Singh}},\
  }\href@noop {} {\bibinfo {title} {{Qubit regularization of asymptotic
  freedom}}} (\bibinfo {year} {2020}),\ \Eprint
  {https://arxiv.org/abs/2012.02153} {arXiv:2012.02153 [hep-lat]} \BibitemShut
  {NoStop}%
\bibitem [{\citenamefont {Kavaki}\ and\ \citenamefont
  {Lewis}(2024)}]{Kavaki:2024ijd}%
  \BibitemOpen
  \bibfield  {author} {\bibinfo {author} {\bibfnamefont {A.~H.~Z.}\
  \bibnamefont {Kavaki}}\ and\ \bibinfo {author} {\bibfnamefont
  {R.}~\bibnamefont {Lewis}},\ }\href@noop {} {\bibinfo {title} {{From square
  plaquettes to triamond lattices for SU(2) gauge theory}}} (\bibinfo {year}
  {2024}),\ \Eprint {https://arxiv.org/abs/2401.14570} {arXiv:2401.14570
  [hep-lat]} \BibitemShut {NoStop}%
\bibitem [{\citenamefont {Murairi}\ \emph
  {et~al.}(2022{\natexlab{a}})\citenamefont {Murairi}, \citenamefont {Cervia},
  \citenamefont {Kumar}, \citenamefont {Bedaque},\ and\ \citenamefont
  {Alexandru}}]{Murairi:2022zdg}%
  \BibitemOpen
  \bibfield  {author} {\bibinfo {author} {\bibfnamefont {E.~M.}\ \bibnamefont
  {Murairi}}, \bibinfo {author} {\bibfnamefont {M.~J.}\ \bibnamefont {Cervia}},
  \bibinfo {author} {\bibfnamefont {H.}~\bibnamefont {Kumar}}, \bibinfo
  {author} {\bibfnamefont {P.~F.}\ \bibnamefont {Bedaque}},\ and\ \bibinfo
  {author} {\bibfnamefont {A.}~\bibnamefont {Alexandru}},\ }\href@noop {}
  {\bibinfo {title} {{How many quantum gates do gauge theories require?}}}
  (\bibinfo {year} {2022}{\natexlab{a}}),\ \Eprint
  {https://arxiv.org/abs/2208.11789} {arXiv:2208.11789 [hep-lat]} \BibitemShut
  {NoStop}%
\bibitem [{\citenamefont {Zohar}\ \emph {et~al.}(2016)\citenamefont {Zohar},
  \citenamefont {Cirac},\ and\ \citenamefont {Reznik}}]{Zohar:2015hwa}%
  \BibitemOpen
  \bibfield  {author} {\bibinfo {author} {\bibfnamefont {E.}~\bibnamefont
  {Zohar}}, \bibinfo {author} {\bibfnamefont {J.~I.}\ \bibnamefont {Cirac}},\
  and\ \bibinfo {author} {\bibfnamefont {B.}~\bibnamefont {Reznik}},\
  }\bibfield  {title} {\bibinfo {title} {{Quantum Simulations of Lattice Gauge
  Theories using Ultracold Atoms in Optical Lattices}},\ }\href
  {https://doi.org/10.1088/0034-4885/79/1/014401} {\bibfield  {journal}
  {\bibinfo  {journal} {Rept. Prog. Phys.}\ }\textbf {\bibinfo {volume} {79}},\
  \bibinfo {pages} {014401} (\bibinfo {year} {2016})},\ \Eprint
  {https://arxiv.org/abs/1503.02312} {arXiv:1503.02312 [quant-ph]} \BibitemShut
  {NoStop}%
\bibitem [{\citenamefont {Zohar}\ \emph
  {et~al.}(2013{\natexlab{a}})\citenamefont {Zohar}, \citenamefont {Cirac},\
  and\ \citenamefont {Reznik}}]{Zohar:2012xf}%
  \BibitemOpen
  \bibfield  {author} {\bibinfo {author} {\bibfnamefont {E.}~\bibnamefont
  {Zohar}}, \bibinfo {author} {\bibfnamefont {J.~I.}\ \bibnamefont {Cirac}},\
  and\ \bibinfo {author} {\bibfnamefont {B.}~\bibnamefont {Reznik}},\
  }\bibfield  {title} {\bibinfo {title} {{Cold-Atom Quantum Simulator for SU(2)
  Yang-Mills Lattice Gauge Theory}},\ }\href
  {https://doi.org/10.1103/PhysRevLett.110.125304} {\bibfield  {journal}
  {\bibinfo  {journal} {Phys. Rev. Lett.}\ }\textbf {\bibinfo {volume} {110}},\
  \bibinfo {pages} {125304} (\bibinfo {year} {2013}{\natexlab{a}})},\ \Eprint
  {https://arxiv.org/abs/1211.2241} {arXiv:1211.2241 [quant-ph]} \BibitemShut
  {NoStop}%
\bibitem [{\citenamefont {Zohar}\ \emph {et~al.}(2012)\citenamefont {Zohar},
  \citenamefont {Cirac},\ and\ \citenamefont {Reznik}}]{Zohar:2012ay}%
  \BibitemOpen
  \bibfield  {author} {\bibinfo {author} {\bibfnamefont {E.}~\bibnamefont
  {Zohar}}, \bibinfo {author} {\bibfnamefont {J.~I.}\ \bibnamefont {Cirac}},\
  and\ \bibinfo {author} {\bibfnamefont {B.}~\bibnamefont {Reznik}},\
  }\bibfield  {title} {\bibinfo {title} {{Simulating Compact Quantum
  Electrodynamics with ultracold atoms: Probing confinement and nonperturbative
  effects}},\ }\href {https://doi.org/10.1103/PhysRevLett.109.125302}
  {\bibfield  {journal} {\bibinfo  {journal} {Phys. Rev. Lett.}\ }\textbf
  {\bibinfo {volume} {109}},\ \bibinfo {pages} {125302} (\bibinfo {year}
  {2012})},\ \Eprint {https://arxiv.org/abs/1204.6574} {arXiv:1204.6574
  [quant-ph]} \BibitemShut {NoStop}%
\bibitem [{\citenamefont {Zohar}\ \emph
  {et~al.}(2013{\natexlab{b}})\citenamefont {Zohar}, \citenamefont {Cirac},\
  and\ \citenamefont {Reznik}}]{Zohar:2013zla}%
  \BibitemOpen
  \bibfield  {author} {\bibinfo {author} {\bibfnamefont {E.}~\bibnamefont
  {Zohar}}, \bibinfo {author} {\bibfnamefont {J.~I.}\ \bibnamefont {Cirac}},\
  and\ \bibinfo {author} {\bibfnamefont {B.}~\bibnamefont {Reznik}},\
  }\bibfield  {title} {\bibinfo {title} {{Quantum simulations of gauge theories
  with ultracold atoms: local gauge invariance from angular momentum
  conservation}},\ }\href {https://doi.org/10.1103/PhysRevA.88.023617}
  {\bibfield  {journal} {\bibinfo  {journal} {Phys. Rev.}\ }\textbf {\bibinfo
  {volume} {A88}},\ \bibinfo {pages} {023617} (\bibinfo {year}
  {2013}{\natexlab{b}})},\ \Eprint {https://arxiv.org/abs/1303.5040}
  {arXiv:1303.5040 [quant-ph]} \BibitemShut {NoStop}%
\bibitem [{\citenamefont {Zache}\ \emph
  {et~al.}(2023{\natexlab{b}})\citenamefont {Zache}, \citenamefont
  {Gonz\'alez-Cuadra},\ and\ \citenamefont {Zoller}}]{Zache:2023dko}%
  \BibitemOpen
  \bibfield  {author} {\bibinfo {author} {\bibfnamefont {T.~V.}\ \bibnamefont
  {Zache}}, \bibinfo {author} {\bibfnamefont {D.}~\bibnamefont
  {Gonz\'alez-Cuadra}},\ and\ \bibinfo {author} {\bibfnamefont
  {P.}~\bibnamefont {Zoller}},\ }\bibfield  {title} {\bibinfo {title} {{Quantum
  and Classical Spin-Network Algorithms for q-Deformed Kogut-Susskind Gauge
  Theories}},\ }\href {https://doi.org/10.1103/PhysRevLett.131.171902}
  {\bibfield  {journal} {\bibinfo  {journal} {Phys. Rev. Lett.}\ }\textbf
  {\bibinfo {volume} {131}},\ \bibinfo {pages} {171902} (\bibinfo {year}
  {2023}{\natexlab{b}})},\ \Eprint {https://arxiv.org/abs/2304.02527}
  {arXiv:2304.02527 [quant-ph]} \BibitemShut {NoStop}%
\bibitem [{\citenamefont {Biedenharn}\ and\ \citenamefont
  {Lohe}(1995)}]{Biedenharn1995}%
  \BibitemOpen
  \bibfield  {author} {\bibinfo {author} {\bibfnamefont {L.~C.}\ \bibnamefont
  {Biedenharn}}\ and\ \bibinfo {author} {\bibfnamefont {M.~A.}\ \bibnamefont
  {Lohe}},\ }\href {https://doi.org/10.1142/2815} {\emph {\bibinfo {title}
  {Quantum Group Symmetry and Q-Tensor Algebras}}}\ (\bibinfo  {publisher}
  {WORLD SCIENTIFIC},\ \bibinfo {year} {1995})\BibitemShut {NoStop}%
\bibitem [{\citenamefont {Davoudi}\ \emph {et~al.}(2024)\citenamefont
  {Davoudi}, \citenamefont {Hsieh},\ and\ \citenamefont
  {Kadam}}]{davoudi2024scattering}%
  \BibitemOpen
  \bibfield  {author} {\bibinfo {author} {\bibfnamefont {Z.}~\bibnamefont
  {Davoudi}}, \bibinfo {author} {\bibfnamefont {C.-C.}\ \bibnamefont {Hsieh}},\
  and\ \bibinfo {author} {\bibfnamefont {S.~V.}\ \bibnamefont {Kadam}},\
  }\href@noop {} {\bibinfo {title} {Scattering wave packets of hadrons in gauge
  theories: Preparation on a quantum computer}} (\bibinfo {year} {2024}),\
  \Eprint {https://arxiv.org/abs/2402.00840} {arXiv:2402.00840 [quant-ph]}
  \BibitemShut {NoStop}%
\bibitem [{\citenamefont {Raychowdhury}\ and\ \citenamefont
  {Stryker}(2018)}]{Raychowdhury:2018osk}%
  \BibitemOpen
  \bibfield  {author} {\bibinfo {author} {\bibfnamefont {I.}~\bibnamefont
  {Raychowdhury}}\ and\ \bibinfo {author} {\bibfnamefont {J.~R.}\ \bibnamefont
  {Stryker}},\ }\href@noop {} {\bibinfo {title} {{Solving Gauss's Law on
  Digital Quantum Computers with Loop-String-Hadron Digitization}}} (\bibinfo
  {year} {2018}),\ \Eprint {https://arxiv.org/abs/1812.07554} {arXiv:1812.07554
  [hep-lat]} \BibitemShut {NoStop}%
\bibitem [{\citenamefont {Kadam}(2023)}]{Kadam:2023gli}%
  \BibitemOpen
  \bibfield  {author} {\bibinfo {author} {\bibfnamefont {S.~V.}\ \bibnamefont
  {Kadam}},\ }\emph {\bibinfo {title} {{Theoretical Developments in Lattice
  Gauge Theory for Applications in Double-beta Decay Processes and Quantum
  Simulation}}},\ \href {https://doi.org/10.13016/dspace/cvbq-c4jt} {Ph.D.
  thesis},\ \bibinfo  {school} {Maryland U., College Park} (\bibinfo {year}
  {2023}),\ \Eprint {https://arxiv.org/abs/2312.00780} {arXiv:2312.00780
  [hep-lat]} \BibitemShut {NoStop}%
\bibitem [{\citenamefont {Davoudi}\ \emph {et~al.}(2020)\citenamefont
  {Davoudi}, \citenamefont {Raychowdhury},\ and\ \citenamefont
  {Shaw}}]{Davoudi:2020yln}%
  \BibitemOpen
  \bibfield  {author} {\bibinfo {author} {\bibfnamefont {Z.}~\bibnamefont
  {Davoudi}}, \bibinfo {author} {\bibfnamefont {I.}~\bibnamefont
  {Raychowdhury}},\ and\ \bibinfo {author} {\bibfnamefont {A.}~\bibnamefont
  {Shaw}},\ }\href@noop {} {\bibinfo {title} {{Search for Efficient
  Formulations for Hamiltonian Simulation of non-Abelian Lattice Gauge
  Theories}}} (\bibinfo {year} {2020}),\ \Eprint
  {https://arxiv.org/abs/2009.11802} {arXiv:2009.11802 [hep-lat]} \BibitemShut
  {NoStop}%
\bibitem [{\citenamefont {Mathew}\ and\ \citenamefont
  {Raychowdhury}(2022)}]{Mathew:2022nep}%
  \BibitemOpen
  \bibfield  {author} {\bibinfo {author} {\bibfnamefont {E.}~\bibnamefont
  {Mathew}}\ and\ \bibinfo {author} {\bibfnamefont {I.}~\bibnamefont
  {Raychowdhury}},\ }\bibfield  {title} {\bibinfo {title} {{Protecting local
  and global symmetries in simulating (1+1)D non-Abelian gauge theories}},\
  }\href {https://doi.org/10.1103/PhysRevD.106.054510} {\bibfield  {journal}
  {\bibinfo  {journal} {Phys. Rev. D}\ }\textbf {\bibinfo {volume} {106}},\
  \bibinfo {pages} {054510} (\bibinfo {year} {2022})},\ \Eprint
  {https://arxiv.org/abs/2206.07444} {arXiv:2206.07444 [hep-lat]} \BibitemShut
  {NoStop}%
\bibitem [{\citenamefont {Kreshchuk}\ \emph
  {et~al.}(2020{\natexlab{a}})\citenamefont {Kreshchuk}, \citenamefont {Kirby},
  \citenamefont {Goldstein}, \citenamefont {Beauchemin},\ and\ \citenamefont
  {Love}}]{Kreshchuk:2020dla}%
  \BibitemOpen
  \bibfield  {author} {\bibinfo {author} {\bibfnamefont {M.}~\bibnamefont
  {Kreshchuk}}, \bibinfo {author} {\bibfnamefont {W.~M.}\ \bibnamefont
  {Kirby}}, \bibinfo {author} {\bibfnamefont {G.}~\bibnamefont {Goldstein}},
  \bibinfo {author} {\bibfnamefont {H.}~\bibnamefont {Beauchemin}},\ and\
  \bibinfo {author} {\bibfnamefont {P.~J.}\ \bibnamefont {Love}},\ }\href@noop
  {} {\bibinfo {title} {{Quantum Simulation of Quantum Field Theory in the
  Light-Front Formulation}}} (\bibinfo {year} {2020}{\natexlab{a}}),\ \Eprint
  {https://arxiv.org/abs/2002.04016} {arXiv:2002.04016 [quant-ph]} \BibitemShut
  {NoStop}%
\bibitem [{\citenamefont {Kreshchuk}\ \emph
  {et~al.}(2020{\natexlab{b}})\citenamefont {Kreshchuk}, \citenamefont {Jia},
  \citenamefont {Kirby}, \citenamefont {Goldstein}, \citenamefont {Vary},\ and\
  \citenamefont {Love}}]{Kreshchuk:2020aiq}%
  \BibitemOpen
  \bibfield  {author} {\bibinfo {author} {\bibfnamefont {M.}~\bibnamefont
  {Kreshchuk}}, \bibinfo {author} {\bibfnamefont {S.}~\bibnamefont {Jia}},
  \bibinfo {author} {\bibfnamefont {W.~M.}\ \bibnamefont {Kirby}}, \bibinfo
  {author} {\bibfnamefont {G.}~\bibnamefont {Goldstein}}, \bibinfo {author}
  {\bibfnamefont {J.~P.}\ \bibnamefont {Vary}},\ and\ \bibinfo {author}
  {\bibfnamefont {P.~J.}\ \bibnamefont {Love}},\ }\href@noop {} {\bibinfo
  {title} {{Simulating Hadronic Physics on NISQ devices using Basis Light-Front
  Quantization}}} (\bibinfo {year} {2020}{\natexlab{b}}),\ \Eprint
  {https://arxiv.org/abs/2011.13443} {arXiv:2011.13443 [quant-ph]} \BibitemShut
  {NoStop}%
\bibitem [{\citenamefont {Kreshchuk}\ \emph
  {et~al.}(2020{\natexlab{c}})\citenamefont {Kreshchuk}, \citenamefont {Jia},
  \citenamefont {Kirby}, \citenamefont {Goldstein}, \citenamefont {Vary},\ and\
  \citenamefont {Love}}]{Kreshchuk:2020kcz}%
  \BibitemOpen
  \bibfield  {author} {\bibinfo {author} {\bibfnamefont {M.}~\bibnamefont
  {Kreshchuk}}, \bibinfo {author} {\bibfnamefont {S.}~\bibnamefont {Jia}},
  \bibinfo {author} {\bibfnamefont {W.~M.}\ \bibnamefont {Kirby}}, \bibinfo
  {author} {\bibfnamefont {G.}~\bibnamefont {Goldstein}}, \bibinfo {author}
  {\bibfnamefont {J.~P.}\ \bibnamefont {Vary}},\ and\ \bibinfo {author}
  {\bibfnamefont {P.~J.}\ \bibnamefont {Love}},\ }\href@noop {} {\bibinfo
  {title} {{Light-Front Field Theory on Current Quantum Computers}}} (\bibinfo
  {year} {2020}{\natexlab{c}}),\ \Eprint {https://arxiv.org/abs/2009.07885}
  {arXiv:2009.07885 [quant-ph]} \BibitemShut {NoStop}%
\bibitem [{\citenamefont {Liu}\ and\ \citenamefont {Xin}(2020)}]{Liu:2020eoa}%
  \BibitemOpen
  \bibfield  {author} {\bibinfo {author} {\bibfnamefont {J.}~\bibnamefont
  {Liu}}\ and\ \bibinfo {author} {\bibfnamefont {Y.}~\bibnamefont {Xin}},\
  }\href@noop {} {\bibinfo {title} {{Quantum simulation of quantum field
  theories as quantum chemistry}}} (\bibinfo {year} {2020}),\ \Eprint
  {https://arxiv.org/abs/2004.13234} {arXiv:2004.13234 [hep-th]} \BibitemShut
  {NoStop}%
\bibitem [{\citenamefont {Fromm}\ \emph {et~al.}(2023)\citenamefont {Fromm},
  \citenamefont {Philipsen}, \citenamefont {Unger},\ and\ \citenamefont
  {Winterowd}}]{Fromm:2023bit}%
  \BibitemOpen
  \bibfield  {author} {\bibinfo {author} {\bibfnamefont {M.}~\bibnamefont
  {Fromm}}, \bibinfo {author} {\bibfnamefont {O.}~\bibnamefont {Philipsen}},
  \bibinfo {author} {\bibfnamefont {W.}~\bibnamefont {Unger}},\ and\ \bibinfo
  {author} {\bibfnamefont {C.}~\bibnamefont {Winterowd}},\ }\href@noop {}
  {\bibinfo {title} {{Quantum Gate Sets for Lattice QCD in the strong coupling
  limit: $N_f=1$}}} (\bibinfo {year} {2023}),\ \Eprint
  {https://arxiv.org/abs/2308.03196} {arXiv:2308.03196 [hep-lat]} \BibitemShut
  {NoStop}%
\bibitem [{\citenamefont {Ciavarella}\ and\ \citenamefont
  {Bauer}(2024)}]{Ciavarella:2024fzw}%
  \BibitemOpen
  \bibfield  {author} {\bibinfo {author} {\bibfnamefont {A.~N.}\ \bibnamefont
  {Ciavarella}}\ and\ \bibinfo {author} {\bibfnamefont {C.~W.}\ \bibnamefont
  {Bauer}},\ }\href@noop {} {\bibinfo {title} {{Quantum Simulation of SU(3)
  Lattice Yang Mills Theory at Leading Order in Large N}}} (\bibinfo {year}
  {2024}),\ \Eprint {https://arxiv.org/abs/2402.10265} {arXiv:2402.10265
  [hep-ph]} \BibitemShut {NoStop}%
\bibitem [{\citenamefont {Horn}(1981)}]{HORN1981149}%
  \BibitemOpen
  \bibfield  {author} {\bibinfo {author} {\bibfnamefont {D.}~\bibnamefont
  {Horn}},\ }\bibfield  {title} {\bibinfo {title} {Finite matrix models with
  continuous local gauge invariance},\ }\href
  {https://doi.org/https://doi.org/10.1016/0370-2693(81)90763-2} {\bibfield
  {journal} {\bibinfo  {journal} {Physics Letters B}\ }\textbf {\bibinfo
  {volume} {100}},\ \bibinfo {pages} {149} (\bibinfo {year}
  {1981})}\BibitemShut {NoStop}%
\bibitem [{\citenamefont {Orland}\ and\ \citenamefont
  {Rohrlich}(1990)}]{ORLAND1990647}%
  \BibitemOpen
  \bibfield  {author} {\bibinfo {author} {\bibfnamefont {P.}~\bibnamefont
  {Orland}}\ and\ \bibinfo {author} {\bibfnamefont {D.}~\bibnamefont
  {Rohrlich}},\ }\bibfield  {title} {\bibinfo {title} {Lattice gauge magnets:
  Local isospin from spin},\ }\href
  {https://doi.org/https://doi.org/10.1016/0550-3213(90)90646-U} {\bibfield
  {journal} {\bibinfo  {journal} {Nuclear Physics B}\ }\textbf {\bibinfo
  {volume} {338}},\ \bibinfo {pages} {647} (\bibinfo {year}
  {1990})}\BibitemShut {NoStop}%
\bibitem [{\citenamefont {Alexandru}\ \emph {et~al.}(2023)\citenamefont
  {Alexandru}, \citenamefont {Bedaque}, \citenamefont {Carosso}, \citenamefont
  {Cervia}, \citenamefont {Murairi},\ and\ \citenamefont
  {Sheng}}]{Alexandru:2023qzd}%
  \BibitemOpen
  \bibfield  {author} {\bibinfo {author} {\bibfnamefont {A.}~\bibnamefont
  {Alexandru}}, \bibinfo {author} {\bibfnamefont {P.~F.}\ \bibnamefont
  {Bedaque}}, \bibinfo {author} {\bibfnamefont {A.}~\bibnamefont {Carosso}},
  \bibinfo {author} {\bibfnamefont {M.~J.}\ \bibnamefont {Cervia}}, \bibinfo
  {author} {\bibfnamefont {E.~M.}\ \bibnamefont {Murairi}},\ and\ \bibinfo
  {author} {\bibfnamefont {A.}~\bibnamefont {Sheng}},\ }\bibfield  {title}
  {\bibinfo {title} {{Fuzzy Gauge Theory for Quantum Computers}},\ }\href@noop
  {} {\bibfield  {journal} {\bibinfo  {journal} {arXiv e-prints}\ } (\bibinfo
  {year} {2023})},\ \Eprint {https://arxiv.org/abs/2308.05253}
  {arXiv:2308.05253 [hep-lat]} \BibitemShut {NoStop}%
\bibitem [{\citenamefont {Brower}\ \emph {et~al.}(1999)\citenamefont {Brower},
  \citenamefont {Chandrasekharan},\ and\ \citenamefont
  {Wiese}}]{Brower:1997ha}%
  \BibitemOpen
  \bibfield  {author} {\bibinfo {author} {\bibfnamefont {R.}~\bibnamefont
  {Brower}}, \bibinfo {author} {\bibfnamefont {S.}~\bibnamefont
  {Chandrasekharan}},\ and\ \bibinfo {author} {\bibfnamefont {U.~J.}\
  \bibnamefont {Wiese}},\ }\bibfield  {title} {\bibinfo {title} {{QCD as a
  quantum link model}},\ }\href {https://doi.org/10.1103/PhysRevD.60.094502}
  {\bibfield  {journal} {\bibinfo  {journal} {Phys. Rev. D}\ }\textbf {\bibinfo
  {volume} {60}},\ \bibinfo {pages} {094502} (\bibinfo {year} {1999})},\
  \Eprint {https://arxiv.org/abs/hep-th/9704106} {arXiv:hep-th/9704106}
  \BibitemShut {NoStop}%
\bibitem [{\citenamefont {Singh}(2019)}]{Singh:2019jog}%
  \BibitemOpen
  \bibfield  {author} {\bibinfo {author} {\bibfnamefont {H.}~\bibnamefont
  {Singh}},\ }\href@noop {} {\bibinfo {title} {{Qubit $O(N)$ nonlinear sigma
  models}}} (\bibinfo {year} {2019}),\ \Eprint
  {https://arxiv.org/abs/1911.12353} {arXiv:1911.12353 [hep-lat]} \BibitemShut
  {NoStop}%
\bibitem [{\citenamefont {Singh}\ and\ \citenamefont
  {Chandrasekharan}(2019)}]{Singh:2019uwd}%
  \BibitemOpen
  \bibfield  {author} {\bibinfo {author} {\bibfnamefont {H.}~\bibnamefont
  {Singh}}\ and\ \bibinfo {author} {\bibfnamefont {S.}~\bibnamefont
  {Chandrasekharan}},\ }\bibfield  {title} {\bibinfo {title} {{Qubit
  regularization of the $O(3)$ sigma model}},\ }\href
  {https://doi.org/10.1103/PhysRevD.100.054505} {\bibfield  {journal} {\bibinfo
   {journal} {Phys. Rev. D}\ }\textbf {\bibinfo {volume} {100}},\ \bibinfo
  {pages} {054505} (\bibinfo {year} {2019})},\ \Eprint
  {https://arxiv.org/abs/1905.13204} {arXiv:1905.13204 [hep-lat]} \BibitemShut
  {NoStop}%
\bibitem [{\citenamefont {Wiese}(2014)}]{Wiese:2014rla}%
  \BibitemOpen
  \bibfield  {author} {\bibinfo {author} {\bibfnamefont {U.-J.}\ \bibnamefont
  {Wiese}},\ }\bibfield  {title} {\bibinfo {title} {{Towards Quantum Simulating
  QCD}},\ }\bibfield  {booktitle} {\emph {\bibinfo {booktitle} {{Proceedings,
  24th International Conference on Ultra-Relativistic Nucleus-Nucleus
  Collisions (Quark Matter 2014): Darmstadt, Germany, May 19-24, 2014}}},\
  }\href {https://doi.org/10.1016/j.nuclphysa.2014.09.102} {\bibfield
  {journal} {\bibinfo  {journal} {Nucl. Phys.}\ }\textbf {\bibinfo {volume}
  {A931}},\ \bibinfo {pages} {246} (\bibinfo {year} {2014})},\ \Eprint
  {https://arxiv.org/abs/1409.7414} {arXiv:1409.7414 [hep-th]} \BibitemShut
  {NoStop}%
\bibitem [{\citenamefont {Brower}\ \emph {et~al.}(2019)\citenamefont {Brower},
  \citenamefont {Berenstein},\ and\ \citenamefont {Kawai}}]{Brower:2020huh}%
  \BibitemOpen
  \bibfield  {author} {\bibinfo {author} {\bibfnamefont {R.~C.}\ \bibnamefont
  {Brower}}, \bibinfo {author} {\bibfnamefont {D.}~\bibnamefont {Berenstein}},\
  and\ \bibinfo {author} {\bibfnamefont {H.}~\bibnamefont {Kawai}},\ }\bibfield
   {title} {\bibinfo {title} {{Lattice Gauge Theory for a Quantum Computer}},\
  }\href@noop {} {\bibfield  {journal} {\bibinfo  {journal} {PoS}\ }\textbf
  {\bibinfo {volume} {LATTICE2019}},\ \bibinfo {pages} {112} (\bibinfo {year}
  {2019})},\ \Eprint {https://arxiv.org/abs/2002.10028} {arXiv:2002.10028
  [hep-lat]} \BibitemShut {NoStop}%
\bibitem [{\citenamefont {Mathis}\ \emph {et~al.}(2020)\citenamefont {Mathis},
  \citenamefont {Mazzola},\ and\ \citenamefont {Tavernelli}}]{Mathis:2020fuo}%
  \BibitemOpen
  \bibfield  {author} {\bibinfo {author} {\bibfnamefont {S.~V.}\ \bibnamefont
  {Mathis}}, \bibinfo {author} {\bibfnamefont {G.}~\bibnamefont {Mazzola}},\
  and\ \bibinfo {author} {\bibfnamefont {I.}~\bibnamefont {Tavernelli}},\
  }\bibfield  {title} {\bibinfo {title} {{Toward scalable simulations of
  Lattice Gauge Theories on quantum computers}},\ }\href
  {https://doi.org/10.1103/PhysRevD.102.094501} {\bibfield  {journal} {\bibinfo
   {journal} {Phys. Rev. D}\ }\textbf {\bibinfo {volume} {102}},\ \bibinfo
  {pages} {094501} (\bibinfo {year} {2020})}\BibitemShut {NoStop}%
\bibitem [{\citenamefont {Halimeh}\ \emph {et~al.}(2020)\citenamefont
  {Halimeh}, \citenamefont {Ott}, \citenamefont {McCulloch}, \citenamefont
  {Yang},\ and\ \citenamefont {Hauke}}]{Halimeh:2020xfd}%
  \BibitemOpen
  \bibfield  {author} {\bibinfo {author} {\bibfnamefont {J.~C.}\ \bibnamefont
  {Halimeh}}, \bibinfo {author} {\bibfnamefont {R.}~\bibnamefont {Ott}},
  \bibinfo {author} {\bibfnamefont {I.~P.}\ \bibnamefont {McCulloch}}, \bibinfo
  {author} {\bibfnamefont {B.}~\bibnamefont {Yang}},\ and\ \bibinfo {author}
  {\bibfnamefont {P.}~\bibnamefont {Hauke}},\ }\bibfield  {title} {\bibinfo
  {title} {{Robustness of gauge-invariant dynamics against defects in
  ultracold-atom gauge theories}},\ }\href
  {https://doi.org/10.1103/PhysRevResearch.2.033361} {\bibfield  {journal}
  {\bibinfo  {journal} {Phys. Rev. Res.}\ }\textbf {\bibinfo {volume} {2}},\
  \bibinfo {pages} {033361} (\bibinfo {year} {2020})},\ \Eprint
  {https://arxiv.org/abs/2005.10249} {arXiv:2005.10249 [cond-mat.quant-gas]}
  \BibitemShut {NoStop}%
\bibitem [{\citenamefont {Budde}\ \emph {et~al.}(2024)\citenamefont {Budde},
  \citenamefont {Marinković},\ and\ \citenamefont
  {Barros}}]{budde2024quantum}%
  \BibitemOpen
  \bibfield  {author} {\bibinfo {author} {\bibfnamefont {T.}~\bibnamefont
  {Budde}}, \bibinfo {author} {\bibfnamefont {M.~K.}\ \bibnamefont
  {Marinković}},\ and\ \bibinfo {author} {\bibfnamefont {J.~C.~P.}\
  \bibnamefont {Barros}},\ }\href@noop {} {\bibinfo {title} {Quantum many-body
  scars for arbitrary integer spin in $2+1$d abelian gauge theories}} (\bibinfo
  {year} {2024}),\ \Eprint {https://arxiv.org/abs/2403.08892} {arXiv:2403.08892
  [hep-lat]} \BibitemShut {NoStop}%
\bibitem [{\citenamefont {Osborne}\ \emph {et~al.}(2024)\citenamefont
  {Osborne}, \citenamefont {McCulloch},\ and\ \citenamefont
  {Halimeh}}]{osborne2024quantum}%
  \BibitemOpen
  \bibfield  {author} {\bibinfo {author} {\bibfnamefont {J.}~\bibnamefont
  {Osborne}}, \bibinfo {author} {\bibfnamefont {I.~P.}\ \bibnamefont
  {McCulloch}},\ and\ \bibinfo {author} {\bibfnamefont {J.~C.}\ \bibnamefont
  {Halimeh}},\ }\href@noop {} {\bibinfo {title} {Quantum many-body scarring in
  $2+1$d gauge theories with dynamical matter}} (\bibinfo {year} {2024}),\
  \Eprint {https://arxiv.org/abs/2403.08858} {arXiv:2403.08858
  [cond-mat.quant-gas]} \BibitemShut {NoStop}%
\bibitem [{\citenamefont {Osborne}\ \emph {et~al.}(2023)\citenamefont
  {Osborne}, \citenamefont {Yang}, \citenamefont {McCulloch}, \citenamefont
  {Hauke},\ and\ \citenamefont {Halimeh}}]{Osborne:2023rzx}%
  \BibitemOpen
  \bibfield  {author} {\bibinfo {author} {\bibfnamefont {J.}~\bibnamefont
  {Osborne}}, \bibinfo {author} {\bibfnamefont {B.}~\bibnamefont {Yang}},
  \bibinfo {author} {\bibfnamefont {I.~P.}\ \bibnamefont {McCulloch}}, \bibinfo
  {author} {\bibfnamefont {P.}~\bibnamefont {Hauke}},\ and\ \bibinfo {author}
  {\bibfnamefont {J.~C.}\ \bibnamefont {Halimeh}},\ }\href@noop {} {\bibinfo
  {title} {{Spin-$S$$\mathrm{U}(1)$ Quantum Link Models with Dynamical Matter
  on a Quantum Simulator}}} (\bibinfo {year} {2023}),\ \Eprint
  {https://arxiv.org/abs/2305.06368} {arXiv:2305.06368 [cond-mat.quant-gas]}
  \BibitemShut {NoStop}%
\bibitem [{\citenamefont {Luo}\ \emph {et~al.}(2019)\citenamefont {Luo},
  \citenamefont {Shen}, \citenamefont {Highman}, \citenamefont {Clark},
  \citenamefont {DeMarco}, \citenamefont {El-Khadra},\ and\ \citenamefont
  {Gadway}}]{Luo:2019vmi}%
  \BibitemOpen
  \bibfield  {author} {\bibinfo {author} {\bibfnamefont {D.}~\bibnamefont
  {Luo}}, \bibinfo {author} {\bibfnamefont {J.}~\bibnamefont {Shen}}, \bibinfo
  {author} {\bibfnamefont {M.}~\bibnamefont {Highman}}, \bibinfo {author}
  {\bibfnamefont {B.~K.}\ \bibnamefont {Clark}}, \bibinfo {author}
  {\bibfnamefont {B.}~\bibnamefont {DeMarco}}, \bibinfo {author} {\bibfnamefont
  {A.~X.}\ \bibnamefont {El-Khadra}},\ and\ \bibinfo {author} {\bibfnamefont
  {B.}~\bibnamefont {Gadway}},\ }\href@noop {} {\bibinfo {title} {{A Framework
  for Simulating Gauge Theories with Dipolar Spin Systems}}} (\bibinfo {year}
  {2019}),\ \Eprint {https://arxiv.org/abs/1912.11488} {arXiv:1912.11488
  [quant-ph]} \BibitemShut {NoStop}%
\bibitem [{\citenamefont {Singh}(2022)}]{PhysRevD.105.114509}%
  \BibitemOpen
  \bibfield  {author} {\bibinfo {author} {\bibfnamefont {H.}~\bibnamefont
  {Singh}},\ }\bibfield  {title} {\bibinfo {title} {Qubit regularized $o(n)$
  nonlinear sigma models},\ }\href
  {https://doi.org/10.1103/PhysRevD.105.114509} {\bibfield  {journal} {\bibinfo
   {journal} {Phys. Rev. D}\ }\textbf {\bibinfo {volume} {105}},\ \bibinfo
  {pages} {114509} (\bibinfo {year} {2022})}\BibitemShut {NoStop}%
\bibitem [{\citenamefont {Bender}\ \emph {et~al.}(2018)\citenamefont {Bender},
  \citenamefont {Zohar}, \citenamefont {Farace},\ and\ \citenamefont
  {Cirac}}]{Bender:2018rdp}%
  \BibitemOpen
  \bibfield  {author} {\bibinfo {author} {\bibfnamefont {J.}~\bibnamefont
  {Bender}}, \bibinfo {author} {\bibfnamefont {E.}~\bibnamefont {Zohar}},
  \bibinfo {author} {\bibfnamefont {A.}~\bibnamefont {Farace}},\ and\ \bibinfo
  {author} {\bibfnamefont {J.~I.}\ \bibnamefont {Cirac}},\ }\bibfield  {title}
  {\bibinfo {title} {{Digital quantum simulation of lattice gauge theories in
  three spatial dimensions}},\ }\href
  {https://doi.org/10.1088/1367-2630/aadb71} {\bibfield  {journal} {\bibinfo
  {journal} {New J. Phys.}\ }\textbf {\bibinfo {volume} {20}},\ \bibinfo
  {pages} {093001} (\bibinfo {year} {2018})},\ \Eprint
  {https://arxiv.org/abs/1804.02082} {arXiv:1804.02082 [quant-ph]} \BibitemShut
  {NoStop}%
\bibitem [{\citenamefont {Hackett}\ \emph {et~al.}(2019)\citenamefont
  {Hackett}, \citenamefont {Howe}, \citenamefont {Hughes}, \citenamefont {Jay},
  \citenamefont {Neil},\ and\ \citenamefont {Simone}}]{Hackett:2018cel}%
  \BibitemOpen
  \bibfield  {author} {\bibinfo {author} {\bibfnamefont {D.~C.}\ \bibnamefont
  {Hackett}}, \bibinfo {author} {\bibfnamefont {K.}~\bibnamefont {Howe}},
  \bibinfo {author} {\bibfnamefont {C.}~\bibnamefont {Hughes}}, \bibinfo
  {author} {\bibfnamefont {W.}~\bibnamefont {Jay}}, \bibinfo {author}
  {\bibfnamefont {E.~T.}\ \bibnamefont {Neil}},\ and\ \bibinfo {author}
  {\bibfnamefont {J.~N.}\ \bibnamefont {Simone}},\ }\bibfield  {title}
  {\bibinfo {title} {{Digitizing Gauge Fields: Lattice Monte Carlo Results for
  Future Quantum Computers}},\ }\href
  {https://doi.org/10.1103/PhysRevA.99.062341} {\bibfield  {journal} {\bibinfo
  {journal} {Phys.\ Rev.\ A}\ }\textbf {\bibinfo {volume} {99}},\ \bibinfo
  {pages} {062341} (\bibinfo {year} {2019})},\ \Eprint
  {https://arxiv.org/abs/1811.03629} {arXiv:1811.03629 [quant-ph]} \BibitemShut
  {NoStop}%
\bibitem [{\citenamefont {Alexandru}\ \emph {et~al.}(2019)\citenamefont
  {Alexandru}, \citenamefont {Bedaque}, \citenamefont {Harmalkar},
  \citenamefont {Lamm}, \citenamefont {Lawrence},\ and\ \citenamefont
  {Warrington}}]{Alexandru:2019nsa}%
  \BibitemOpen
  \bibfield  {author} {\bibinfo {author} {\bibfnamefont {A.}~\bibnamefont
  {Alexandru}}, \bibinfo {author} {\bibfnamefont {P.~F.}\ \bibnamefont
  {Bedaque}}, \bibinfo {author} {\bibfnamefont {S.}~\bibnamefont {Harmalkar}},
  \bibinfo {author} {\bibfnamefont {H.}~\bibnamefont {Lamm}}, \bibinfo {author}
  {\bibfnamefont {S.}~\bibnamefont {Lawrence}},\ and\ \bibinfo {author}
  {\bibfnamefont {N.~C.}\ \bibnamefont {Warrington}} (\bibinfo {collaboration}
  {NuQS}),\ }\bibfield  {title} {\bibinfo {title} {Gluon field digitization for
  quantum computers},\ }\href {https://doi.org/10.1103/PhysRevD.100.114501}
  {\bibfield  {journal} {\bibinfo  {journal} {Phys.Rev.D}\ }\textbf {\bibinfo
  {volume} {100}},\ \bibinfo {pages} {114501} (\bibinfo {year} {2019})},\
  \Eprint {https://arxiv.org/abs/1906.11213} {arXiv:1906.11213 [hep-lat]}
  \BibitemShut {NoStop}%
\bibitem [{\citenamefont {Yamamoto}(2021)}]{Yamamoto:2020eqi}%
  \BibitemOpen
  \bibfield  {author} {\bibinfo {author} {\bibfnamefont {A.}~\bibnamefont
  {Yamamoto}},\ }\bibfield  {title} {\bibinfo {title} {{Real-time simulation of
  (2+1)-dimensional lattice gauge theory on qubits}},\ }\href
  {https://doi.org/10.1093/ptep/ptaa171} {\bibfield  {journal} {\bibinfo
  {journal} {PTEP}\ }\textbf {\bibinfo {volume} {2021}},\ \bibinfo {pages}
  {013B06} (\bibinfo {year} {2021})},\ \Eprint
  {https://arxiv.org/abs/2008.11395} {arXiv:2008.11395 [hep-lat]} \BibitemShut
  {NoStop}%
\bibitem [{\citenamefont {Ji}\ \emph {et~al.}(2020)\citenamefont {Ji},
  \citenamefont {Lamm},\ and\ \citenamefont {Zhu}}]{Ji:2020kjk}%
  \BibitemOpen
  \bibfield  {author} {\bibinfo {author} {\bibfnamefont {Y.}~\bibnamefont
  {Ji}}, \bibinfo {author} {\bibfnamefont {H.}~\bibnamefont {Lamm}},\ and\
  \bibinfo {author} {\bibfnamefont {S.}~\bibnamefont {Zhu}} (\bibinfo
  {collaboration} {NuQS}),\ }\bibfield  {title} {\bibinfo {title} {{Gluon Field
  Digitization via Group Space Decimation for Quantum Computers}},\ }\href
  {https://doi.org/10.1103/PhysRevD.102.114513} {\bibfield  {journal} {\bibinfo
   {journal} {Phys. Rev. D}\ }\textbf {\bibinfo {volume} {102}},\ \bibinfo
  {pages} {114513} (\bibinfo {year} {2020})},\ \Eprint
  {https://arxiv.org/abs/2005.14221} {arXiv:2005.14221 [hep-lat]} \BibitemShut
  {NoStop}%
\bibitem [{\citenamefont {Haase}\ \emph {et~al.}(2021)\citenamefont {Haase},
  \citenamefont {Dellantonio}, \citenamefont {Celi}, \citenamefont {Paulson},
  \citenamefont {Kan}, \citenamefont {Jansen},\ and\ \citenamefont
  {Muschik}}]{Haase:2020kaj}%
  \BibitemOpen
  \bibfield  {author} {\bibinfo {author} {\bibfnamefont {J.~F.}\ \bibnamefont
  {Haase}}, \bibinfo {author} {\bibfnamefont {L.}~\bibnamefont {Dellantonio}},
  \bibinfo {author} {\bibfnamefont {A.}~\bibnamefont {Celi}}, \bibinfo {author}
  {\bibfnamefont {D.}~\bibnamefont {Paulson}}, \bibinfo {author} {\bibfnamefont
  {A.}~\bibnamefont {Kan}}, \bibinfo {author} {\bibfnamefont {K.}~\bibnamefont
  {Jansen}},\ and\ \bibinfo {author} {\bibfnamefont {C.~A.}\ \bibnamefont
  {Muschik}},\ }\bibfield  {title} {\bibinfo {title} {{A resource efficient
  approach for quantum and classical simulations of gauge theories in particle
  physics}},\ }\href {https://doi.org/10.22331/q-2021-02-04-393} {\bibfield
  {journal} {\bibinfo  {journal} {Quantum}\ }\textbf {\bibinfo {volume} {5}},\
  \bibinfo {pages} {393} (\bibinfo {year} {2021})},\ \Eprint
  {https://arxiv.org/abs/2006.14160} {arXiv:2006.14160 [quant-ph]} \BibitemShut
  {NoStop}%
\bibitem [{\citenamefont {Carena}\ \emph {et~al.}(2021)\citenamefont {Carena},
  \citenamefont {Lamm}, \citenamefont {Li},\ and\ \citenamefont
  {Liu}}]{Carena:2021ltu}%
  \BibitemOpen
  \bibfield  {author} {\bibinfo {author} {\bibfnamefont {M.}~\bibnamefont
  {Carena}}, \bibinfo {author} {\bibfnamefont {H.}~\bibnamefont {Lamm}},
  \bibinfo {author} {\bibfnamefont {Y.-Y.}\ \bibnamefont {Li}},\ and\ \bibinfo
  {author} {\bibfnamefont {W.}~\bibnamefont {Liu}},\ }\bibfield  {title}
  {\bibinfo {title} {{Lattice renormalization of quantum simulations}},\ }\href
  {https://doi.org/10.1103/PhysRevD.104.094519} {\bibfield  {journal} {\bibinfo
   {journal} {Phys. Rev. D}\ }\textbf {\bibinfo {volume} {104}},\ \bibinfo
  {pages} {094519} (\bibinfo {year} {2021})},\ \Eprint
  {https://arxiv.org/abs/2107.01166} {arXiv:2107.01166 [hep-lat]} \BibitemShut
  {NoStop}%
\bibitem [{\citenamefont {Armon}\ \emph {et~al.}(2021)\citenamefont {Armon},
  \citenamefont {Ashkenazi}, \citenamefont {Garc\'\i{}a-Moreno}, \citenamefont
  {Gonz\'alez-Tudela},\ and\ \citenamefont {Zohar}}]{Armon:2021uqr}%
  \BibitemOpen
  \bibfield  {author} {\bibinfo {author} {\bibfnamefont {T.}~\bibnamefont
  {Armon}}, \bibinfo {author} {\bibfnamefont {S.}~\bibnamefont {Ashkenazi}},
  \bibinfo {author} {\bibfnamefont {G.}~\bibnamefont {Garc\'\i{}a-Moreno}},
  \bibinfo {author} {\bibfnamefont {A.}~\bibnamefont {Gonz\'alez-Tudela}},\
  and\ \bibinfo {author} {\bibfnamefont {E.}~\bibnamefont {Zohar}},\
  }\href@noop {} {\bibinfo {title} {{Photon-mediated Stroboscopic Quantum
  Simulation of a $\mathbb{Z}_{2}$ Lattice Gauge Theory}}} (\bibinfo {year}
  {2021}),\ \Eprint {https://arxiv.org/abs/2107.13024} {arXiv:2107.13024
  [quant-ph]} \BibitemShut {NoStop}%
\bibitem [{\citenamefont {Charles}\ \emph {et~al.}(2024)\citenamefont
  {Charles}, \citenamefont {Gustafson}, \citenamefont {Hardt}, \citenamefont
  {Herren}, \citenamefont {Hogan}, \citenamefont {Lamm}, \citenamefont
  {Starecheski}, \citenamefont {Van~de Water},\ and\ \citenamefont
  {Wagman}}]{Charles:2023zbl}%
  \BibitemOpen
  \bibfield  {author} {\bibinfo {author} {\bibfnamefont {C.}~\bibnamefont
  {Charles}}, \bibinfo {author} {\bibfnamefont {E.~J.}\ \bibnamefont
  {Gustafson}}, \bibinfo {author} {\bibfnamefont {E.}~\bibnamefont {Hardt}},
  \bibinfo {author} {\bibfnamefont {F.}~\bibnamefont {Herren}}, \bibinfo
  {author} {\bibfnamefont {N.}~\bibnamefont {Hogan}}, \bibinfo {author}
  {\bibfnamefont {H.}~\bibnamefont {Lamm}}, \bibinfo {author} {\bibfnamefont
  {S.}~\bibnamefont {Starecheski}}, \bibinfo {author} {\bibfnamefont {R.~S.}\
  \bibnamefont {Van~de Water}},\ and\ \bibinfo {author} {\bibfnamefont {M.~L.}\
  \bibnamefont {Wagman}},\ }\bibfield  {title} {\bibinfo {title} {{Simulating
  Z2 lattice gauge theory on a quantum computer}},\ }\href
  {https://doi.org/10.1103/PhysRevE.109.015307} {\bibfield  {journal} {\bibinfo
   {journal} {Phys. Rev. E}\ }\textbf {\bibinfo {volume} {109}},\ \bibinfo
  {pages} {015307} (\bibinfo {year} {2024})},\ \Eprint
  {https://arxiv.org/abs/2305.02361} {arXiv:2305.02361 [hep-lat]} \BibitemShut
  {NoStop}%
\bibitem [{\citenamefont {Irmejs}\ \emph {et~al.}(2023)\citenamefont {Irmejs},
  \citenamefont {Banuls},\ and\ \citenamefont {Cirac}}]{Irmejs:2022gwv}%
  \BibitemOpen
  \bibfield  {author} {\bibinfo {author} {\bibfnamefont {R.}~\bibnamefont
  {Irmejs}}, \bibinfo {author} {\bibfnamefont {M.~C.}\ \bibnamefont {Banuls}},\
  and\ \bibinfo {author} {\bibfnamefont {J.~I.}\ \bibnamefont {Cirac}},\
  }\bibfield  {title} {\bibinfo {title} {{Quantum simulation of Z2 lattice
  gauge theory with minimal resources}},\ }\href
  {https://doi.org/10.1103/PhysRevD.108.074503} {\bibfield  {journal} {\bibinfo
   {journal} {Phys. Rev. D}\ }\textbf {\bibinfo {volume} {108}},\ \bibinfo
  {pages} {074503} (\bibinfo {year} {2023})}\BibitemShut {NoStop}%
\bibitem [{\citenamefont {Gustafson}\ and\ \citenamefont
  {Lamm}(2021)}]{Gustafson:2020yfe}%
  \BibitemOpen
  \bibfield  {author} {\bibinfo {author} {\bibfnamefont {E.~J.}\ \bibnamefont
  {Gustafson}}\ and\ \bibinfo {author} {\bibfnamefont {H.}~\bibnamefont
  {Lamm}},\ }\bibfield  {title} {\bibinfo {title} {{Toward quantum simulations
  of $\mathbb{Z}_2$ gauge theory without state preparation}},\ }\href
  {https://doi.org/10.1103/PhysRevD.103.054507} {\bibfield  {journal} {\bibinfo
   {journal} {Phys. Rev. D}\ }\textbf {\bibinfo {volume} {103}},\ \bibinfo
  {pages} {054507} (\bibinfo {year} {2021})},\ \Eprint
  {https://arxiv.org/abs/2011.11677} {arXiv:2011.11677 [hep-lat]} \BibitemShut
  {NoStop}%
\bibitem [{\citenamefont {Hartung}\ \emph {et~al.}(2022)\citenamefont
  {Hartung}, \citenamefont {Jakobs}, \citenamefont {Jansen}, \citenamefont
  {Ostmeyer},\ and\ \citenamefont {Urbach}}]{Hartung:2022hoz}%
  \BibitemOpen
  \bibfield  {author} {\bibinfo {author} {\bibfnamefont {T.}~\bibnamefont
  {Hartung}}, \bibinfo {author} {\bibfnamefont {T.}~\bibnamefont {Jakobs}},
  \bibinfo {author} {\bibfnamefont {K.}~\bibnamefont {Jansen}}, \bibinfo
  {author} {\bibfnamefont {J.}~\bibnamefont {Ostmeyer}},\ and\ \bibinfo
  {author} {\bibfnamefont {C.}~\bibnamefont {Urbach}},\ }\bibfield  {title}
  {\bibinfo {title} {{Digitising SU(2) gauge fields and the freezing
  transition}},\ }\href {https://doi.org/10.1140/epjc/s10052-022-10192-5}
  {\bibfield  {journal} {\bibinfo  {journal} {Eur. Phys. J. C}\ }\textbf
  {\bibinfo {volume} {82}},\ \bibinfo {pages} {237} (\bibinfo {year}
  {2022})}\BibitemShut {NoStop}%
\bibitem [{\citenamefont {Carena}\ \emph {et~al.}(2024)\citenamefont {Carena},
  \citenamefont {Lamm}, \citenamefont {Li},\ and\ \citenamefont
  {Liu}}]{Carena:2024dzu}%
  \BibitemOpen
  \bibfield  {author} {\bibinfo {author} {\bibfnamefont {M.}~\bibnamefont
  {Carena}}, \bibinfo {author} {\bibfnamefont {H.}~\bibnamefont {Lamm}},
  \bibinfo {author} {\bibfnamefont {Y.-Y.}\ \bibnamefont {Li}},\ and\ \bibinfo
  {author} {\bibfnamefont {W.}~\bibnamefont {Liu}},\ }\href@noop {} {\bibinfo
  {title} {{Quantum error thresholds for gauge-redundant digitizations of
  lattice field theories}}} (\bibinfo {year} {2024}),\ \Eprint
  {https://arxiv.org/abs/2402.16780} {arXiv:2402.16780 [hep-lat]} \BibitemShut
  {NoStop}%
\bibitem [{\citenamefont {Zohar}\ and\ \citenamefont
  {Burrello}(2015)}]{Zohar:2014qma}%
  \BibitemOpen
  \bibfield  {author} {\bibinfo {author} {\bibfnamefont {E.}~\bibnamefont
  {Zohar}}\ and\ \bibinfo {author} {\bibfnamefont {M.}~\bibnamefont
  {Burrello}},\ }\bibfield  {title} {\bibinfo {title} {{Formulation of lattice
  gauge theories for quantum simulations}},\ }\href
  {https://doi.org/10.1103/PhysRevD.91.054506} {\bibfield  {journal} {\bibinfo
  {journal} {Phys. Rev.}\ }\textbf {\bibinfo {volume} {D91}},\ \bibinfo {pages}
  {054506} (\bibinfo {year} {2015})},\ \Eprint
  {https://arxiv.org/abs/1409.3085} {arXiv:1409.3085 [quant-ph]} \BibitemShut
  {NoStop}%
\bibitem [{\citenamefont {Zohar}\ \emph {et~al.}(2017)\citenamefont {Zohar},
  \citenamefont {Farace}, \citenamefont {Reznik},\ and\ \citenamefont
  {Cirac}}]{Zohar:2016iic}%
  \BibitemOpen
  \bibfield  {author} {\bibinfo {author} {\bibfnamefont {E.}~\bibnamefont
  {Zohar}}, \bibinfo {author} {\bibfnamefont {A.}~\bibnamefont {Farace}},
  \bibinfo {author} {\bibfnamefont {B.}~\bibnamefont {Reznik}},\ and\ \bibinfo
  {author} {\bibfnamefont {J.~I.}\ \bibnamefont {Cirac}},\ }\bibfield  {title}
  {\bibinfo {title} {{Digital lattice gauge theories}},\ }\href
  {https://doi.org/10.1103/PhysRevA.95.023604} {\bibfield  {journal} {\bibinfo
  {journal} {Phys. Rev.}\ }\textbf {\bibinfo {volume} {A95}},\ \bibinfo {pages}
  {023604} (\bibinfo {year} {2017})},\ \Eprint
  {https://arxiv.org/abs/1607.08121} {arXiv:1607.08121 [quant-ph]} \BibitemShut
  {NoStop}%
\bibitem [{\citenamefont {Creutz}\ \emph {et~al.}(1979)\citenamefont {Creutz},
  \citenamefont {Jacobs},\ and\ \citenamefont {Rebbi}}]{Creutz:1979zg}%
  \BibitemOpen
  \bibfield  {author} {\bibinfo {author} {\bibfnamefont {M.}~\bibnamefont
  {Creutz}}, \bibinfo {author} {\bibfnamefont {L.}~\bibnamefont {Jacobs}},\
  and\ \bibinfo {author} {\bibfnamefont {C.}~\bibnamefont {Rebbi}},\ }\bibfield
   {title} {\bibinfo {title} {{Monte Carlo Study of Abelian Lattice Gauge
  Theories}},\ }\href {https://doi.org/10.1103/PhysRevD.20.1915} {\bibfield
  {journal} {\bibinfo  {journal} {Phys. Rev.}\ }\textbf {\bibinfo {volume}
  {D20}},\ \bibinfo {pages} {1915} (\bibinfo {year} {1979})}\BibitemShut
  {NoStop}%
\bibitem [{\citenamefont {Creutz}\ and\ \citenamefont
  {Okawa}(1983)}]{Creutz:1982dn}%
  \BibitemOpen
  \bibfield  {author} {\bibinfo {author} {\bibfnamefont {M.}~\bibnamefont
  {Creutz}}\ and\ \bibinfo {author} {\bibfnamefont {M.}~\bibnamefont {Okawa}},\
  }\bibfield  {title} {\bibinfo {title} {{Generalized Actions in $Z(p$) Lattice
  Gauge Theory}},\ }\href {https://doi.org/10.1016/0550-3213(83)90220-1}
  {\bibfield  {journal} {\bibinfo  {journal} {Nucl. Phys.}\ }\textbf {\bibinfo
  {volume} {B220}},\ \bibinfo {pages} {149} (\bibinfo {year}
  {1983})}\BibitemShut {NoStop}%
\bibitem [{\citenamefont {Bhanot}\ and\ \citenamefont
  {Rebbi}(1981)}]{Bhanot:1981xp}%
  \BibitemOpen
  \bibfield  {author} {\bibinfo {author} {\bibfnamefont {G.}~\bibnamefont
  {Bhanot}}\ and\ \bibinfo {author} {\bibfnamefont {C.}~\bibnamefont {Rebbi}},\
  }\bibfield  {title} {\bibinfo {title} {{Monte Carlo Simulations of Lattice
  Models With Finite Subgroups of SU(3) as Gauge Groups}},\ }\href
  {https://doi.org/10.1103/PhysRevD.24.3319} {\bibfield  {journal} {\bibinfo
  {journal} {Phys. Rev.}\ }\textbf {\bibinfo {volume} {D24}},\ \bibinfo {pages}
  {3319} (\bibinfo {year} {1981})}\BibitemShut {NoStop}%
\bibitem [{\citenamefont {Petcher}\ and\ \citenamefont
  {Weingarten}(1980)}]{Petcher:1980cq}%
  \BibitemOpen
  \bibfield  {author} {\bibinfo {author} {\bibfnamefont {D.}~\bibnamefont
  {Petcher}}\ and\ \bibinfo {author} {\bibfnamefont {D.~H.}\ \bibnamefont
  {Weingarten}},\ }\bibfield  {title} {\bibinfo {title} {{Monte Carlo
  Calculations and a Model of the Phase Structure for Gauge Theories on
  Discrete Subgroups of SU(2)}},\ }\href
  {https://doi.org/10.1103/PhysRevD.22.2465} {\bibfield  {journal} {\bibinfo
  {journal} {Phys. Rev.}\ }\textbf {\bibinfo {volume} {D22}},\ \bibinfo {pages}
  {2465} (\bibinfo {year} {1980})}\BibitemShut {NoStop}%
\bibitem [{\citenamefont {Bhanot}(1982)}]{Bhanot:1981pj}%
  \BibitemOpen
  \bibfield  {author} {\bibinfo {author} {\bibfnamefont {G.}~\bibnamefont
  {Bhanot}},\ }\bibfield  {title} {\bibinfo {title} {{SU(3) Lattice Gauge
  Theory in Four-dimensions With a Modified Wilson Action}},\ }\href
  {https://doi.org/10.1016/0370-2693(82)91207-2} {\bibfield  {journal}
  {\bibinfo  {journal} {Phys. Lett.}\ }\textbf {\bibinfo {volume} {108B}},\
  \bibinfo {pages} {337} (\bibinfo {year} {1982})}\BibitemShut {NoStop}%
\bibitem [{\citenamefont {Ji}\ \emph {et~al.}(2023)\citenamefont {Ji},
  \citenamefont {Lamm},\ and\ \citenamefont {Zhu}}]{Ji:2022qvr}%
  \BibitemOpen
  \bibfield  {author} {\bibinfo {author} {\bibfnamefont {Y.}~\bibnamefont
  {Ji}}, \bibinfo {author} {\bibfnamefont {H.}~\bibnamefont {Lamm}},\ and\
  \bibinfo {author} {\bibfnamefont {S.}~\bibnamefont {Zhu}} (\bibinfo
  {collaboration} {NuQS Collaboration}),\ }\bibfield  {title} {\bibinfo {title}
  {Gluon digitization via character expansion for quantum computers},\ }\href
  {https://doi.org/10.1103/PhysRevD.107.114503} {\bibfield  {journal} {\bibinfo
   {journal} {Phys. Rev. D}\ }\textbf {\bibinfo {volume} {107}},\ \bibinfo
  {pages} {114503} (\bibinfo {year} {2023})}\BibitemShut {NoStop}%
\bibitem [{\citenamefont {Alexandru}\ \emph {et~al.}(2021)\citenamefont
  {Alexandru}, \citenamefont {Bedaque}, \citenamefont {Brett},\ and\
  \citenamefont {Lamm}}]{Alexandru:2021jpm}%
  \BibitemOpen
  \bibfield  {author} {\bibinfo {author} {\bibfnamefont {A.}~\bibnamefont
  {Alexandru}}, \bibinfo {author} {\bibfnamefont {P.~F.}\ \bibnamefont
  {Bedaque}}, \bibinfo {author} {\bibfnamefont {R.}~\bibnamefont {Brett}},\
  and\ \bibinfo {author} {\bibfnamefont {H.}~\bibnamefont {Lamm}},\ }\href@noop
  {} {\bibinfo {title} {{The spectrum of qubitized QCD: glueballs in a
  $S(1080)$ gauge theory}}} (\bibinfo {year} {2021}),\ \Eprint
  {https://arxiv.org/abs/2112.08482} {arXiv:2112.08482 [hep-lat]} \BibitemShut
  {NoStop}%
\bibitem [{\citenamefont {Carena}\ \emph
  {et~al.}(2022{\natexlab{a}})\citenamefont {Carena}, \citenamefont
  {Gustafson}, \citenamefont {Lamm}, \citenamefont {Li},\ and\ \citenamefont
  {Liu}}]{Carena:2022hpz}%
  \BibitemOpen
  \bibfield  {author} {\bibinfo {author} {\bibfnamefont {M.}~\bibnamefont
  {Carena}}, \bibinfo {author} {\bibfnamefont {E.~J.}\ \bibnamefont
  {Gustafson}}, \bibinfo {author} {\bibfnamefont {H.}~\bibnamefont {Lamm}},
  \bibinfo {author} {\bibfnamefont {Y.-Y.}\ \bibnamefont {Li}},\ and\ \bibinfo
  {author} {\bibfnamefont {W.}~\bibnamefont {Liu}},\ }\bibfield  {title}
  {\bibinfo {title} {{Gauge theory couplings on anisotropic lattices}},\ }\href
  {https://doi.org/10.1103/PhysRevD.106.114504} {\bibfield  {journal} {\bibinfo
   {journal} {Phys. Rev. D}\ }\textbf {\bibinfo {volume} {106}},\ \bibinfo
  {pages} {114504} (\bibinfo {year} {2022}{\natexlab{a}})},\ \Eprint
  {https://arxiv.org/abs/2208.10417} {arXiv:2208.10417 [hep-lat]} \BibitemShut
  {NoStop}%
\bibitem [{\citenamefont {Weingarten}\ and\ \citenamefont
  {Petcher}(1981)}]{Weingarten:1980hx}%
  \BibitemOpen
  \bibfield  {author} {\bibinfo {author} {\bibfnamefont {D.~H.}\ \bibnamefont
  {Weingarten}}\ and\ \bibinfo {author} {\bibfnamefont {D.~N.}\ \bibnamefont
  {Petcher}},\ }\bibfield  {title} {\bibinfo {title} {{Monte Carlo Integration
  for Lattice Gauge Theories with Fermions}},\ }\href
  {https://doi.org/10.1016/0370-2693(81)90112-X} {\bibfield  {journal}
  {\bibinfo  {journal} {Phys. Lett.}\ }\textbf {\bibinfo {volume} {99B}},\
  \bibinfo {pages} {333} (\bibinfo {year} {1981})}\BibitemShut {NoStop}%
\bibitem [{\citenamefont {Weingarten}(1982)}]{Weingarten:1981jy}%
  \BibitemOpen
  \bibfield  {author} {\bibinfo {author} {\bibfnamefont {D.}~\bibnamefont
  {Weingarten}},\ }\bibfield  {title} {\bibinfo {title} {{Monte Carlo
  Evaluation of Hadron Masses in Lattice Gauge Theories with Fermions}},\
  }\href {https://doi.org/10.1016/0370-2693(82)90463-4} {\bibfield  {journal}
  {\bibinfo  {journal} {Phys. Lett.}\ }\textbf {\bibinfo {volume} {109B}},\
  \bibinfo {pages} {57} (\bibinfo {year} {1982})},\ \bibinfo {note}
  {[,631(1981)]}\BibitemShut {NoStop}%
\bibitem [{\citenamefont {Kogut}(1980)}]{Kogut:1980qb}%
  \BibitemOpen
  \bibfield  {author} {\bibinfo {author} {\bibfnamefont {J.~B.}\ \bibnamefont
  {Kogut}},\ }\bibfield  {title} {\bibinfo {title} {{1/n Expansions and the
  Phase Diagram of Discrete Lattice Gauge Theories With Matter Fields}},\
  }\href {https://doi.org/10.1103/PhysRevD.21.2316} {\bibfield  {journal}
  {\bibinfo  {journal} {Phys. Rev. D}\ }\textbf {\bibinfo {volume} {21}},\
  \bibinfo {pages} {2316} (\bibinfo {year} {1980})}\BibitemShut {NoStop}%
\bibitem [{\citenamefont {Romers}(2007)}]{romers2007discrete}%
  \BibitemOpen
  \bibfield  {author} {\bibinfo {author} {\bibfnamefont {J.}~\bibnamefont
  {Romers}},\ }\emph {\bibinfo {title} {Discrete gauge theories in two spatial
  dimensions}},\ \href@noop {} {Ph.D. thesis},\ \bibinfo  {school} {Master’s
  thesis, Universiteit van Amsterdam} (\bibinfo {year} {2007})\BibitemShut
  {NoStop}%
\bibitem [{\citenamefont {Fradkin}\ and\ \citenamefont
  {Shenker}(1979)}]{Fradkin:1978dv}%
  \BibitemOpen
  \bibfield  {author} {\bibinfo {author} {\bibfnamefont {E.~H.}\ \bibnamefont
  {Fradkin}}\ and\ \bibinfo {author} {\bibfnamefont {S.~H.}\ \bibnamefont
  {Shenker}},\ }\bibfield  {title} {\bibinfo {title} {{Phase Diagrams of
  Lattice Gauge Theories with Higgs Fields}},\ }\href
  {https://doi.org/10.1103/PhysRevD.19.3682} {\bibfield  {journal} {\bibinfo
  {journal} {Phys. Rev. D}\ }\textbf {\bibinfo {volume} {19}},\ \bibinfo
  {pages} {3682} (\bibinfo {year} {1979})}\BibitemShut {NoStop}%
\bibitem [{\citenamefont {Harlow}\ and\ \citenamefont
  {Ooguri}(2018)}]{Harlow:2018tng}%
  \BibitemOpen
  \bibfield  {author} {\bibinfo {author} {\bibfnamefont {D.}~\bibnamefont
  {Harlow}}\ and\ \bibinfo {author} {\bibfnamefont {H.}~\bibnamefont
  {Ooguri}},\ }\href@noop {} {\bibinfo {title} {{Symmetries in quantum field
  theory and quantum gravity}}} (\bibinfo {year} {2018}),\ \Eprint
  {https://arxiv.org/abs/1810.05338} {arXiv:1810.05338 [hep-th]} \BibitemShut
  {NoStop}%
\bibitem [{\citenamefont {Horn}\ \emph {et~al.}(1979)\citenamefont {Horn},
  \citenamefont {Weinstein},\ and\ \citenamefont {Yankielowicz}}]{Horn:1979fy}%
  \BibitemOpen
  \bibfield  {author} {\bibinfo {author} {\bibfnamefont {D.}~\bibnamefont
  {Horn}}, \bibinfo {author} {\bibfnamefont {M.}~\bibnamefont {Weinstein}},\
  and\ \bibinfo {author} {\bibfnamefont {S.}~\bibnamefont {Yankielowicz}},\
  }\bibfield  {title} {\bibinfo {title} {{Hamiltonian Approach to Z(N) Lattice
  Gauge Theories}},\ }\href {https://doi.org/10.1103/PhysRevD.19.3715}
  {\bibfield  {journal} {\bibinfo  {journal} {Phys. Rev. D}\ }\textbf {\bibinfo
  {volume} {19}},\ \bibinfo {pages} {3715} (\bibinfo {year}
  {1979})}\BibitemShut {NoStop}%
\bibitem [{\citenamefont {Rajput}\ \emph {et~al.}(2021)\citenamefont {Rajput},
  \citenamefont {Roggero},\ and\ \citenamefont {Wiebe}}]{rajput2021quantum}%
  \BibitemOpen
  \bibfield  {author} {\bibinfo {author} {\bibfnamefont {A.}~\bibnamefont
  {Rajput}}, \bibinfo {author} {\bibfnamefont {A.}~\bibnamefont {Roggero}},\
  and\ \bibinfo {author} {\bibfnamefont {N.}~\bibnamefont {Wiebe}},\
  }\href@noop {} {\bibinfo {title} {Quantum error correction with gauge
  symmetries}} (\bibinfo {year} {2021}),\ \Eprint
  {https://arxiv.org/abs/2112.05186} {arXiv:2112.05186 [quant-ph]} \BibitemShut
  {NoStop}%
\bibitem [{\citenamefont {Carena}\ \emph
  {et~al.}(2022{\natexlab{b}})\citenamefont {Carena}, \citenamefont {Lamm},
  \citenamefont {Li},\ and\ \citenamefont {Liu}}]{Carena:2022kpg}%
  \BibitemOpen
  \bibfield  {author} {\bibinfo {author} {\bibfnamefont {M.}~\bibnamefont
  {Carena}}, \bibinfo {author} {\bibfnamefont {H.}~\bibnamefont {Lamm}},
  \bibinfo {author} {\bibfnamefont {Y.-Y.}\ \bibnamefont {Li}},\ and\ \bibinfo
  {author} {\bibfnamefont {W.}~\bibnamefont {Liu}},\ }\bibfield  {title}
  {\bibinfo {title} {{Improved Hamiltonians for Quantum Simulations of Gauge
  Theories}},\ }\href {https://doi.org/10.1103/PhysRevLett.129.051601}
  {\bibfield  {journal} {\bibinfo  {journal} {Phys. Rev. Lett.}\ }\textbf
  {\bibinfo {volume} {129}},\ \bibinfo {pages} {051601} (\bibinfo {year}
  {2022}{\natexlab{b}})}\BibitemShut {NoStop}%
\bibitem [{\citenamefont {Lamm}\ \emph {et~al.}(2019)\citenamefont {Lamm},
  \citenamefont {Lawrence},\ and\ \citenamefont {Yamauchi}}]{Lamm:2019bik}%
  \BibitemOpen
  \bibfield  {author} {\bibinfo {author} {\bibfnamefont {H.}~\bibnamefont
  {Lamm}}, \bibinfo {author} {\bibfnamefont {S.}~\bibnamefont {Lawrence}},\
  and\ \bibinfo {author} {\bibfnamefont {Y.}~\bibnamefont {Yamauchi}} (\bibinfo
  {collaboration} {NuQS}),\ }\bibfield  {title} {\bibinfo {title} {{General
  Methods for Digital Quantum Simulation of Gauge Theories}},\ }\href
  {https://doi.org/10.1103/PhysRevD.100.034518} {\bibfield  {journal} {\bibinfo
   {journal} {Phys. Rev.}\ }\textbf {\bibinfo {volume} {D100}},\ \bibinfo
  {pages} {034518} (\bibinfo {year} {2019})}\BibitemShut {NoStop}%
\bibitem [{\citenamefont {Alam}\ \emph {et~al.}(2022)\citenamefont {Alam},
  \citenamefont {Hadfield}, \citenamefont {Lamm},\ and\ \citenamefont
  {Li}}]{Alam:2021uuq}%
  \BibitemOpen
  \bibfield  {author} {\bibinfo {author} {\bibfnamefont {M.~S.}\ \bibnamefont
  {Alam}}, \bibinfo {author} {\bibfnamefont {S.}~\bibnamefont {Hadfield}},
  \bibinfo {author} {\bibfnamefont {H.}~\bibnamefont {Lamm}},\ and\ \bibinfo
  {author} {\bibfnamefont {A.~C.~Y.}\ \bibnamefont {Li}} (\bibinfo
  {collaboration} {SQMS}),\ }\bibfield  {title} {\bibinfo {title} {{Primitive
  quantum gates for dihedral gauge theories}},\ }\href
  {https://doi.org/10.1103/PhysRevD.105.114501} {\bibfield  {journal} {\bibinfo
   {journal} {Phys. Rev. D}\ }\textbf {\bibinfo {volume} {105}},\ \bibinfo
  {pages} {114501} (\bibinfo {year} {2022})}\BibitemShut {NoStop}%
\bibitem [{\citenamefont {Fromm}\ \emph {et~al.}(2022)\citenamefont {Fromm},
  \citenamefont {Philipsen},\ and\ \citenamefont {Winterowd}}]{Fromm:2022vaj}%
  \BibitemOpen
  \bibfield  {author} {\bibinfo {author} {\bibfnamefont {M.}~\bibnamefont
  {Fromm}}, \bibinfo {author} {\bibfnamefont {O.}~\bibnamefont {Philipsen}},\
  and\ \bibinfo {author} {\bibfnamefont {C.}~\bibnamefont {Winterowd}},\
  }\href@noop {} {\bibinfo {title} {{Dihedral Lattice Gauge Theories on a
  Quantum Annealer}}} (\bibinfo {year} {2022}),\ \Eprint
  {https://arxiv.org/abs/2206.14679} {arXiv:2206.14679 [hep-lat]} \BibitemShut
  {NoStop}%
\bibitem [{\citenamefont {Gustafson}\ \emph {et~al.}(2022)\citenamefont
  {Gustafson}, \citenamefont {Lamm}, \citenamefont {Lovelace},\ and\
  \citenamefont {Musk}}]{Gustafson:2022xdt}%
  \BibitemOpen
  \bibfield  {author} {\bibinfo {author} {\bibfnamefont {E.~J.}\ \bibnamefont
  {Gustafson}}, \bibinfo {author} {\bibfnamefont {H.}~\bibnamefont {Lamm}},
  \bibinfo {author} {\bibfnamefont {F.}~\bibnamefont {Lovelace}},\ and\
  \bibinfo {author} {\bibfnamefont {D.}~\bibnamefont {Musk}},\ }\bibfield
  {title} {\bibinfo {title} {{Primitive quantum gates for an SU(2) discrete
  subgroup: Binary tetrahedral}},\ }\href
  {https://doi.org/10.1103/PhysRevD.106.114501} {\bibfield  {journal} {\bibinfo
   {journal} {Phys. Rev. D}\ }\textbf {\bibinfo {volume} {106}},\ \bibinfo
  {pages} {114501} (\bibinfo {year} {2022})},\ \Eprint
  {https://arxiv.org/abs/2208.12309} {arXiv:2208.12309 [quant-ph]} \BibitemShut
  {NoStop}%
\bibitem [{\citenamefont {Gustafson}\ \emph {et~al.}(2024)\citenamefont
  {Gustafson}, \citenamefont {Lamm},\ and\ \citenamefont
  {Lovelace}}]{Gustafson:2023kvd}%
  \BibitemOpen
  \bibfield  {author} {\bibinfo {author} {\bibfnamefont {E.~J.}\ \bibnamefont
  {Gustafson}}, \bibinfo {author} {\bibfnamefont {H.}~\bibnamefont {Lamm}},\
  and\ \bibinfo {author} {\bibfnamefont {F.}~\bibnamefont {Lovelace}},\
  }\bibfield  {title} {\bibinfo {title} {{Primitive quantum gates for an SU(2)
  discrete subgroup: Binary octahedral}},\ }\href
  {https://doi.org/10.1103/PhysRevD.109.054503} {\bibfield  {journal} {\bibinfo
   {journal} {Phys. Rev. D}\ }\textbf {\bibinfo {volume} {109}},\ \bibinfo
  {pages} {054503} (\bibinfo {year} {2024})},\ \Eprint
  {https://arxiv.org/abs/2312.10285} {arXiv:2312.10285 [hep-lat]} \BibitemShut
  {NoStop}%
\bibitem [{\citenamefont {P{\"u}schel}\ \emph {et~al.}(1999)\citenamefont
  {P{\"u}schel}, \citenamefont {R{\"o}tteler},\ and\ \citenamefont
  {Beth}}]{Pueschel:1998zzo}%
  \BibitemOpen
  \bibfield  {author} {\bibinfo {author} {\bibfnamefont {M.}~\bibnamefont
  {P{\"u}schel}}, \bibinfo {author} {\bibfnamefont {M.}~\bibnamefont
  {R{\"o}tteler}},\ and\ \bibinfo {author} {\bibfnamefont {T.}~\bibnamefont
  {Beth}},\ }\bibfield  {title} {\bibinfo {title} {Fast quantum {F}ourier
  transforms for a class of non-abelian groups},\ }in\ \href@noop {} {\emph
  {\bibinfo {booktitle} {International Symposium on Applied Algebra, Algebraic
  Algorithms, and Error-Correcting Codes}}}\ (\bibinfo {organization}
  {Springer},\ \bibinfo {year} {1999})\ pp.\ \bibinfo {pages} {148--159},\
  \Eprint {https://arxiv.org/abs/quant-ph/9807064} {arXiv:quant-ph/9807064}
  \BibitemShut {NoStop}%
\bibitem [{\citenamefont {Grimus}\ and\ \citenamefont
  {Ludl}(2010)}]{grimus2010principal}%
  \BibitemOpen
  \bibfield  {author} {\bibinfo {author} {\bibfnamefont {W.}~\bibnamefont
  {Grimus}}\ and\ \bibinfo {author} {\bibfnamefont {P.~O.}\ \bibnamefont
  {Ludl}},\ }\bibfield  {title} {\bibinfo {title} {Principal series of finite
  subgroups of su (3)},\ }\href@noop {} {\bibfield  {journal} {\bibinfo
  {journal} {Journal of Physics A: Mathematical and Theoretical}\ }\textbf
  {\bibinfo {volume} {43}},\ \bibinfo {pages} {445209} (\bibinfo {year}
  {2010})}\BibitemShut {NoStop}%
\bibitem [{\citenamefont {Mariani}\ \emph {et~al.}(2023)\citenamefont
  {Mariani}, \citenamefont {Pradhan},\ and\ \citenamefont
  {Ercolessi}}]{PhysRevD.107.114513}%
  \BibitemOpen
  \bibfield  {author} {\bibinfo {author} {\bibfnamefont {A.}~\bibnamefont
  {Mariani}}, \bibinfo {author} {\bibfnamefont {S.}~\bibnamefont {Pradhan}},\
  and\ \bibinfo {author} {\bibfnamefont {E.}~\bibnamefont {Ercolessi}},\
  }\bibfield  {title} {\bibinfo {title} {Hamiltonians and gauge-invariant
  hilbert space for lattice yang-mills-like theories with finite gauge group},\
  }\href {https://doi.org/10.1103/PhysRevD.107.114513} {\bibfield  {journal}
  {\bibinfo  {journal} {Phys. Rev. D}\ }\textbf {\bibinfo {volume} {107}},\
  \bibinfo {pages} {114513} (\bibinfo {year} {2023})}\BibitemShut {NoStop}%
\bibitem [{\citenamefont {Chuang}\ and\ \citenamefont
  {Nielsen}(1997)}]{Chuang:1996hw}%
  \BibitemOpen
  \bibfield  {author} {\bibinfo {author} {\bibfnamefont {I.~L.}\ \bibnamefont
  {Chuang}}\ and\ \bibinfo {author} {\bibfnamefont {M.~A.}\ \bibnamefont
  {Nielsen}},\ }\bibfield  {title} {\bibinfo {title} {{Prescription for
  experimental determination of the dynamics of a quantum black box}},\ }\href
  {https://doi.org/10.1080/09500349708231894} {\bibfield  {journal} {\bibinfo
  {journal} {J. Mod. Opt.}\ }\textbf {\bibinfo {volume} {44}},\ \bibinfo
  {pages} {2455} (\bibinfo {year} {1997})},\ \Eprint
  {https://arxiv.org/abs/quant-ph/9610001} {arXiv:quant-ph/9610001}
  \BibitemShut {NoStop}%
\bibitem [{\citenamefont {Jordan}\ \emph {et~al.}(2014)\citenamefont {Jordan},
  \citenamefont {Lee},\ and\ \citenamefont {Preskill}}]{Jordan:2011ci}%
  \BibitemOpen
  \bibfield  {author} {\bibinfo {author} {\bibfnamefont {S.~P.}\ \bibnamefont
  {Jordan}}, \bibinfo {author} {\bibfnamefont {K.~S.~M.}\ \bibnamefont {Lee}},\
  and\ \bibinfo {author} {\bibfnamefont {J.}~\bibnamefont {Preskill}},\
  }\bibfield  {title} {\bibinfo {title} {{Quantum Computation of Scattering in
  Scalar Quantum Field Theories}},\ }\href@noop {} {\bibfield  {journal}
  {\bibinfo  {journal} {Quant. Inf. Comput.}\ }\textbf {\bibinfo {volume}
  {14}},\ \bibinfo {pages} {1014} (\bibinfo {year} {2014})},\ \Eprint
  {https://arxiv.org/abs/1112.4833} {arXiv:1112.4833 [hep-th]} \BibitemShut
  {NoStop}%
\bibitem [{\citenamefont {Jordan}\ \emph {et~al.}(2012)\citenamefont {Jordan},
  \citenamefont {Lee},\ and\ \citenamefont {Preskill}}]{Jordan:2011ne}%
  \BibitemOpen
  \bibfield  {author} {\bibinfo {author} {\bibfnamefont {S.~P.}\ \bibnamefont
  {Jordan}}, \bibinfo {author} {\bibfnamefont {K.~S.~M.}\ \bibnamefont {Lee}},\
  and\ \bibinfo {author} {\bibfnamefont {J.}~\bibnamefont {Preskill}},\
  }\bibfield  {title} {\bibinfo {title} {{Quantum Algorithms for Quantum Field
  Theories}},\ }\href {https://doi.org/10.1126/science.1217069} {\bibfield
  {journal} {\bibinfo  {journal} {Science}\ }\textbf {\bibinfo {volume}
  {336}},\ \bibinfo {pages} {1130} (\bibinfo {year} {2012})},\ \Eprint
  {https://arxiv.org/abs/1111.3633} {arXiv:1111.3633 [quant-ph]} \BibitemShut
  {NoStop}%
\bibitem [{\citenamefont {Di}\ and\ \citenamefont {Wei}(2011)}]{Di:2011cvl}%
  \BibitemOpen
  \bibfield  {author} {\bibinfo {author} {\bibfnamefont {Y.-M.}\ \bibnamefont
  {Di}}\ and\ \bibinfo {author} {\bibfnamefont {H.-R.}\ \bibnamefont {Wei}},\
  }\bibfield  {title} {\bibinfo {title} {{Elementary gates for ternary quantum
  logic circuit}},\ }\href@noop {} {\bibfield  {journal} {\bibinfo  {journal}
  {arXiv e-prints}\ } (\bibinfo {year} {2011})},\ \Eprint
  {https://arxiv.org/abs/1105.5485} {arXiv:1105.5485 [quant-ph]} \BibitemShut
  {NoStop}%
\bibitem [{\citenamefont {{Baker}}\ \emph {et~al.}(2019)\citenamefont
  {{Baker}}, \citenamefont {{Duckering}}, \citenamefont {{Hoover}},\ and\
  \citenamefont {{Chong}}}]{2019arXiv190401671B}%
  \BibitemOpen
  \bibfield  {author} {\bibinfo {author} {\bibfnamefont {J.~M.}\ \bibnamefont
  {{Baker}}}, \bibinfo {author} {\bibfnamefont {C.}~\bibnamefont
  {{Duckering}}}, \bibinfo {author} {\bibfnamefont {A.}~\bibnamefont
  {{Hoover}}},\ and\ \bibinfo {author} {\bibfnamefont {F.~T.}\ \bibnamefont
  {{Chong}}},\ }\bibfield  {title} {\bibinfo {title} {{Decomposing Quantum
  Generalized Toffoli with an Arbitrary Number of Ancilla}},\ }\href@noop {}
  {\bibfield  {journal} {\bibinfo  {journal} {arXiv e-prints}\ } (\bibinfo
  {year} {2019})},\ \Eprint {https://arxiv.org/abs/1904.01671}
  {arXiv:1904.01671 [quant-ph]} \BibitemShut {NoStop}%
\bibitem [{\citenamefont {Barenco}\ \emph {et~al.}(1995)\citenamefont
  {Barenco}, \citenamefont {Bennett}, \citenamefont {Cleve}, \citenamefont
  {DiVincenzo}, \citenamefont {Margolus}, \citenamefont {Shor}, \citenamefont
  {Sleator}, \citenamefont {Smolin},\ and\ \citenamefont
  {Weinfurter}}]{PhysRevA.52.3457}%
  \BibitemOpen
  \bibfield  {author} {\bibinfo {author} {\bibfnamefont {A.}~\bibnamefont
  {Barenco}}, \bibinfo {author} {\bibfnamefont {C.~H.}\ \bibnamefont
  {Bennett}}, \bibinfo {author} {\bibfnamefont {R.}~\bibnamefont {Cleve}},
  \bibinfo {author} {\bibfnamefont {D.~P.}\ \bibnamefont {DiVincenzo}},
  \bibinfo {author} {\bibfnamefont {N.}~\bibnamefont {Margolus}}, \bibinfo
  {author} {\bibfnamefont {P.}~\bibnamefont {Shor}}, \bibinfo {author}
  {\bibfnamefont {T.}~\bibnamefont {Sleator}}, \bibinfo {author} {\bibfnamefont
  {J.~A.}\ \bibnamefont {Smolin}},\ and\ \bibinfo {author} {\bibfnamefont
  {H.}~\bibnamefont {Weinfurter}},\ }\bibfield  {title} {\bibinfo {title}
  {Elementary gates for quantum computation},\ }\href
  {https://doi.org/10.1103/PhysRevA.52.3457} {\bibfield  {journal} {\bibinfo
  {journal} {Phys. Rev. A}\ }\textbf {\bibinfo {volume} {52}},\ \bibinfo
  {pages} {3457} (\bibinfo {year} {1995})}\BibitemShut {NoStop}%
\bibitem [{\citenamefont {Murairi}\ \emph {et~al.}(2024)\citenamefont
  {Murairi}, \citenamefont {Alam}, \citenamefont {Hadfield}, \citenamefont
  {Lamm},\ and\ \citenamefont {Gustafson}}]{ffttocome}%
  \BibitemOpen
  \bibfield  {author} {\bibinfo {author} {\bibfnamefont {E.}~\bibnamefont
  {Murairi}}, \bibinfo {author} {\bibfnamefont {S.}~\bibnamefont {Alam}},
  \bibinfo {author} {\bibfnamefont {S.}~\bibnamefont {Hadfield}}, \bibinfo
  {author} {\bibfnamefont {H.}~\bibnamefont {Lamm}},\ and\ \bibinfo {author}
  {\bibfnamefont {E.~J.}\ \bibnamefont {Gustafson}},\ }\bibfield  {title}
  {\bibinfo {title} {{Subgroup adapted quantum Fourier transforms for a set of
  nonabelian finite groups}},\ }\href@noop {} {\bibfield  {journal} {\bibinfo
  {journal} {in preparation}\ } (\bibinfo {year} {2024})}\BibitemShut {NoStop}%
\bibitem [{\citenamefont {Hoyer}(1997)}]{Hoyer:1997qc}%
  \BibitemOpen
  \bibfield  {author} {\bibinfo {author} {\bibfnamefont {P.}~\bibnamefont
  {Hoyer}},\ }\href@noop {} {\bibinfo {title} {{Efficient quantum transforms}}}
  (\bibinfo {year} {1997}),\ \Eprint {https://arxiv.org/abs/quant-ph/9702028}
  {arXiv:quant-ph/9702028} \BibitemShut {NoStop}%
\bibitem [{\citenamefont {Beals}(1997)}]{10.1145/258533.258548}%
  \BibitemOpen
  \bibfield  {author} {\bibinfo {author} {\bibfnamefont {R.}~\bibnamefont
  {Beals}},\ }\bibfield  {title} {\bibinfo {title} {Quantum computation of
  fourier transforms over symmetric groups},\ }in\ \href
  {https://doi.org/10.1145/258533.258548} {\emph {\bibinfo {booktitle}
  {Proceedings of the Twenty-Ninth Annual ACM Symposium on Theory of
  Computing}}},\ \bibinfo {series and number} {STOC '97}\ (\bibinfo {year}
  {1997})\ p.\ \bibinfo {pages} {48–53}\BibitemShut {NoStop}%
\bibitem [{\citenamefont {Kawano}\ and\ \citenamefont
  {Sekigawa}(2016)}]{KAWANO2016219}%
  \BibitemOpen
  \bibfield  {author} {\bibinfo {author} {\bibfnamefont {Y.}~\bibnamefont
  {Kawano}}\ and\ \bibinfo {author} {\bibfnamefont {H.}~\bibnamefont
  {Sekigawa}},\ }\bibfield  {title} {\bibinfo {title} {Quantum fourier
  transform over symmetric groups — improved result},\ }\href
  {https://doi.org/https://doi.org/10.1016/j.jsc.2015.11.016} {\bibfield
  {journal} {\bibinfo  {journal} {Journal of Symbolic Computation}\ }\textbf
  {\bibinfo {volume} {75}},\ \bibinfo {pages} {219} (\bibinfo {year}
  {2016})}\BibitemShut {NoStop}%
\bibitem [{\citenamefont {Moore}\ \emph {et~al.}(2006)\citenamefont {Moore},
  \citenamefont {Rockmore},\ and\ \citenamefont {Russell}}]{moore2006generic}%
  \BibitemOpen
  \bibfield  {author} {\bibinfo {author} {\bibfnamefont {C.}~\bibnamefont
  {Moore}}, \bibinfo {author} {\bibfnamefont {D.}~\bibnamefont {Rockmore}},\
  and\ \bibinfo {author} {\bibfnamefont {A.}~\bibnamefont {Russell}},\
  }\bibfield  {title} {\bibinfo {title} {Generic quantum {F}ourier
  transforms},\ }\href@noop {} {\bibfield  {journal} {\bibinfo  {journal} {ACM
  Transactions on Algorithms (TALG)}\ }\textbf {\bibinfo {volume} {2}},\
  \bibinfo {pages} {707} (\bibinfo {year} {2006})}\BibitemShut {NoStop}%
\bibitem [{\citenamefont {Murairi}\ \emph
  {et~al.}(2022{\natexlab{b}})\citenamefont {Murairi}, \citenamefont {Cervia},
  \citenamefont {Kumar}, \citenamefont {Bedaque},\ and\ \citenamefont
  {Alexandru}}]{Murairi:2022}%
  \BibitemOpen
  \bibfield  {author} {\bibinfo {author} {\bibfnamefont {E.~M.}\ \bibnamefont
  {Murairi}}, \bibinfo {author} {\bibfnamefont {M.~J.}\ \bibnamefont {Cervia}},
  \bibinfo {author} {\bibfnamefont {H.}~\bibnamefont {Kumar}}, \bibinfo
  {author} {\bibfnamefont {P.~F.}\ \bibnamefont {Bedaque}},\ and\ \bibinfo
  {author} {\bibfnamefont {A.}~\bibnamefont {Alexandru}},\ }\bibfield  {title}
  {\bibinfo {title} {How many quantum gates do gauge theories require?},\
  }\href {https://doi.org/10.1103/PhysRevD.106.094504} {\bibfield  {journal}
  {\bibinfo  {journal} {Phys. Rev. D}\ }\textbf {\bibinfo {volume} {106}},\
  \bibinfo {pages} {094504} (\bibinfo {year} {2022}{\natexlab{b}})}\BibitemShut
  {NoStop}%
\bibitem [{\citenamefont {Hadfield}(2021)}]{Hadfield_2021}%
  \BibitemOpen
  \bibfield  {author} {\bibinfo {author} {\bibfnamefont {S.}~\bibnamefont
  {Hadfield}},\ }\bibfield  {title} {\bibinfo {title} {On the representation of
  boolean and real functions as hamiltonians for quantum computing},\ }\href
  {https://doi.org/10.1145/3478519} {\bibfield  {journal} {\bibinfo  {journal}
  {ACM Transactions on Quantum Computing}\ }\textbf {\bibinfo {volume} {2}},\
  \bibinfo {pages} {1–21} (\bibinfo {year} {2021})}\BibitemShut {NoStop}%
\bibitem [{\citenamefont {Hirayama}\ \emph {et~al.}(2006)\citenamefont
  {Hirayama}, \citenamefont {Takahashi},\ and\ \citenamefont
  {Nishitani}}]{4145683}%
  \BibitemOpen
  \bibfield  {author} {\bibinfo {author} {\bibfnamefont {T.}~\bibnamefont
  {Hirayama}}, \bibinfo {author} {\bibfnamefont {M.}~\bibnamefont
  {Takahashi}},\ and\ \bibinfo {author} {\bibfnamefont {Y.}~\bibnamefont
  {Nishitani}},\ }\bibfield  {title} {\bibinfo {title} {Simplification of
  exclusive-or sum-of-products expressions through function transformation},\
  }in\ \href {https://doi.org/10.1109/APCCAS.2006.342502} {\emph {\bibinfo
  {booktitle} {APCCAS 2006 - 2006 IEEE Asia Pacific Conference on Circuits and
  Systems}}}\ (\bibinfo {year} {2006})\ pp.\ \bibinfo {pages}
  {1480--1483}\BibitemShut {NoStop}%
\bibitem [{\citenamefont {Stergiou}\ \emph {et~al.}(2004)\citenamefont
  {Stergiou}, \citenamefont {Daskalakis},\ and\ \citenamefont
  {Papakonstantinou}}]{10.1145/988952.988971}%
  \BibitemOpen
  \bibfield  {author} {\bibinfo {author} {\bibfnamefont {S.}~\bibnamefont
  {Stergiou}}, \bibinfo {author} {\bibfnamefont {K.}~\bibnamefont
  {Daskalakis}},\ and\ \bibinfo {author} {\bibfnamefont {G.}~\bibnamefont
  {Papakonstantinou}},\ }\bibfield  {title} {\bibinfo {title} {A fast and
  efficient heuristic esop minimization algorithm},\ }in\ \href
  {https://doi.org/10.1145/988952.988971} {\emph {\bibinfo {booktitle}
  {Proceedings of the 14th ACM Great Lakes Symposium on VLSI}}},\ \bibinfo
  {series and number} {GLSVLSI '04}\ (\bibinfo {year} {2004})\ p.\ \bibinfo
  {pages} {78–81}\BibitemShut {NoStop}%
\bibitem [{\citenamefont {Mishchenko}\ and\ \citenamefont
  {Perkowski}(2001)}]{mishchenko2001fast}%
  \BibitemOpen
  \bibfield  {author} {\bibinfo {author} {\bibfnamefont {A.}~\bibnamefont
  {Mishchenko}}\ and\ \bibinfo {author} {\bibfnamefont {M.}~\bibnamefont
  {Perkowski}},\ }\bibfield  {title} {\bibinfo {title} {Fast heuristic
  minimization of exclusive-sums-of-products},\ }in\ \href@noop {} {\emph
  {\bibinfo {booktitle} {5th International Workshop on Applications of the Reed
  Muller Expansion in Circuit Design}}}\ (\bibinfo {year} {2001})\BibitemShut
  {NoStop}%
\bibitem [{\citenamefont {Nielsen}\ and\ \citenamefont
  {Chuang}(2000)}]{nielsen2000quantum}%
  \BibitemOpen
  \bibfield  {author} {\bibinfo {author} {\bibfnamefont {M.~A.}\ \bibnamefont
  {Nielsen}}\ and\ \bibinfo {author} {\bibfnamefont {I.~L.}\ \bibnamefont
  {Chuang}},\ }\href {https://doi.org/10.1017/CBO9780511976667} {\emph
  {\bibinfo {title} {Quantum computation and quantum information}}}\ (\bibinfo
  {publisher} {Cambridge University Press, Cambridge},\ \bibinfo {year}
  {2000})\BibitemShut {NoStop}%
\bibitem [{\citenamefont {Childs}\ and\ \citenamefont
  {Van~Dam}(2010)}]{childs2010quantum}%
  \BibitemOpen
  \bibfield  {author} {\bibinfo {author} {\bibfnamefont {A.~M.}\ \bibnamefont
  {Childs}}\ and\ \bibinfo {author} {\bibfnamefont {W.}~\bibnamefont
  {Van~Dam}},\ }\bibfield  {title} {\bibinfo {title} {Quantum algorithms for
  algebraic problems},\ }\href@noop {} {\bibfield  {journal} {\bibinfo
  {journal} {Reviews of Modern Physics}\ }\textbf {\bibinfo {volume} {82}},\
  \bibinfo {pages} {1} (\bibinfo {year} {2010})}\BibitemShut {NoStop}%
\bibitem [{\citenamefont {Ji}\ \emph {et~al.}()\citenamefont {Ji},
  \citenamefont {Lamm}, \citenamefont {Murairi},\ and\ \citenamefont
  {Zhu}}]{hcompiler}%
  \BibitemOpen
  \bibfield  {author} {\bibinfo {author} {\bibfnamefont {Y.}~\bibnamefont
  {Ji}}, \bibinfo {author} {\bibfnamefont {H.}~\bibnamefont {Lamm}}, \bibinfo
  {author} {\bibfnamefont {E.~M.}\ \bibnamefont {Murairi}},\ and\ \bibinfo
  {author} {\bibfnamefont {S.}~\bibnamefont {Zhu}},\ }\href
  {https://github.com/emm71201/Qubit-Qutrit-Compiler} {\bibinfo {title}
  {Qubits-qutrits heterogeneous quantum circuit compiler}}\BibitemShut
  {NoStop}%
\bibitem [{\citenamefont {Eastin}\ and\ \citenamefont
  {Knill}(2009)}]{Eastin_2009}%
  \BibitemOpen
  \bibfield  {author} {\bibinfo {author} {\bibfnamefont {B.}~\bibnamefont
  {Eastin}}\ and\ \bibinfo {author} {\bibfnamefont {E.}~\bibnamefont {Knill}},\
  }\bibfield  {title} {\bibinfo {title} {Restrictions on transversal encoded
  quantum gate sets},\ }\href {https://doi.org/10.1103/physrevlett.102.110502}
  {\bibfield  {journal} {\bibinfo  {journal} {Physical Review Letters}\
  }\textbf {\bibinfo {volume} {102}},\ \bibinfo {pages} {110502} (\bibinfo
  {year} {2009})}\BibitemShut {NoStop}%
\bibitem [{\citenamefont {{Calderbank}}\ and\ \citenamefont
  {{Shor}}(1996)}]{1996PhRvA..54.1098C}%
  \BibitemOpen
  \bibfield  {author} {\bibinfo {author} {\bibfnamefont {A.~R.}\ \bibnamefont
  {{Calderbank}}}\ and\ \bibinfo {author} {\bibfnamefont {P.~W.}\ \bibnamefont
  {{Shor}}},\ }\bibfield  {title} {\bibinfo {title} {{Good quantum
  error-correcting codes exist}},\ }\href
  {https://doi.org/10.1103/PhysRevA.54.1098} {\bibfield  {journal} {\bibinfo
  {journal} {\pra}\ }\textbf {\bibinfo {volume} {54}},\ \bibinfo {pages} {1098}
  (\bibinfo {year} {1996})}\BibitemShut {NoStop}%
\bibitem [{\citenamefont {Steane}(1996)}]{PhysRevLett.77.793}%
  \BibitemOpen
  \bibfield  {author} {\bibinfo {author} {\bibfnamefont {A.~M.}\ \bibnamefont
  {Steane}},\ }\bibfield  {title} {\bibinfo {title} {Error correcting codes in
  quantum theory},\ }\href {https://doi.org/10.1103/PhysRevLett.77.793}
  {\bibfield  {journal} {\bibinfo  {journal} {Phys. Rev. Lett.}\ }\textbf
  {\bibinfo {volume} {77}},\ \bibinfo {pages} {793} (\bibinfo {year}
  {1996})}\BibitemShut {NoStop}%
\bibitem [{\citenamefont {{Steane}}(1996)}]{1996RSPSA.452.2551S}%
  \BibitemOpen
  \bibfield  {author} {\bibinfo {author} {\bibfnamefont {A.}~\bibnamefont
  {{Steane}}},\ }\bibfield  {title} {\bibinfo {title} {{Multiple-Particle
  Interference and Quantum Error Correction}},\ }\href
  {https://doi.org/10.1098/rspa.1996.0136} {\bibfield  {journal} {\bibinfo
  {journal} {Proceedings of the Royal Society of London Series A}\ }\textbf
  {\bibinfo {volume} {452}},\ \bibinfo {pages} {2551} (\bibinfo {year}
  {1996})},\ \Eprint {https://arxiv.org/abs/quant-ph/9601029}
  {arXiv:quant-ph/9601029 [quant-ph]} \BibitemShut {NoStop}%
\bibitem [{\citenamefont {Steane}(1996)}]{PhysRevA.54.4741}%
  \BibitemOpen
  \bibfield  {author} {\bibinfo {author} {\bibfnamefont {A.~M.}\ \bibnamefont
  {Steane}},\ }\bibfield  {title} {\bibinfo {title} {Simple quantum
  error-correcting codes},\ }\href {https://doi.org/10.1103/PhysRevA.54.4741}
  {\bibfield  {journal} {\bibinfo  {journal} {Phys. Rev. A}\ }\textbf {\bibinfo
  {volume} {54}},\ \bibinfo {pages} {4741} (\bibinfo {year}
  {1996})}\BibitemShut {NoStop}%
\bibitem [{\citenamefont {{Kitaev}}(1997)}]{1997RuMaS..52.1191K}%
  \BibitemOpen
  \bibfield  {author} {\bibinfo {author} {\bibfnamefont {A.~Y.}\ \bibnamefont
  {{Kitaev}}},\ }\bibfield  {title} {\bibinfo {title} {{Quantum computations:
  algorithms and error correction}},\ }\href
  {https://doi.org/10.1070/RM1997v052n06ABEH002155} {\bibfield  {journal}
  {\bibinfo  {journal} {Russian Mathematical Surveys}\ }\textbf {\bibinfo
  {volume} {52}},\ \bibinfo {pages} {1191} (\bibinfo {year}
  {1997})}\BibitemShut {NoStop}%
\bibitem [{\citenamefont {Kubischta}\ and\ \citenamefont
  {Teixeira}(2023)}]{Kubischta:2023nlb}%
  \BibitemOpen
  \bibfield  {author} {\bibinfo {author} {\bibfnamefont {E.}~\bibnamefont
  {Kubischta}}\ and\ \bibinfo {author} {\bibfnamefont {I.}~\bibnamefont
  {Teixeira}},\ }\href@noop {} {\bibinfo {title} {{A Family of Quantum Codes
  with Exotic Transversal Gates}}} (\bibinfo {year} {2023}),\ \Eprint
  {https://arxiv.org/abs/2305.07023} {arXiv:2305.07023 [quant-ph]} \BibitemShut
  {NoStop}%
\bibitem [{\citenamefont {Denys}\ and\ \citenamefont
  {Leverrier}(2023{\natexlab{a}})}]{Denys:2023syu}%
  \BibitemOpen
  \bibfield  {author} {\bibinfo {author} {\bibfnamefont {A.}~\bibnamefont
  {Denys}}\ and\ \bibinfo {author} {\bibfnamefont {A.}~\bibnamefont
  {Leverrier}},\ }\href@noop {} {\bibinfo {title} {{Multimode bosonic cat codes
  with an easily implementable universal gate set}}} (\bibinfo {year}
  {2023}{\natexlab{a}}),\ \Eprint {https://arxiv.org/abs/2306.11621}
  {arXiv:2306.11621 [quant-ph]} \BibitemShut {NoStop}%
\bibitem [{\citenamefont {Jain}\ \emph {et~al.}(2023)\citenamefont {Jain},
  \citenamefont {Iosue}, \citenamefont {Barg},\ and\ \citenamefont
  {Albert}}]{Jain:2023deu}%
  \BibitemOpen
  \bibfield  {author} {\bibinfo {author} {\bibfnamefont {S.~P.}\ \bibnamefont
  {Jain}}, \bibinfo {author} {\bibfnamefont {J.~T.}\ \bibnamefont {Iosue}},
  \bibinfo {author} {\bibfnamefont {A.}~\bibnamefont {Barg}},\ and\ \bibinfo
  {author} {\bibfnamefont {V.~V.}\ \bibnamefont {Albert}},\ }\href@noop {}
  {\bibinfo {title} {{Quantum spherical codes}}} (\bibinfo {year} {2023}),\
  \Eprint {https://arxiv.org/abs/2302.11593} {arXiv:2302.11593 [quant-ph]}
  \BibitemShut {NoStop}%
\bibitem [{\citenamefont {Denys}\ and\ \citenamefont
  {Leverrier}(2023{\natexlab{b}})}]{Denys:2022iyj}%
  \BibitemOpen
  \bibfield  {author} {\bibinfo {author} {\bibfnamefont {A.}~\bibnamefont
  {Denys}}\ and\ \bibinfo {author} {\bibfnamefont {A.}~\bibnamefont
  {Leverrier}},\ }\bibfield  {title} {\bibinfo {title} {{The $2T$-qutrit, a
  two-mode bosonic qutrit}},\ }\href
  {https://doi.org/10.22331/q-2023-06-05-1032} {\bibfield  {journal} {\bibinfo
  {journal} {Quantum}\ }\textbf {\bibinfo {volume} {7}},\ \bibinfo {pages}
  {1032} (\bibinfo {year} {2023}{\natexlab{b}})}\BibitemShut {NoStop}%
\bibitem [{\citenamefont {Herbert}\ \emph {et~al.}(2023)\citenamefont
  {Herbert}, \citenamefont {Gross},\ and\ \citenamefont
  {Newman}}]{Herbert:2023qgu}%
  \BibitemOpen
  \bibfield  {author} {\bibinfo {author} {\bibfnamefont {X.}~\bibnamefont
  {Herbert}}, \bibinfo {author} {\bibfnamefont {J.}~\bibnamefont {Gross}},\
  and\ \bibinfo {author} {\bibfnamefont {M.}~\bibnamefont {Newman}},\
  }\href@noop {} {\bibinfo {title} {{Qutrit codes within representations of
  SU(3)}}} (\bibinfo {year} {2023}),\ \Eprint
  {https://arxiv.org/abs/2312.00162} {arXiv:2312.00162 [quant-ph]} \BibitemShut
  {NoStop}%
\bibitem [{\citenamefont {Bocharov}\ \emph {et~al.}(2015)\citenamefont
  {Bocharov}, \citenamefont {Roetteler},\ and\ \citenamefont
  {Svore}}]{PhysRevLett.114.080502}%
  \BibitemOpen
  \bibfield  {author} {\bibinfo {author} {\bibfnamefont {A.}~\bibnamefont
  {Bocharov}}, \bibinfo {author} {\bibfnamefont {M.}~\bibnamefont
  {Roetteler}},\ and\ \bibinfo {author} {\bibfnamefont {K.~M.}\ \bibnamefont
  {Svore}},\ }\bibfield  {title} {\bibinfo {title} {Efficient synthesis of
  universal repeat-until-success quantum circuits},\ }\href
  {https://doi.org/10.1103/PhysRevLett.114.080502} {\bibfield  {journal}
  {\bibinfo  {journal} {Phys. Rev. Lett.}\ }\textbf {\bibinfo {volume} {114}},\
  \bibinfo {pages} {080502} (\bibinfo {year} {2015})}\BibitemShut {NoStop}%
\bibitem [{\citenamefont {Selinger}(2015)}]{10.5555/2685188.2685198}%
  \BibitemOpen
  \bibfield  {author} {\bibinfo {author} {\bibfnamefont {P.}~\bibnamefont
  {Selinger}},\ }\bibfield  {title} {\bibinfo {title} {Efficient clifford+t
  approximation of single-qubit operators},\ }\href@noop {} {\bibfield
  {journal} {\bibinfo  {journal} {Quantum Info. Comput.}\ }\textbf {\bibinfo
  {volume} {15}},\ \bibinfo {pages} {159–180} (\bibinfo {year}
  {2015})}\BibitemShut {NoStop}%
\bibitem [{\citenamefont {Kan}\ and\ \citenamefont {Nam}(2021)}]{Kan:2021xfc}%
  \BibitemOpen
  \bibfield  {author} {\bibinfo {author} {\bibfnamefont {A.}~\bibnamefont
  {Kan}}\ and\ \bibinfo {author} {\bibfnamefont {Y.}~\bibnamefont {Nam}},\
  }\href@noop {} {\bibinfo {title} {{Lattice Quantum Chromodynamics and
  Electrodynamics on a Universal Quantum Computer}}} (\bibinfo {year} {2021}),\
  \Eprint {https://arxiv.org/abs/2107.12769} {arXiv:2107.12769 [quant-ph]}
  \BibitemShut {NoStop}%
\bibitem [{\citenamefont {Shaw}\ \emph {et~al.}(2020)\citenamefont {Shaw},
  \citenamefont {Lougovski}, \citenamefont {Stryker},\ and\ \citenamefont
  {Wiebe}}]{Shaw:2020udc}%
  \BibitemOpen
  \bibfield  {author} {\bibinfo {author} {\bibfnamefont {A.~F.}\ \bibnamefont
  {Shaw}}, \bibinfo {author} {\bibfnamefont {P.}~\bibnamefont {Lougovski}},
  \bibinfo {author} {\bibfnamefont {J.~R.}\ \bibnamefont {Stryker}},\ and\
  \bibinfo {author} {\bibfnamefont {N.}~\bibnamefont {Wiebe}},\ }\bibfield
  {title} {\bibinfo {title} {{Quantum Algorithms for Simulating the Lattice
  Schwinger Model}},\ }\href {https://doi.org/10.22331/q-2020-08-10-306}
  {\bibfield  {journal} {\bibinfo  {journal} {Quantum}\ }\textbf {\bibinfo
  {volume} {4}},\ \bibinfo {pages} {306} (\bibinfo {year} {2020})}\BibitemShut
  {NoStop}%
\bibitem [{\citenamefont {Kogut}\ and\ \citenamefont
  {Susskind}(1975)}]{PhysRevD.11.395}%
  \BibitemOpen
  \bibfield  {author} {\bibinfo {author} {\bibfnamefont {J.}~\bibnamefont
  {Kogut}}\ and\ \bibinfo {author} {\bibfnamefont {L.}~\bibnamefont
  {Susskind}},\ }\bibfield  {title} {\bibinfo {title} {Hamiltonian formulation
  of {W}ilson's lattice gauge theories},\ }\href
  {https://doi.org/10.1103/PhysRevD.11.395} {\bibfield  {journal} {\bibinfo
  {journal} {Phys. Rev. D}\ }\textbf {\bibinfo {volume} {11}},\ \bibinfo
  {pages} {395} (\bibinfo {year} {1975})}\BibitemShut {NoStop}%
\bibitem [{\citenamefont {Cohen}\ \emph {et~al.}(2021)\citenamefont {Cohen},
  \citenamefont {Lamm}, \citenamefont {Lawrence},\ and\ \citenamefont
  {Yamauchi}}]{Cohen:2021imf}%
  \BibitemOpen
  \bibfield  {author} {\bibinfo {author} {\bibfnamefont {T.~D.}\ \bibnamefont
  {Cohen}}, \bibinfo {author} {\bibfnamefont {H.}~\bibnamefont {Lamm}},
  \bibinfo {author} {\bibfnamefont {S.}~\bibnamefont {Lawrence}},\ and\
  \bibinfo {author} {\bibfnamefont {Y.}~\bibnamefont {Yamauchi}} (\bibinfo
  {collaboration} {NuQS}),\ }\bibfield  {title} {\bibinfo {title} {{Quantum
  algorithms for transport coefficients in gauge theories}},\ }\href
  {https://doi.org/10.1103/PhysRevD.104.094514} {\bibfield  {journal} {\bibinfo
   {journal} {Phys. Rev. D}\ }\textbf {\bibinfo {volume} {104}},\ \bibinfo
  {pages} {094514} (\bibinfo {year} {2021})}\BibitemShut {NoStop}%
\bibitem [{\citenamefont {Xie}\ \emph {et~al.}(2022)\citenamefont {Xie},
  \citenamefont {Guo}, \citenamefont {Xing}, \citenamefont {Xue}, \citenamefont
  {Zhang},\ and\ \citenamefont {Zhu}}]{Xie:2022jgj}%
  \BibitemOpen
  \bibfield  {author} {\bibinfo {author} {\bibfnamefont {X.-D.}\ \bibnamefont
  {Xie}}, \bibinfo {author} {\bibfnamefont {X.}~\bibnamefont {Guo}}, \bibinfo
  {author} {\bibfnamefont {H.}~\bibnamefont {Xing}}, \bibinfo {author}
  {\bibfnamefont {Z.-Y.}\ \bibnamefont {Xue}}, \bibinfo {author} {\bibfnamefont
  {D.-B.}\ \bibnamefont {Zhang}},\ and\ \bibinfo {author} {\bibfnamefont
  {S.-L.}\ \bibnamefont {Zhu}} (\bibinfo {collaboration} {QuNu}),\ }\bibfield
  {title} {\bibinfo {title} {{Variational thermal quantum simulation of the
  lattice Schwinger model}},\ }\href
  {https://doi.org/10.1103/PhysRevD.106.054509} {\bibfield  {journal} {\bibinfo
   {journal} {Phys. Rev. D}\ }\textbf {\bibinfo {volume} {106}},\ \bibinfo
  {pages} {054509} (\bibinfo {year} {2022})},\ \Eprint
  {https://arxiv.org/abs/2205.12767} {arXiv:2205.12767 [quant-ph]} \BibitemShut
  {NoStop}%
\bibitem [{\citenamefont {Davoudi}\ \emph {et~al.}(2023)\citenamefont
  {Davoudi}, \citenamefont {Mueller},\ and\ \citenamefont
  {Powers}}]{Davoudi:2022uzo}%
  \BibitemOpen
  \bibfield  {author} {\bibinfo {author} {\bibfnamefont {Z.}~\bibnamefont
  {Davoudi}}, \bibinfo {author} {\bibfnamefont {N.}~\bibnamefont {Mueller}},\
  and\ \bibinfo {author} {\bibfnamefont {C.}~\bibnamefont {Powers}},\
  }\bibfield  {title} {\bibinfo {title} {{Towards Quantum Computing Phase
  Diagrams of Gauge Theories with Thermal Pure Quantum States}},\ }\href
  {https://doi.org/10.1103/PhysRevLett.131.081901} {\bibfield  {journal}
  {\bibinfo  {journal} {Phys. Rev. Lett.}\ }\textbf {\bibinfo {volume} {131}},\
  \bibinfo {pages} {081901} (\bibinfo {year} {2023})}\BibitemShut {NoStop}%
\bibitem [{\citenamefont {Avkhadiev}\ \emph {et~al.}(2020)\citenamefont
  {Avkhadiev}, \citenamefont {Shanahan},\ and\ \citenamefont
  {Young}}]{Avkhadiev:2019niu}%
  \BibitemOpen
  \bibfield  {author} {\bibinfo {author} {\bibfnamefont {A.}~\bibnamefont
  {Avkhadiev}}, \bibinfo {author} {\bibfnamefont {P.~E.}\ \bibnamefont
  {Shanahan}},\ and\ \bibinfo {author} {\bibfnamefont {R.~D.}\ \bibnamefont
  {Young}},\ }\bibfield  {title} {\bibinfo {title} {{Accelerating Lattice
  Quantum Field Theory Calculations via Interpolator Optimization Using Noisy
  Intermediate-Scale Quantum Computing}},\ }\href
  {https://doi.org/10.1103/PhysRevLett.124.080501} {\bibfield  {journal}
  {\bibinfo  {journal} {Phys. Rev. Lett.}\ }\textbf {\bibinfo {volume} {124}},\
  \bibinfo {pages} {080501} (\bibinfo {year} {2020})}\BibitemShut {NoStop}%
\bibitem [{\citenamefont {Gustafson}(2022)}]{Gustafson:2022hjf}%
  \BibitemOpen
  \bibfield  {author} {\bibinfo {author} {\bibfnamefont {E.~J.}\ \bibnamefont
  {Gustafson}},\ }\href@noop {} {\bibinfo {title} {{Stout Smearing on a Quantum
  Computer}}} (\bibinfo {year} {2022}),\ \Eprint
  {https://arxiv.org/abs/2211.05607} {arXiv:2211.05607 [hep-lat]} \BibitemShut
  {NoStop}%
\bibitem [{\citenamefont {Peruzzo}\ \emph {et~al.}(2014)\citenamefont
  {Peruzzo}, \citenamefont {McClean}, \citenamefont {Shadbolt}, \citenamefont
  {Yung}, \citenamefont {Zhou}, \citenamefont {Love}, \citenamefont
  {Aspuru-Guzik},\ and\ \citenamefont {O’brien}}]{peruzzo2014variational}%
  \BibitemOpen
  \bibfield  {author} {\bibinfo {author} {\bibfnamefont {A.}~\bibnamefont
  {Peruzzo}}, \bibinfo {author} {\bibfnamefont {J.}~\bibnamefont {McClean}},
  \bibinfo {author} {\bibfnamefont {P.}~\bibnamefont {Shadbolt}}, \bibinfo
  {author} {\bibfnamefont {M.-H.}\ \bibnamefont {Yung}}, \bibinfo {author}
  {\bibfnamefont {X.-Q.}\ \bibnamefont {Zhou}}, \bibinfo {author}
  {\bibfnamefont {P.~J.}\ \bibnamefont {Love}}, \bibinfo {author}
  {\bibfnamefont {A.}~\bibnamefont {Aspuru-Guzik}},\ and\ \bibinfo {author}
  {\bibfnamefont {J.~L.}\ \bibnamefont {O’brien}},\ }\bibfield  {title}
  {\bibinfo {title} {A variational eigenvalue solver on a photonic quantum
  processor},\ }\href@noop {} {\bibfield  {journal} {\bibinfo  {journal}
  {Nature communications}\ }\textbf {\bibinfo {volume} {5}},\ \bibinfo {pages}
  {4213} (\bibinfo {year} {2014})}\BibitemShut {NoStop}%
\bibitem [{\citenamefont {Gustafson}\ \emph
  {et~al.}(2019{\natexlab{a}})\citenamefont {Gustafson}, \citenamefont
  {Meurice},\ and\ \citenamefont {Unmuth-Yockey}}]{Gustafson:2019mpk}%
  \BibitemOpen
  \bibfield  {author} {\bibinfo {author} {\bibfnamefont {E.}~\bibnamefont
  {Gustafson}}, \bibinfo {author} {\bibfnamefont {Y.}~\bibnamefont {Meurice}},\
  and\ \bibinfo {author} {\bibfnamefont {J.}~\bibnamefont {Unmuth-Yockey}},\
  }\href@noop {} {\bibinfo {title} {{Quantum simulation of scattering in the
  quantum Ising model}}} (\bibinfo {year} {2019}{\natexlab{a}}),\ \Eprint
  {https://arxiv.org/abs/1901.05944} {arXiv:1901.05944 [hep-lat]} \BibitemShut
  {NoStop}%
\bibitem [{\citenamefont {Gustafson}\ \emph
  {et~al.}(2019{\natexlab{b}})\citenamefont {Gustafson}, \citenamefont
  {Dreher}, \citenamefont {Hang},\ and\ \citenamefont
  {Meurice}}]{Gustafson:2019vsd}%
  \BibitemOpen
  \bibfield  {author} {\bibinfo {author} {\bibfnamefont {E.}~\bibnamefont
  {Gustafson}}, \bibinfo {author} {\bibfnamefont {P.}~\bibnamefont {Dreher}},
  \bibinfo {author} {\bibfnamefont {Z.}~\bibnamefont {Hang}},\ and\ \bibinfo
  {author} {\bibfnamefont {Y.}~\bibnamefont {Meurice}},\ }\href@noop {}
  {\bibinfo {title} {Real time evolution of a one-dimensional field theory on a
  20 qubit machine}} (\bibinfo {year} {2019}{\natexlab{b}}),\ \Eprint
  {https://arxiv.org/abs/1910.09478} {arXiv:1910.09478 [hep-lat]} \BibitemShut
  {NoStop}%
\bibitem [{\citenamefont {Harmalkar}\ \emph {et~al.}(2020)\citenamefont
  {Harmalkar}, \citenamefont {Lamm},\ and\ \citenamefont
  {Lawrence}}]{Harmalkar:2020mpd}%
  \BibitemOpen
  \bibfield  {author} {\bibinfo {author} {\bibfnamefont {S.}~\bibnamefont
  {Harmalkar}}, \bibinfo {author} {\bibfnamefont {H.}~\bibnamefont {Lamm}},\
  and\ \bibinfo {author} {\bibfnamefont {S.}~\bibnamefont {Lawrence}} (\bibinfo
  {collaboration} {NuQS}),\ }\href@noop {} {\bibinfo {title} {{Quantum
  Simulation of Field Theories Without State Preparation}}} (\bibinfo {year}
  {2020}),\ \Eprint {https://arxiv.org/abs/2001.11490} {arXiv:2001.11490
  [hep-lat]} \BibitemShut {NoStop}%
\bibitem [{\citenamefont {Jordan}\ \emph {et~al.}(2018)\citenamefont {Jordan},
  \citenamefont {Krovi}, \citenamefont {Lee},\ and\ \citenamefont
  {Preskill}}]{Jordan:2017lea}%
  \BibitemOpen
  \bibfield  {author} {\bibinfo {author} {\bibfnamefont {S.~P.}\ \bibnamefont
  {Jordan}}, \bibinfo {author} {\bibfnamefont {H.}~\bibnamefont {Krovi}},
  \bibinfo {author} {\bibfnamefont {K.~S.}\ \bibnamefont {Lee}},\ and\ \bibinfo
  {author} {\bibfnamefont {J.}~\bibnamefont {Preskill}},\ }\bibfield  {title}
  {\bibinfo {title} {{BQP-completeness of Scattering in Scalar Quantum Field
  Theory}},\ }\href {https://doi.org/10.22331/q-2018-01-08-44} {\bibfield
  {journal} {\bibinfo  {journal} {Quantum}\ }\textbf {\bibinfo {volume} {2}},\
  \bibinfo {pages} {44} (\bibinfo {year} {2018})}\BibitemShut {NoStop}%
\bibitem [{\citenamefont {Riera}\ \emph {et~al.}(2012)\citenamefont {Riera},
  \citenamefont {Gogolin},\ and\ \citenamefont
  {Eisert}}]{PhysRevLett.108.080402}%
  \BibitemOpen
  \bibfield  {author} {\bibinfo {author} {\bibfnamefont {A.}~\bibnamefont
  {Riera}}, \bibinfo {author} {\bibfnamefont {C.}~\bibnamefont {Gogolin}},\
  and\ \bibinfo {author} {\bibfnamefont {J.}~\bibnamefont {Eisert}},\
  }\bibfield  {title} {\bibinfo {title} {Thermalization in nature and on a
  quantum computer},\ }\href {https://doi.org/10.1103/PhysRevLett.108.080402}
  {\bibfield  {journal} {\bibinfo  {journal} {Phys. Rev. Lett.}\ }\textbf
  {\bibinfo {volume} {108}},\ \bibinfo {pages} {080402} (\bibinfo {year}
  {2012})}\BibitemShut {NoStop}%
\bibitem [{\citenamefont {Motta}\ \emph {et~al.}(2020)\citenamefont {Motta},
  \citenamefont {Sun}, \citenamefont {Tan}, \citenamefont {O’Rourke},
  \citenamefont {Ye}, \citenamefont {Minnich}, \citenamefont {Brandao},\ and\
  \citenamefont {Chan}}]{motta2020determining}%
  \BibitemOpen
  \bibfield  {author} {\bibinfo {author} {\bibfnamefont {M.}~\bibnamefont
  {Motta}}, \bibinfo {author} {\bibfnamefont {C.}~\bibnamefont {Sun}}, \bibinfo
  {author} {\bibfnamefont {A.~T.}\ \bibnamefont {Tan}}, \bibinfo {author}
  {\bibfnamefont {M.~J.}\ \bibnamefont {O’Rourke}}, \bibinfo {author}
  {\bibfnamefont {E.}~\bibnamefont {Ye}}, \bibinfo {author} {\bibfnamefont
  {A.~J.}\ \bibnamefont {Minnich}}, \bibinfo {author} {\bibfnamefont {F.~G.}\
  \bibnamefont {Brandao}},\ and\ \bibinfo {author} {\bibfnamefont {G.~K.-L.}\
  \bibnamefont {Chan}},\ }\bibfield  {title} {\bibinfo {title} {Determining
  eigenstates and thermal states on a quantum computer using quantum imaginary
  time evolution},\ }\href@noop {} {\bibfield  {journal} {\bibinfo  {journal}
  {Nature Physics}\ }\textbf {\bibinfo {volume} {16}},\ \bibinfo {pages} {205}
  (\bibinfo {year} {2020})}\BibitemShut {NoStop}%
\bibitem [{\citenamefont {Clemente}\ \emph {et~al.}(2020)\citenamefont
  {Clemente} \emph {et~al.}}]{Clemente:2020lpr}%
  \BibitemOpen
  \bibfield  {author} {\bibinfo {author} {\bibfnamefont {G.}~\bibnamefont
  {Clemente}} \emph {et~al.} (\bibinfo {collaboration} {QuBiPF}),\ }\bibfield
  {title} {\bibinfo {title} {{Quantum computation of thermal averages in the
  presence of a sign problem}},\ }\href
  {https://doi.org/10.1103/PhysRevD.101.074510} {\bibfield  {journal} {\bibinfo
   {journal} {Phys. Rev. D}\ }\textbf {\bibinfo {volume} {101}},\ \bibinfo
  {pages} {074510} (\bibinfo {year} {2020})},\ \Eprint
  {https://arxiv.org/abs/2001.05328} {arXiv:2001.05328 [hep-lat]} \BibitemShut
  {NoStop}%
\bibitem [{\citenamefont {de~Jong}\ \emph {et~al.}(2021)\citenamefont
  {de~Jong}, \citenamefont {Lee}, \citenamefont {Mulligan}, \citenamefont
  {P\l{}osko\'n}, \citenamefont {Ringer},\ and\ \citenamefont
  {Yao}}]{deJong:2021wsd}%
  \BibitemOpen
  \bibfield  {author} {\bibinfo {author} {\bibfnamefont {W.~A.}\ \bibnamefont
  {de~Jong}}, \bibinfo {author} {\bibfnamefont {K.}~\bibnamefont {Lee}},
  \bibinfo {author} {\bibfnamefont {J.}~\bibnamefont {Mulligan}}, \bibinfo
  {author} {\bibfnamefont {M.}~\bibnamefont {P\l{}osko\'n}}, \bibinfo {author}
  {\bibfnamefont {F.}~\bibnamefont {Ringer}},\ and\ \bibinfo {author}
  {\bibfnamefont {X.}~\bibnamefont {Yao}},\ }\href@noop {} {\bibinfo {title}
  {{Quantum simulation of non-equilibrium dynamics and thermalization in the
  Schwinger model}}} (\bibinfo {year} {2021}),\ \Eprint
  {https://arxiv.org/abs/2106.08394} {arXiv:2106.08394 [quant-ph]} \BibitemShut
  {NoStop}%
\bibitem [{\citenamefont {Gustafson}\ \emph {et~al.}(2023)\citenamefont
  {Gustafson}, \citenamefont {Lamm},\ and\ \citenamefont
  {Unmuth-Yockey}}]{Gustafson:2023ayr}%
  \BibitemOpen
  \bibfield  {author} {\bibinfo {author} {\bibfnamefont {E.~J.}\ \bibnamefont
  {Gustafson}}, \bibinfo {author} {\bibfnamefont {H.}~\bibnamefont {Lamm}},\
  and\ \bibinfo {author} {\bibfnamefont {J.}~\bibnamefont {Unmuth-Yockey}},\
  }\bibfield  {title} {\bibinfo {title} {{Quantum mean estimation for lattice
  field theory}},\ }\href {https://doi.org/10.1103/PhysRevD.107.114511}
  {\bibfield  {journal} {\bibinfo  {journal} {Phys. Rev. D}\ }\textbf {\bibinfo
  {volume} {107}},\ \bibinfo {pages} {114511} (\bibinfo {year} {2023})},\
  \Eprint {https://arxiv.org/abs/2303.00094} {arXiv:2303.00094 [hep-lat]}
  \BibitemShut {NoStop}%
\bibitem [{\citenamefont {Kane}\ \emph {et~al.}(2023)\citenamefont {Kane},
  \citenamefont {Gomes},\ and\ \citenamefont {Kreshchuk}}]{Kane:2023jdo}%
  \BibitemOpen
  \bibfield  {author} {\bibinfo {author} {\bibfnamefont {C.~F.}\ \bibnamefont
  {Kane}}, \bibinfo {author} {\bibfnamefont {N.}~\bibnamefont {Gomes}},\ and\
  \bibinfo {author} {\bibfnamefont {M.}~\bibnamefont {Kreshchuk}},\ }\href@noop
  {} {\bibinfo {title} {{Nearly-optimal state preparation for quantum
  simulations of lattice gauge theories}}} (\bibinfo {year} {2023}),\ \Eprint
  {https://arxiv.org/abs/2310.13757} {arXiv:2310.13757 [quant-ph]} \BibitemShut
  {NoStop}%
\bibitem [{\citenamefont {Florio}\ \emph {et~al.}(2023)\citenamefont {Florio},
  \citenamefont {Weichselbaum}, \citenamefont {Valgushev},\ and\ \citenamefont
  {Pisarski}}]{Florio:2023kel}%
  \BibitemOpen
  \bibfield  {author} {\bibinfo {author} {\bibfnamefont {A.}~\bibnamefont
  {Florio}}, \bibinfo {author} {\bibfnamefont {A.}~\bibnamefont
  {Weichselbaum}}, \bibinfo {author} {\bibfnamefont {S.}~\bibnamefont
  {Valgushev}},\ and\ \bibinfo {author} {\bibfnamefont {R.~D.}\ \bibnamefont
  {Pisarski}},\ }\href@noop {} {\bibinfo {title} {{Mass gaps of a
  $\mathbb{Z}_3$ gauge theory with three fermion flavors in 1 + 1 dimensions}}}
  (\bibinfo {year} {2023}),\ \Eprint {https://arxiv.org/abs/2310.18312}
  {arXiv:2310.18312 [hep-th]} \BibitemShut {NoStop}%
\bibitem [{\citenamefont {Emonts}\ and\ \citenamefont
  {Zohar}(2018)}]{Zohar:2018nvl}%
  \BibitemOpen
  \bibfield  {author} {\bibinfo {author} {\bibfnamefont {P.}~\bibnamefont
  {Emonts}}\ and\ \bibinfo {author} {\bibfnamefont {E.}~\bibnamefont {Zohar}},\
  }\bibfield  {title} {\bibinfo {title} {{Gauss law, Minimal Coupling and
  Fermionic PEPS for Lattice Gauge Theories}},\ }in\ \href
  {https://doi.org/10.21468/SciPostPhysLectNotes.12} {\emph {\bibinfo
  {booktitle} {{Tensor Network and entanglement}}}}\ (\bibinfo {year} {2018})\
  \Eprint {https://arxiv.org/abs/1807.01294} {arXiv:1807.01294 [quant-ph]}
  \BibitemShut {NoStop}%
\bibitem [{\citenamefont {Gui}\ \emph {et~al.}(2024)\citenamefont {Gui},
  \citenamefont {Dalzell}, \citenamefont {Achille}, \citenamefont {Suchara},\
  and\ \citenamefont {Chong}}]{Gui:2023coj}%
  \BibitemOpen
  \bibfield  {author} {\bibinfo {author} {\bibfnamefont {K.}~\bibnamefont
  {Gui}}, \bibinfo {author} {\bibfnamefont {A.~M.}\ \bibnamefont {Dalzell}},
  \bibinfo {author} {\bibfnamefont {A.}~\bibnamefont {Achille}}, \bibinfo
  {author} {\bibfnamefont {M.}~\bibnamefont {Suchara}},\ and\ \bibinfo {author}
  {\bibfnamefont {F.~T.}\ \bibnamefont {Chong}},\ }\bibfield  {title} {\bibinfo
  {title} {{Spacetime-Efficient Low-Depth Quantum State Preparation with
  Applications}},\ }\href {https://doi.org/10.22331/q-2024-02-15-1257}
  {\bibfield  {journal} {\bibinfo  {journal} {Quantum}\ }\textbf {\bibinfo
  {volume} {8}},\ \bibinfo {pages} {1257} (\bibinfo {year} {2024})},\ \Eprint
  {https://arxiv.org/abs/2303.02131} {arXiv:2303.02131 [quant-ph]} \BibitemShut
  {NoStop}%
\bibitem [{\citenamefont {Shende}\ \emph {et~al.}(2006)\citenamefont {Shende},
  \citenamefont {Bullock},\ and\ \citenamefont {Markov}}]{Shende2006}%
  \BibitemOpen
  \bibfield  {author} {\bibinfo {author} {\bibfnamefont {V.}~\bibnamefont
  {Shende}}, \bibinfo {author} {\bibfnamefont {S.}~\bibnamefont {Bullock}},\
  and\ \bibinfo {author} {\bibfnamefont {I.}~\bibnamefont {Markov}},\
  }\bibfield  {title} {\bibinfo {title} {Synthesis of quantum-logic circuits},\
  }\href {https://doi.org/10.1109/tcad.2005.855930} {\bibfield  {journal}
  {\bibinfo  {journal} {IEEE Transactions on Computer-Aided Design of
  Integrated Circuits and Systems}\ }\textbf {\bibinfo {volume} {25}},\
  \bibinfo {pages} {1000–1010} (\bibinfo {year} {2006})}\BibitemShut
  {NoStop}%
\bibitem [{\citenamefont {Drury}\ and\ \citenamefont {Love}(2008)}]{Drury2008}%
  \BibitemOpen
  \bibfield  {author} {\bibinfo {author} {\bibfnamefont {B.}~\bibnamefont
  {Drury}}\ and\ \bibinfo {author} {\bibfnamefont {P.}~\bibnamefont {Love}},\
  }\bibfield  {title} {\bibinfo {title} {Constructive quantum shannon
  decomposition from cartan involutions},\ }\href
  {https://doi.org/10.1088/1751-8113/41/39/395305} {\bibfield  {journal}
  {\bibinfo  {journal} {Journal of Physics A: Mathematical and Theoretical}\
  }\textbf {\bibinfo {volume} {41}},\ \bibinfo {pages} {395305} (\bibinfo
  {year} {2008})}\BibitemShut {NoStop}%
\bibitem [{\citenamefont {Paige}\ and\ \citenamefont {Wei}(1994)}]{Paige1994}%
  \BibitemOpen
  \bibfield  {author} {\bibinfo {author} {\bibfnamefont {C.}~\bibnamefont
  {Paige}}\ and\ \bibinfo {author} {\bibfnamefont {M.}~\bibnamefont {Wei}},\
  }\bibfield  {title} {\bibinfo {title} {History and generality of the cs
  decomposition},\ }\href {https://doi.org/10.1016/0024-3795(94)90446-4}
  {\bibfield  {journal} {\bibinfo  {journal} {Linear Algebra and its
  Applications}\ }\textbf {\bibinfo {volume} {208–209}},\ \bibinfo {pages}
  {303–326} (\bibinfo {year} {1994})}\BibitemShut {NoStop}%
\bibitem [{\citenamefont {Chen}\ and\ \citenamefont {Wang}(2013)}]{Chen2013}%
  \BibitemOpen
  \bibfield  {author} {\bibinfo {author} {\bibfnamefont {Y.}~\bibnamefont
  {Chen}}\ and\ \bibinfo {author} {\bibfnamefont {J.}~\bibnamefont {Wang}},\
  }\bibfield  {title} {\bibinfo {title} {Qcompiler: Quantum compilation with
  the csd method},\ }\href {https://doi.org/10.1016/j.cpc.2012.10.019}
  {\bibfield  {journal} {\bibinfo  {journal} {Computer Physics Communications}\
  }\textbf {\bibinfo {volume} {184}},\ \bibinfo {pages} {853–865} (\bibinfo
  {year} {2013})}\BibitemShut {NoStop}%
\bibitem [{\citenamefont {Di}\ and\ \citenamefont
  {Wei}(2015)}]{PhysRevA.92.062317}%
  \BibitemOpen
  \bibfield  {author} {\bibinfo {author} {\bibfnamefont {Y.-M.}\ \bibnamefont
  {Di}}\ and\ \bibinfo {author} {\bibfnamefont {H.-R.}\ \bibnamefont {Wei}},\
  }\bibfield  {title} {\bibinfo {title} {Optimal synthesis of multivalued
  quantum circuits},\ }\href {https://doi.org/10.1103/PhysRevA.92.062317}
  {\bibfield  {journal} {\bibinfo  {journal} {Phys. Rev. A}\ }\textbf {\bibinfo
  {volume} {92}},\ \bibinfo {pages} {062317} (\bibinfo {year}
  {2015})}\BibitemShut {NoStop}%
\bibitem [{\citenamefont {Khan}\ and\ \citenamefont
  {Perkowski}(2006)}]{Khan2006}%
  \BibitemOpen
  \bibfield  {author} {\bibinfo {author} {\bibfnamefont {F.~S.}\ \bibnamefont
  {Khan}}\ and\ \bibinfo {author} {\bibfnamefont {M.}~\bibnamefont
  {Perkowski}},\ }\bibfield  {title} {\bibinfo {title} {Synthesis of
  multi-qudit hybrid and d-valued quantum logic circuits by decomposition},\
  }\href {https://doi.org/10.1016/j.tcs.2006.09.006} {\bibfield  {journal}
  {\bibinfo  {journal} {Theoretical Computer Science}\ }\textbf {\bibinfo
  {volume} {367}},\ \bibinfo {pages} {336–346} (\bibinfo {year}
  {2006})}\BibitemShut {NoStop}%
\bibitem [{\citenamefont {Di}\ and\ \citenamefont
  {Wei}(2013)}]{PhysRevA.87.012325}%
  \BibitemOpen
  \bibfield  {author} {\bibinfo {author} {\bibfnamefont {Y.-M.}\ \bibnamefont
  {Di}}\ and\ \bibinfo {author} {\bibfnamefont {H.-R.}\ \bibnamefont {Wei}},\
  }\bibfield  {title} {\bibinfo {title} {Synthesis of multivalued quantum logic
  circuits by elementary gates},\ }\href
  {https://doi.org/10.1103/PhysRevA.87.012325} {\bibfield  {journal} {\bibinfo
  {journal} {Phys. Rev. A}\ }\textbf {\bibinfo {volume} {87}},\ \bibinfo
  {pages} {012325} (\bibinfo {year} {2013})}\BibitemShut {NoStop}%
\bibitem [{\citenamefont {Vatan}\ and\ \citenamefont
  {Williams}(2004)}]{PhysRevA.69.032315}%
  \BibitemOpen
  \bibfield  {author} {\bibinfo {author} {\bibfnamefont {F.}~\bibnamefont
  {Vatan}}\ and\ \bibinfo {author} {\bibfnamefont {C.}~\bibnamefont
  {Williams}},\ }\bibfield  {title} {\bibinfo {title} {Optimal quantum circuits
  for general two-qubit gates},\ }\href
  {https://doi.org/10.1103/PhysRevA.69.032315} {\bibfield  {journal} {\bibinfo
  {journal} {Phys. Rev. A}\ }\textbf {\bibinfo {volume} {69}},\ \bibinfo
  {pages} {032315} (\bibinfo {year} {2004})}\BibitemShut {NoStop}%
\end{thebibliography}%

\end{document}